\documentclass[11pt]{article}

\usepackage[T1]{fontenc}
\usepackage[utf8]{inputenc}
\usepackage{lmodern}
\usepackage{graphicx}
\usepackage{subcaption}
\usepackage{amsmath,amssymb,amsfonts,amsthm}
\usepackage{booktabs}
\usepackage{dsfont}
\usepackage{color}
\usepackage{hyperref}
\usepackage{tabularx}
\usepackage{etoolbox}
\usepackage{latexsym}
\usepackage{makecell}
\usepackage{chngcntr}
\usepackage{cancel}

\counterwithin*{equation}{section}

\usepackage{algorithm}
\usepackage{algorithmic}

\usepackage{tikz}

\usepackage[backend=biber]{biblatex}
\newtheorem{Theorem}{Theorem}[part]
\newtheorem{Definition}{Definition}[part]
\newtheorem{Proposition}{Proposition}[part]

\newtheorem{Assumption}{Assumption}[part]
\newtheorem{Lemma}{Lemma}[part]
\newtheorem{Corollary}{Corollary}[part]

\newcommand{\Sum}[2]{\sum\limits_{#1}^{#2}}

\newcommand{\Int}[2]{\int_{#1}^{#2}}

\newcommand{\Ind}[1]{\mathds{1}_{\left\{ #1 \right\}}}
\newcommand{\Expect}[1]{\mathbb{E}\left[ #1 \right]}

\newcommand{\N}{\mathbb{N}}
\newcommand{\R}{\mathbb{R}}

\newcommand{\Z}{\mathbb{Z}}
\newcommand{\E}{\mathbb{E}}

\newcommand{\T}{\mathbb{T}}

\newcommand{\Ac}{\mathcal{A}}

\newcommand{\Fc}{\mathcal{F}}

\usepackage{geometry}
\title{Option market making with hedging-induced market impact}

\author{
	Paulin Aubert\thanks{Laboratoire de Mathématiques et Modélisation d'Évry, Université Évry Paris-Saclay, Exiom Partners, France, \sf paulin.aubert@univ-evry.fr.} $\:$ 
	Etienne Chevalier\thanks{Laboratoire de Mathématiques et Modélisation d'Évry, Université Évry Paris-Saclay, UMR 8071 CNRS, France, \sf etienne.chevalier@univ-evry.fr.} $\:$ 
	Vathana Ly Vath\thanks{Laboratoire de Mathématiques et Modélisation d'Évry, Université Paris-Saclay, ENSIIE, UMR 8071 CNRS, France, \sf vathana.lyvath@ensiie.fr.}
}

\date{\today}

\begin{document}
	
	\maketitle
	
	\begin{abstract} 
		This paper develops a model for option market making in which the hedging activity of the market maker generates price impact on the underlying asset. The option order flow is modeled by Cox processes, with intensities depending on the state of the underlying and on the market maker's quoted prices. The resulting dynamics combine stochastic option demand with both permanent and transient impact on the underlying, leading to a coupled evolution of inventory and price. We first study market manipulation and arbitrage phenomena that may arise from the feedback between option trading and underlying impact. We then establish the well-posedness of the mixed control problem, which involves continuous quoting decisions and impulsive hedging actions. Finally, we implement a numerical method based on policy optimization to approximate optimal strategies and illustrate the interplay between option market liquidity, inventory risk, and underlying impact.
	\end{abstract}
	
	\vspace{7mm}
	
	\par \bigskip
	
	
	\section{Introduction} 
	\label{SEC:optimal_market_making_introduction}
	
	Academic research on market making has progressively shifted from stylized microstructure models toward dynamic frameworks able to capture high-frequency trading, inventory risk, and market impact. A key milestone in this evolution is the stochastic-control approach of Avellaneda and Stoikov \cite{AvellanedaStoikov}, which formulates the quoting problem in continuous time and solves it through Hamilton–Jacobi–Bellman equations. This framework has been extended in various directions, for instance by Guilbaud and Pham \cite{GuilbaudPham11}, who incorporated both limit and market orders, or by Guéant et al. \cite{Gueant2013}, who derived explicit solutions in an affine setting, enabling practical calibration. Cartea and coauthors \cite{Cartea14,CarteaJaimungal15} further broadened the perspective by linking market making to optimal execution and risk management.  
	
	In parallel, a large literature has developed on optimal execution and market impact, starting with Almgren and Chriss \cite{AlmgrenChriss2001} and enriched by contributions that distinguish temporary and permanent impact \cite{GatheralSchied13} or account for the self-exciting nature of order flow via Hawkes processes \cite{AlfonsiBlanc16}. These advances are highly relevant for option market making, where quoting decisions affect not only order flow but also the underlying dynamics through hedging. Models explicitly integrating such feedback effects have emerged more recently. For example, Stoikov and Saglam \cite{StoikovSaglam09} introduced delta-aware option market making, El Aoud and Abergel \cite{ElAoudAbergel2015} modeled joint underlying–option dynamics in a stochastic-control framework, and Baldacci, Bergault and Guéant \cite{Gueant2020} developed a multi-option setting where order intensities depend on aggregate sensitivities.  
	
	While these models provide valuable insight, classical solution techniques become rapidly intractable as soon as one accounts for coupled markets, feedback loops, and state-dependent intensities. This has motivated the use of simulation-based and machine-learning approaches. Spooner et al. \cite{Spooner2018} applied reinforcement learning to learn quoting strategies without an explicit parametric model, and Buehler et al. \cite{Buehler2019} introduced the Deep Hedging framework, where neural networks optimize strategies directly on simulated paths.
	
	\medskip 
	
	In this paper, we consider a market maker mandated to ensure the liquidity of a European option, while complying with hedging constraints imposed by a third party (regulator or principal). Unlike other approaches in the literature, notably \cite{StoikovSaglam09, ElAoudAbergel2015, Gueant2020}, we make no explicit assumptions about the dynamics of the option mid-price. Instead, we model order arrivals in the options market by Cox processes whose intensities depend on both the quotes offered by the market maker and the state of the underlying, allowing the option price to emerge endogenously as a result of supply and demand.
	These intensities are modeled in a flexible way, allowing the framework to reflect key market effects such as the dependence of order-flow activity on option moneyness and time to expiry.
	Furthermore, we assume that the market maker has an impact on the price of the underlying via the consumption of the order book, whose shape is sufficiently general to represent low-liquidity assets. This impact is modeled by a permanent and a resilient component. Finally, we develop a numerical method based on neural networks to approximate the market maker's optimal quoting strategy, while taking into account the associated liquidity and transaction risks.
	
	\medskip 
	
	The remainder of the paper is structured as follows. 
	Section~\ref{SEC:optimal_market_making_problem_description} introduces the modeling framework and formulates the market-making problem. 
	Section~\ref{SEC:model_consistency_well_posedness} establishes the model’s consistency by examining market integrity and the well-posedness of the control problem. 
	It demonstrates the absence of instantaneous and dynamic arbitrage opportunities, rules out transaction-triggered price manipulation, derives a terminal bound in the coupled setting, and proves quadratic-growth estimates ensuring the finiteness of the value function. 
	Finally, Section~\ref{SEC:numerical_method} develops the numerical methodology and reports experiments. 
	We describe the discrete-time simulator and the neural policy training procedure, then present empirical results in a baseline configuration, two asymmetric scenarios, and a low-liquidity regime.
	
	\section{Market modeling and the market maker problem}
	\label{SEC:optimal_market_making_problem_description}
	\subsection{Modeling the underlying market}	
	Let $(\Omega, \Fc, \mathbb{P})$ be a probability space equipped with a filtration $(\Fc_t)_{t \geq 0}$ satisfying the usual conditions. Throughout this work, all random variables and stochastic processes are defined on $(\Omega, \Fc, \mathbb{P})$.
	\subsubsection{Order book representation} 
	
	In line with the approach of Predoiu, Shaikhet, and Shreve~\cite{PredoiuShaikhetShreve11}, we aim to represent the order book of the underlying asset while maintaining full flexibility regarding its structure. In particular, the framework accommodates order books with missing price levels, which may result in a discontinuous shape with gaps.
	While empirical evidence indicates that order book depth can be highly stochastic, we assume for simplicity that the shape of the order book is time-invariant. 
	This assumption could be relaxed, for example by allowing regime-switching or stochastic variations in the order book shape.
	It is characterized by two piecewise continuous functions: $f_B$ for the bid side and $f_A$ for the ask side. These functions are required to satisfy Assumptions~\ref{ASS:market_making_orderbook_integrability} and \ref{ASS:market_making_right_truncated_depth}.
	
	\begin{Assumption}[Integrability of the order book shape]
		\label{ASS:market_making_orderbook_integrability}  
		The order book densities satisfy
		\begin{equation*}
			f_B : \R^- \to [0,\infty), \qquad f_A : \R^+ \to [0,\infty),
		\end{equation*}
		together with the integrability conditions:
		\begin{equation*}
			\int_{-\infty}^{0} (1+|u|) f_B(u)du < \infty,
			\qquad 
			\int_{0}^{\infty} (1+u) f_A(u)du < \infty.
		\end{equation*}
	\end{Assumption}
	Assumption~\ref{ASS:market_making_orderbook_integrability} ensures that the total quantity and value represented in the order book are finite. This condition reflects the idea that market participants do not perceive the firm’s fundamental value as unbounded.   
	We also impose that beyond a certain price level no liquidity is available, which is consistent with empirical observations that liquidity does not extend indefinitely. This requirement is formalized in Assumption~\ref{ASS:market_making_right_truncated_depth}
	\begin{Assumption}[Right–truncated depth]
		\label{ASS:market_making_right_truncated_depth}
		There exist $U_A,U_B\in(0,\infty)$ such that
		\begin{equation*}
			f_B(u)=0\ \text{for a.e. }u<-U_B,
			\qquad
			f_A(u)=0\ \text{for a.e. }u>U_A.
		\end{equation*}
	\end{Assumption}
	
	To further characterize the order book, we introduce cumulative distribution functions that quantify the total volume available when liquidity is consumed up to a given depth. For the bid side, given a current best bid price $b > 0$, we set
	\begin{equation}
		\label{EQN:market_making_cumulative_dist_function_bid}
		\Phi_B(x) := \int_{-x}^{0} f_B(u)du, \qquad x \in [0,b].
	\end{equation}
	The restriction to $x \in [0,b]$ ensures that the bid side does not extend below zero prices, where liquidity is absent.
	On the ask side we define
	\begin{equation}
		\label{EQN:market_making_cumulative_dist_function_ask}
		\Phi_A(x) := \int_{0}^{x} f_A(u)du, \qquad x \in \R^+.
	\end{equation}
	In this case, no analogous restriction is required, since the ask side can in principle extend arbitrarily far above the current price, with the truncation guaranteed only by Assumption~\ref{ASS:market_making_right_truncated_depth}.
	By Assumptions~\ref{ASS:market_making_orderbook_integrability} and 
	\ref{ASS:market_making_right_truncated_depth}, the cumulative functions $\Phi_B$ and $\Phi_A$ are finite, nondecreasing, and satisfy the bounds
	\begin{align*}
		&\forall b>0,\ \forall x \in [0,b]: 
		\quad 0 \le \Phi_B(x) \le \Phi_B(b) \le \Phi_B(U_B) < \infty, \\[0.3em]
		&\forall x \in \R^+ \cup \{+\infty\}: 
		\quad 0 \le \Phi_A(x) \le \Phi_A(+\infty) = \Phi_A(U_A) < \infty.
	\end{align*}
	
	The functions $\Phi_B$ and $\Phi_A$ quantify cumulative liquidity from the best quote up to a given depth: if a market order consumes all liquidity up to level $x$, the executed volume is $\Phi_B(x)$ or $\Phi_A(x)$.  
	Conversely, given a target volume $y$, one asks which price interval must be traversed. Since plateaus may occur in $\Phi_B,\Phi_A$ when there are gaps in the order book, they are not strictly increasing. We therefore use the generalized inverse introduced in Definition~\ref{DEF:market_making_inverse}.
	
	\begin{Definition}[Generalized inverse of cumulative liquidity functions] 
		\label{DEF:market_making_inverse} 
		Let $b>0$ denote the current best bid price. The generalized inverses of the cumulative liquidity functions $\Phi_B$ and $\Phi_A$, denoted respectively by $\Phi_B^{-1}$ and $\Phi_A^{-1}$, are defined by
		\begin{equation*}
			\Phi_B^{-1}(y) = \inf\{ x \geq 0 | \Phi_B(x) \geq y \}, \qquad \forall y \in [0, \Phi_B(b)],
		\end{equation*}
		and
		\begin{equation*}
			\Phi_A^{-1}(y) = \inf \{ x \geq 0 | \Phi_A(x) \geq y \}, \qquad \forall y \in [0, \Phi_A(+\infty)].
		\end{equation*}
	\end{Definition}
	The generalized inverse functions $\Phi_B^{-1}$ and $\Phi_A^{-1}$ are nondecreasing on their respective domains and bounded, with 
	\begin{equation}
		\label{EQN:market_making_phi_inv}
		0 \leq \Phi_B^{-1}(y) \leq U_B \quad \text{for } y \in [0, \Phi_B(b)], 
		\qquad
		0 \leq \Phi_A^{-1}(y) \leq U_A \quad \text{for } y \in [0, \Phi_A(+\infty)].
	\end{equation}
	Moreover, they have finite total variation on their domains and are therefore integrable.
	
	\subsubsection{Execution price and cost functions} 
	
	We now define execution cost functions, which link order size to the monetary amount required to consume the book up to a given depth.
	Let $b>0$ denote the current best bid price. When an agent sells a quantity $0 \leq q^- \leq \Phi_B(b)$, the order book is consumed starting from $b$ down to the level $b - \Phi_B^{-1}(q^-)$. 
	The total revenue obtained (or, equivalently, the monetary value of the trade from the buyer’s perspective) is given by
	\begin{equation*} 
		P_B(b,q^-) = \Int{b-\Phi_B^{-1}(q^-)}{b} y f_B(y-b)dy.
	\end{equation*}
	Introducing the change of variables $u = y-b$, so that $y=b+u$ and $u \in [-\Phi_B^{-1}(q^-),0]$, this expression becomes
	\begin{equation}\label{EQN:underyling_bid_price}
		P_B(b,q^-) = \Int{-\Phi_B^{-1}(q^-)}{0} (b+u) f_B(u)du.
	\end{equation}
	By symmetry, let $a>0$ denote the best ask price. If an agent buys a quantity $0 \leq q^+ \leq \Phi_A(+\infty)$, the book is consumed from $a$ up to the level $a + \Phi_A^{-1}(q^+)$. The corresponding execution cost is
	\begin{equation*} 
		P_A(a,q^+) = \Int{a}{a+\Phi_A^{-1}(q^+)} y f_A(y-a) dy.
	\end{equation*}
	With the substitution $u = y-a$, so that $y=a+u$ and $u \in [0,\Phi_A^{-1}(q^+)]$, this rewrites as
	\begin{equation} \label{EQN:underyling_ask_price}
		P_A(a,q^+) = \Int{0}{\Phi_A^{-1}(q^+)} (a+u) f_A(u) du.
	\end{equation}
	By Assumption~\ref{ASS:market_making_orderbook_integrability}, the execution costs \eqref{EQN:underyling_bid_price}–\eqref{EQN:underyling_ask_price} are
	finite for all admissible order sizes $0 \le q^- \le \Phi_B(b)$ and $0 \le q^+ \le \Phi_A(+\infty)$. 
	
	We next examine structural properties of execution costs, showing that they admit integral representations and satisfy convexity/concavity.  
	\begin{Lemma}[Analytical structure of execution costs]
		\label{PROP:market_making_convex_concave_cost_functions}
		Let $b>0$ the best bid price and $a>0$ denote the best ask price. For every admissible order size
		\begin{equation*}
			0\le q^-\le \Phi_B(b)\qquad\text{and}\qquad 0\le q^+\le \Phi_A(+\infty),
		\end{equation*}
		the cost functionals \eqref{EQN:underyling_bid_price} and \eqref{EQN:underyling_ask_price} admit the representations
		\begin{align}
			P_B(b,q^-) &= \int_{-\Phi_B^{-1}(q^-)}^{0} (b+u)f_B(u)du 
			= bq^- - \int_{0}^{q^-} \Phi_B^{-1}(y)dy, \label{EQN:market_making_PB_rep} \\[0.3em]
			P_A(a,q^+) &= \int_{0}^{\Phi_A^{-1}(q^+)} (a+u)f_A(u)du 
			= aq^+ + \int_{0}^{q^+} \Phi_A^{-1}(y)dy. \label{EQN:market_making_PA_rep} 
		\end{align}
		Consequently, $q\mapsto P_B(b,q)$ is concave and $q\mapsto P_A(a,q)$ is convex  on their domains. 
		Moreover, both maps are absolutely continuous and $\mathcal C^1$ a.e., with a.e. derivatives
		\begin{equation*}
			\partial_q P_B(b,q)=b-\Phi_B^{-1}(q),\qquad 
			\partial_q P_A(a,q)=a+\Phi_A^{-1}(q).
		\end{equation*}
		In particular, the marginal execution price remains bounded between $b-U_B$ and $b$ on the bid side, and between $a$ and $a+U_A$ on the ask side.
	\end{Lemma}
	
	\begin{proof}
		The equalities \eqref{EQN:market_making_PB_rep} and \eqref{EQN:market_making_PA_rep} follow from the following identities: 
		\begin{equation*}
			\int_{-x}^{0} uf_B(u)du=-\int_{0}^{\Phi_B(x)} \Phi_B^{-1}(y)dy,
			\qquad
			\int_{0}^{x} ud\Phi_A(u)=\int_{0}^{\Phi_A(x)} \Phi_A^{-1}(y)dy.
		\end{equation*}
		Here $q^- = \Phi_B(x)$ and $q^+ = \Phi_A(x)$. Moreover, $d\Phi_A(u) = f_A(u)du$ and $d\Phi_B(u) = f_B(u)du$.
		Since $\Phi_B^{-1}$ and $\Phi_A^{-1}$ are nondecreasing, the primitives
		$q\mapsto -\int_0^q \Phi_B^{-1}(y)dy$ and $q\mapsto \int_0^q \Phi_A^{-1}(y)dy$ are, respectively, convex and concave. Adding the linear terms $bq$ and $aq$ preserves these properties. Absolute continuity and the a.e. derivative formulas are immediate.
		Equation~\ref{EQN:market_making_phi_inv} gives, $\Phi_B^{-1}(y)\le U_B$ and $\Phi_A^{-1}(y)\le U_A$, which yields the stated bounds on the a.e. derivatives.
	\end{proof}
	
	From an economic perspective, the concavity of $P_B$ captures that the marginal revenue from selling decreases with order size, since progressively lower bid quotes are reached when larger quantities are executed.
	Conversely, the convexity of $P_A$ reflects that the marginal cost of buying increases with order size, as deeper layers of the ask book are consumed at higher prices.
	
	\subsubsection{Dynamics of the underlying} 
	\label{SUBSEC:market_making_dynamics_underlying}
	\paragraph*{Impact of market orders on quotes} ~\\
	Let $(P_t)_{t \geq 0}$ denote the mid-price process and $(S_t)_{t \geq 0}$ the spread process. At any time $t \geq 0$, the best bid and ask prices are given by
	\begin{equation*}
		B_t = P_t - \frac{1}{2} S_t, \qquad A_t = P_t + \frac{1}{2} S_t.
	\end{equation*}
	
	Consider a state $(p, s)\in (\R^+)^2$, with corresponding best quotes $b = p - s/2$ and $a = p + s/2$.  
	The submission of market orders consumes liquidity and shifts the best quotes.  Specifically, a sell order of size $q^- \in [0,\Phi_B(b)]$ moves the bid down to $b - \Phi_B^{-1}(q^-)$, while a buy order of size $q^+ \in [0,\Phi_A(+\infty))$ pushes the ask up to $a + \Phi_A^{-1}(q^+)$. This consumption widens the spread, which becomes
	\begin{align*}
		S_{t^+} &= (p + s/2) + \Phi_A^{-1}(q^+) - (p - s/2) + \Phi_B^{-1}(q^-) \\
		&= s + \Phi_A^{-1}(q^+) + \Phi_B^{-1}(q^-).
	\end{align*}
	At the same time, the mid-price is updated as the average of the new best bid and ask:
	\begin{align*}
		P_{t^+} &= \frac{1}{2} \left(  (p + s/2) + \Phi_A^{-1}(q^+) + (p - s/2) + \Phi_B^{-1}(q^-) \right) \\
		&= p + \frac{1}{2} \left( \Phi_A^{-1}(q^+) - \Phi_B^{-1}(q^-) \right).
	\end{align*}
	These formulas highlight two key effects of order flow on price formation. 
	First, the updated mid-price incorporates the imbalance between buying and selling pressure, shifting upward when buy orders dominate and downward when sell orders prevail. 
	Second, the spread increases as deeper layers of the order book are consumed, reflecting the greater uncertainty and reduced liquidity that follow large market orders.
	
	\paragraph*{Order-flow sources and intensity dynamics} ~\\
	The dynamics of the mid-price are driven by two sources of order flow: exogenous trades submitted by other market participants, and the market maker’s own interventions used to hedge option exposures.  
	The latter are modeled as impulse trades occurring at stopping times $(\nu_i)_{i\in\N}$ for sales on the bid side and $(\tau_i)_{i\in\N}$ for purchases on the ask side, with traded sizes $\xi_i^-\ge 0$ and $\xi_i^+\ge 0$, where $\xi_i^-$ is $\Fc_{\nu_i}$-measurable and $\xi_i^+$ is $\Fc_{\tau_i}$-measurable.
	The corresponding counting processes of interventions are defined by
	\begin{equation} 
		\label{EQN:market_making_impulse_counting_process}
		H_t^- := \sum_{i\ge 1}\mathbf{1}_{\{\nu_i\le t\}}, 
		\qquad
		H_t^+ := \sum_{i\ge 1}\mathbf{1}_{\{\tau_i\le t\}}.
	\end{equation}
	with $H^-_0 = H^+_0 = 0$.
	
	Exogenous arrivals on each side are modeled by marked Hawkes processes with exponential kernels.
	On the bid side (sell arrivals), let $(N_t^-)_{t\ge0}$ be the counting process of exogenous sales hitting the bid. 
	Each jump time $T_k^-$ of $N^-$ carries an i.i.d. mark $M_k^- \in [0,1]$ with density $f^-$. The mark represents the fraction of the locally available depth consumed by that trade.
	On the bid side, the available depth is stochastic, so modelling orders as fractions ensures the dynamics remain well-defined.
	Similarly, on the ask side (buy arrivals), $(N_t^+)_{t\ge0}$ counts exogenous buys hitting the ask, each jump time $T_\ell^+$ being endowed with an i.i.d. mark $M_\ell^+ \in [0,1]$ with density $f^+$.
	The intensities $(\lambda_t^-)_{t\ge0}$ and $(\lambda_t^+)_{t\ge0}$ evolve according to
	\begin{equation}
		\label{EQN:market_making_full_intensity_bid}
		d\lambda_t^- = \theta^-\big(\mu^- - \lambda_t^-\big)dt + \kappa^- dN_t^- + \kappa^- dH_t^-,
	\end{equation}
	and
	\begin{equation}
		\label{EQN:market_making_full_intensity_ask}
		d\lambda_t^+ = \theta^+\big(\mu^+ - \lambda_t^+\big)dt + \kappa^+ dN_t^+ + \kappa^+ dH_t^+.
	\end{equation}
	with initial conditions $\lambda_0^-\ge 0$ and $\lambda_0^+\ge 0$, where $\mu^\pm\ge0$ are baseline levels, $\theta^\pm>0$ are mean-reversion rates, and $\kappa^\pm\ge0$ are excitation heights. \\
	When $H^-\equiv 0$ and $H^+\equiv 0$, the dynamics reduce to standard Hawkes  processes which is assumed to respect Assumption~\ref{ASS:market_making_hawkes_stability}.
	\begin{Assumption}[Non-explosivity of the Hawkes intensities]
		\label{ASS:market_making_hawkes_stability} 
		In the absence of interventions, the Hawkes components are subcritical:
		\begin{equation*}
			\frac{\kappa^-}{\theta^-} < 1
			\qquad\text{and}\qquad 
			\frac{\kappa^+}{\theta^+} < 1,
		\end{equation*}
		so that $(\lambda_t^\pm)$ are finite almost surely.
	\end{Assumption}
	This stability condition rules out explosion, guarantees the well-posedness of the controlled dynamics, and will be instrumental in proving the well-posedness of the value function and of the optimization problem.
	
	\paragraph*{Modeling mid-price and spread dynamics} ~\\
	In line with Alfonsi and Blanc~\cite{AlfonsiBlanc16}, each trade has a permanent impact of fraction $\eta\in[0,1]$ that durably shifts the mid-price, and a transient impact of fraction $1-\eta$ that decays over time through a mean-reverting process.
	Let $(P_t)_{t \geq 0}$ denote the mid-price process. It evolves under the combined effect of exogenous buy and sell orders, as well as the market maker’s interventions. The dynamics take the form
	\begin{equation}
		\label{EQN:mid_price_dynamics}
		\left\{
		\begin{array}{l}
			\displaystyle dP_t = \frac{\eta}{2} \left( \Phi_A^{-1}\left(M_t^+  \Phi_A(+\infty) \right) dN_t^+ - \Phi_B^{-1}\left(M_t^- \Phi_B(B_t) \right) dN_t^- \right) + dD_t, \\ [0.5em]
			\displaystyle P_{\nu_i^+} = P_{\nu_i} - \frac{\eta}{2} \Phi_B^{-1} (\xi_i^-), \\ [0.5em]
			\displaystyle P_{\tau_i^+} = P_{\tau_i} + \frac{\eta}{2}  \Phi_A^{-1} (\xi_i^+),
		\end{array}
		\right.
	\end{equation} 
	with initial value $P_0 = p \in \R^+$. 
	$(D_t)_{t \geq 0}$ is the transient component of the impact, which satisfies
	\begin{equation}
		\label{EQN:resilient_impact_dynamics}
		\left\{
		\begin{array}{l}
			\displaystyle dD_t = -r D_t dt + \frac{1 - \eta}{2} \left( \Phi_A^{-1}\left(M_t^+ \Phi_A(+\infty) \right) dN_t^+ - \Phi_B^{-1}\left(M_t^- \Phi_B(B_t) \right) dN_t^- \right), \\ [0.5em]
			\displaystyle D_{\nu_i^+} = D_{\nu_i} - \frac{1 - \eta}{2} \Phi_B^{-1} (\xi_i^-), \\ [0.5em]
			\displaystyle D_{\tau_i^+} = D_{\tau_i} + \frac{1 - \eta}{2} \Phi_A^{-1} (\xi_i^+),
		\end{array}
		\right.
	\end{equation}
	where $D_0 = d \in \R$ and $r > 0$ denotes the rate of resilience.    
	
	The spread process $(S_t)_{t \geq 0}$ models the distance between bid and ask. 
	It widens when liquidity is consumed on either side of the book and narrows gradually due to resilience effects. 
	Assuming a minimum spread $\delta > 0$ and a mean-reversion rate $\rho > 0$, we set
	\begin{equation}
		\label{EQN:spread_dynamics}
		\left\{
		\begin{array}{l}
			dS_t = - \rho (S_t - \delta) dt + \Phi_A^{-1}\left(M_t^+ \Phi_A(+\infty) \right) dN_t^+ + \Phi_B^{-1}\left(M_t^- \Phi_B(B_t) \right) dN_t^-, \\ [0.5em]
			S_{\nu_i^+} = S_{\nu_i} + \Phi_B^{-1} (\xi_i^-), \\ [0.5em]
			S_{\tau_i^+} = S_{\tau_i} + \Phi_A^{-1}(\xi_i^+),
		\end{array}
		\right.
	\end{equation}
	with initial condition $S_0 = s \geq \delta$.  
	
	Equations \eqref{EQN:mid_price_dynamics}–\eqref{EQN:spread_dynamics} describe how permanent shifts, transient effects, and spread adjustments drive the underlying dynamics.
	The conditions ensuring strictly positive quotes under admissible strategies are
	given in Lemma~\ref{LEM:market_making_positivity_quotes}.
	In practice, the market could also receive limit orders on the underlying. 
	Explicitly representing these flows would be costly, as it would require modelling both exogenous limit order arrivals and the market maker’s own limit orders. 
	To keep the focus on the option market-making dimension of the problem, interventions on the underlying are therefore restricted to market orders. 
	The effects of limit orders on mid-price and spread are captured implicitly through the mean-reversion of the transient impact $D_t$ in \eqref{EQN:resilient_impact_dynamics} and of the spread $S_t$ in \eqref{EQN:spread_dynamics}, following the modeling approach of Alfonsi and Blanc~\cite{AlfonsiBlanc16}, providing a tractable representation of order book resilience while remaining consistent with the microstructure interpretation.
	
	\subsection{Modeling the options market}
	
	We consider a European option with maturity $T \in \R^+$ and payoff function $\varphi$ of linear growth, written on the previously described underlying asset. The market maker is mandated to provide liquidity for this derivative product.
	Let $e := (p, d, s, \lambda_0^-, \lambda_0^+)$ denote the state vector of the underlying market, where $p$ is the mid-price, $d$ the resilient part of the mid-price, $s$ the spread and $\lambda_0^\pm$ the initial intensities. We assume that all market participants are able to estimate the same reference price for the option, denoted by $b$, which is a deterministic function:
	\begin{equation*}
		b: (t, e) \in [0, T] \times (\R^+)^5 \mapsto b(t, e) \in \R^+
	\end{equation*}
	Function $b$ is chosen to respect Assumption~\ref{ASS:market_making_ref_price_linear_growth}.
	\begin{Assumption}[Linear growth of the option reference price]
		\label{ASS:market_making_ref_price_linear_growth}
		There exists a constant $C^{(b)} > 0$ such that, for all $(t,e)\in[0,T]\times(\R^+)^5$,
		\begin{equation*}
			0 \leq b(t,e) \leq C^{(b)} \big(1+\|e\|\big).
		\end{equation*}
	\end{Assumption}
	Depending on the intended hedging strategy, further assumptions may be made about the regularity of $b$, such as differentiability with respect to certain variables.
	
	The market maker sets bid and ask quotes on the option, denoted by $\beta$ and $\alpha$. Arrivals of option orders are modeled by two Cox processes, $N^b$ for buy orders at the bid and $N^a$ for sell orders at the ask. Their intensities are given by bounded deterministic functions
	\begin{align*}
		&\lambda^b: (t, \alpha, \beta, e) \in [0,T] \times (R^+)^2 \times (\R^+)^5 \mapsto \lambda^b(t, \alpha, \beta, e), \\ 
		&\lambda^a: (t, \alpha, \beta, e) \in [0,T] \times (R^+)^2 \times (\R^+)^5 \mapsto \lambda^a(t, \alpha, \beta, e),
	\end{align*}
	in accordance with Assumption~\ref{ASS:market_making_option_intensities}.
	\begin{Assumption}[Bounded option trade intensities] 
		\label{ASS:market_making_option_intensities} 
		The Cox intensities for option trades are positive and bounded by finite constants 
		$\overline\lambda^b, \overline\lambda^a > 0$, leading to
		\begin{equation*}
			0 \leq \lambda^b(t,\alpha,\beta,e) \leq \overline\lambda^b, \qquad
			0 \leq \lambda^a(t,\alpha,\beta,e) \leq \overline\lambda^a.
		\end{equation*}
		Both intensities are assumed to satisfy the linear growth condition
		\begin{equation*}
			\sup_{\alpha,\beta} \Big\{ \lambda^a(t,\alpha,\beta,e)(\alpha - b(t,e)) 
			- \lambda^b(t,\alpha,\beta,e)(\beta - b(t,e)) \Big\} 
			\leq C^{(\lambda)} (1+\|e\|),
		\end{equation*}
		for some constant $C^{(\lambda)}>0$. This ensures that the expected P\&L contribution of option trades remains uniformly bounded, with at most linear growth in the state variables.
	\end{Assumption}
	
	Assumption~\ref{ASS:market_making_option_intensities} formalizes how order intensities depend on the market maker’s quotes relative to the reference price. Uncompetitive quotes reduce order flow as investors turn to alternative venues, boundedness expresses the finite size of the market, and the growth condition ensures consistency with the benchmark price while preventing unbounded gains or losses. Together, these conditions provide a coherent and realistic description of option order flow.
	
	\subsection{The market maker’s optimization problem}
	\label{SUBSEC:market_making_objective_market_maker}
	\subsubsection{Market maker inventories}
	
	By setting bid and ask quotes $(\beta_t, \alpha_t)$ on the option market, the market maker generates order arrivals modeled by the counting processes $(N_t^b)_{t \geq 0}$ and $(N_t^a)_{t \geq 0}$. The resulting option inventory process $(I_t)_{t \geq 0}$ records the net position in derivative contracts and evolves according to
	\begin{equation}
		\label{EQN:market_maker_option_inventory_dynamics}
		\begin{cases}
			dI_t = dN_t^b - dN_t^a, \\[0.3em]
			I_0 = i \in \Z.
		\end{cases}
	\end{equation}
	
	To hedge this exposure, the market maker trades in the underlying through sequences of sell and buy orders $(\nu_i)_{i\in\N}$ and $(\tau_i)_{i\in\N}$ at mid-price $p$ and spread $s$. Each intervention involves a quantity $\xi_i^-\in[0,\Phi_B(p-s/2)]$, 
	$\Fc_{\nu_i}$-measurable, or $\xi_i^+\in[0,\Phi_A(+\infty)]$, $\Fc_{\tau_i}$-measurable. 
	The underlying inventory process $(Q_t)_{t \ge 0}$ is piecewise constant, updated only at these intervention times
	\begin{equation}
		\label{EQN:market_maker_underlying_inventory_dynamics}
		\begin{cases}
			dQ_t = 0, \\[0.3em]
			Q_{\nu_i^+} = Q_{\nu_i} - \xi_i^-, \\[0.3em]
			Q_{\tau_i^+} = Q_{\tau_i} + \xi_i^+, \\[0.3em]
			Q_0 = q \in \R.
		\end{cases}
	\end{equation}
	
	Finally, these activities generate cash flows. Each underlying trade incurs a fixed cost $c \geq 0$, while option transactions occur at the bid and ask quotes $(\alpha_t,\beta_t)_{t \geq 0}$. 
	The cash balance process $(X_t)_{t\ge0}$ combines these option trades with the execution costs of underlying interventions:
	\begin{equation}
		\label{EQN:market_maker_cash_dynamics}
		\begin{cases}
			dX_t = \alpha_t dN_t^a - \beta_t dN_t^b, \\[0.5em]
			X_{\nu_i^+} = X_{\nu_i} + P_B\!\left(P_{\nu_i} - \tfrac{1}{2} S_{\nu_i}, \xi_i^-\right) - c, \\[0.5em]
			X_{\tau_i^+} = X_{\tau_i} - P_A\!\left(P_{\tau_i} + \tfrac{1}{2} S_{\tau_i}, \xi_i^+\right) - c, \\[0.5em]
			X_0 = x \in \R.
		\end{cases}
	\end{equation}
	$P_B$ and $P_A$ denote the execution cost functionals defined in \eqref{EQN:underyling_bid_price}–\eqref{EQN:underyling_ask_price}. 
	
	\subsubsection{Admissible strategies}
	\label{SUBSEC:market_making_admissible_strategies}
	
	Admissible strategies must satisfy both economic and technical requirements. 
	On the option market, quotes must be ordered so that $\beta_t \le \alpha_t$ at all times. 
	On the underlying market, trades are constrained by liquidity and short-selling. 
	Executed quantities must lie within the available depth and inventories cannot be driven below $-\Phi_A(+\infty)$, ensuring positions remain coverable. 
	Finally, technical conditions guarantee well-posedness: quotes $(\beta_t,\alpha_t)$ must be $\Fc$-predictable, intervention times strictly increasing stopping times, and traded sizes $\Fc$-measurable. 
	We also require square integrability of the number of interventions. 
	Together, these conditions define the admissible set $\Ac(p,s,q)$ which is described in Definition~\eqref{DEF:admissible_strategies}.
	\begin{Definition}[Admissible strategies]
		\label{DEF:admissible_strategies}
		Given a state $(p,s,q)$ with mid-price $p$, spread $s$ and inventory $q$, 
		the set of admissible strategies is defined by
		\begin{equation}
			\begin{aligned}
				\Ac(p,s,q) := \big\{ &(\alpha,\beta,(\nu_i,\xi_i^-)_{i\ge1},(\tau_i,\xi_i^+)_{i\ge1}) \ : \\
				& (\alpha,\beta) \ \Fc\text{-predictable, with } 0 \le \beta \le \alpha, \\[4pt]
				& (\nu_i), (\tau_i) \ \text{increasing sequences of stopping times}, \\[4pt]
				& \xi_i^- \ \Fc_{\nu_i}\text{-measurable}, \quad 
				\xi_i^+ \ \Fc_{\tau_i}\text{-measurable}, \\[4pt]
				& 0 \le \xi_i^- \le \Phi_B(p - s/2), \quad 
				0 \le \xi_i^+ \le \Phi_A(+\infty), \\[4pt]
				& q - \xi_i^- > -\Phi_A(+\infty) \ \text{for all } i, \\[4pt]
				& \E[H_T^\pm] < \infty, \quad \E[(H_T^\pm)^2] < \infty \big\}.
			\end{aligned}
		\end{equation}
	\end{Definition}
	The next lemma shows that, under suitable initial conditions, bid and ask quotes remain nonnegative for all times and all admissible strategies.
	
	\begin{Lemma}[Non-negativity preservation of bid–ask quotes]
		\label{LEM:market_making_positivity_quotes}
		Let $(P_t,S_t)_{t\ge0}$ be the mid-price and spread defined in
		\eqref{EQN:mid_price_dynamics}–\eqref{EQN:spread_dynamics}, and set
		$B_t := P_t - \tfrac12 S_t$ and $A_t := P_t + \tfrac12 S_t$. 
		For any admissible strategy in $\Ac(p,s,q)$ with initial conditions
		\begin{equation*}
			B_0 = p-\tfrac{s}{2} \ge 0 
			\qquad \text{and}
			\qquad
			B_0 \ge D_0,
		\end{equation*}
		we have
		\begin{equation*}
			\mathbb{P}\big(B_t\ge0, A_t\ge0, \forall t\ge0\big)=1.
		\end{equation*}
	\end{Lemma}
	
	\begin{proof}
		By \eqref{EQN:spread_dynamics}, we have $S_t \geq \delta \geq 0$ for all $t \geq 0$. Since $A_t = B_t + S_t$, it is enough to prove that $B_t \geq 0$ for all $t$. We first show that jumps cannot drive the bid below zero, and then verify that the inter-jump dynamics preserve non-negativity. 
		From \eqref{EQN:mid_price_dynamics}–\eqref{EQN:spread_dynamics}, an ask-side jump of size $q$ produces $\Delta P_t = +\tfrac12 \Phi_A^{-1}(q)$ and $\Delta S_t = +\Phi_A^{-1}(q)$, which implies $\Delta B_t = 0$. A bid-side jump of size $q$ yields $\Delta P_t = -\tfrac12 \Phi_B^{-1}(q)$ and $\Delta S_t = +\Phi_B^{-1}(q)$, so that $\Delta B_t = -\Phi_B^{-1}(q) \leq 0$. Moreover, by construction of the model and by admissibility of the strategy, the consumed quantity on the bid side is bounded above by the available depth: $q \leq \Phi_B(B_{t^-})$. Using the monotonicity of the generalized inverse, this leads to
		\begin{equation*}
			B_{t^+} = B_{t^-}-\Phi_B^{-1}(q) \ge B_{t^-}-\Phi_B^{-1}\!\big(\Phi_B(B_{t^-})\big)
			\ge 0,
		\end{equation*}
		which shows that jumps cannot make the bid negative.  
		
		On any interval $[u,t)$ with no jumps, the dynamics are given by $dP_t = dD_t = -r D_tdt$ and $dS_t = -\rho(S_t - \delta)dt$. Integrating yields
		\begin{equation*}
			D_t = D_u e^{-r(t-u)}, 
			\qquad 
			S_t = \delta + (S_u-\delta)e^{-\rho(t-u)},
		\end{equation*}
		and therefore
		\begin{equation*}
			B_t = P_t-\tfrac12 S_t 
			= \big(P_u-D_u\big) + D_u e^{-r(t-u)} - \tfrac12\delta - \tfrac12(S_u-\delta)e^{-\rho(t-u)}.
		\end{equation*}
		This yields the following explicit inter-jump representation:
		\begin{equation}\label{eq:interjump-B}
			B_t = B_u - D_u\big(1-e^{-r(t-u)}\big) + \tfrac12(S_u-\delta)\big(1-e^{-\rho(t-u)}\big).
		\end{equation}
		Since $S_u\ge\delta$ and $1-e^{-\rho(\cdot)}\ge0$, the last term is nonnegative, and we obtain
		\begin{equation*}
			B_t \ge B_u - D_u\big(1-e^{-r(t-u)}\big) \ge B_u - D_u.
		\end{equation*}
		In particular, on the initial inter-jump interval $[0,T_1)$, the assumption $B_0 \ge D_0$ implies $B_t \ge B_0 - D_0 \ge0$.
		
		Taken together, these arguments show that the bid remains nonnegative both at jump times and during inter-jump periods. As a result, $B_t \geq 0$ holds for all $t \geq 0$, and consequently $A_t = B_t + S_t \geq 0$ as well.
	\end{proof}
	Lemma~\ref{LEM:market_making_positivity_quotes} ensures that the model never generates negative quotes. The condition $B_0 \geq D_0$ excludes initial states where a residual transient impact would not be reflected in the mid-price, which could drive the bid below zero during resilience.
	
	\subsubsection{Performance criterion and value function}
	
	Given an initial state $(t,x,q,i,e)\in[0,T]\times\R^3\times(\R^+)^5$, the market maker chooses an admissible strategy $\gamma$ to maximize expected cash holdings plus the liquidation value of residual positions, adjusted by running penalties. The penalty function $g$ penalizes deviations from hedging, and the incentive function $h$ enforces market-making activity. 
	We assume that the penalty and incentive functions $g$ and $h$ satisfy Assumption~\ref{ASS:market_making_penalties}.
	The function $g$ enforces the desired hedging strategy for the market maker, while $h$ represents a performance-based component requiring the market maker to maintain a minimum level of trading activity, ensuring that contractual liquidity targets are met. 
	Together, these functions provide a tractable way to capture contractual aspects of the market maker’s incentives within the model.
	
	\begin{Assumption}[Quadratic growth bounds for penalty and incentive functions] 
		\label{ASS:market_making_penalties}
		There exist constants $C^{(g)}, C^{(h)} > 0$ such that for all $(t,q,i,e)$,
		\begin{equation*}
			0 \leq g(t,q,i,e) \leq C^{(g)} \big( 1 + q^2 + i^2 + \|e\|^2 \big),
			\qquad 
			0 \leq h(t,q,i,e) \leq C^{(h)} \big( 1 + q^2 + i^2 + \|e\|^2 \big).
		\end{equation*}
	\end{Assumption}
	
	Let $E_t=(P_t,D_t,S_t,\lambda_t^-,\lambda_t^+)$ with $E_0=e$ and $\gamma \in \Ac(p,s,q)$. The objective function is defined by
	\begin{equation}
		\label{EQN:objective_function}
		J_\gamma \left(t, x, q, i, e \right) = \Expect{X_T + L(T, Q_T, I_T, P_T, S_T) - \Int{t}{T} \left(g + h \right)(u, Q_u, I_u, E_u) du},
	\end{equation}
	where the liquidation function is given by
	\begin{align}
		\label{EQN:liquidation_function}
		L(T, q, i, p, s) &= \Ind{q \geq 0} P_B\left(p - s/2, \min\{q, \Phi_B(p-s/2) \}\right) \nonumber \\
		&\quad- \Ind{q < 0} P_A\left(p + s/2, -q\right) + i\varphi(p)  .
	\end{align}
	By admissibility, a short position can always be covered at maturity, while excess long inventory beyond $\Phi_B$ is liquidated at zero value, discouraging overly aggressive hedging. The option inventory settles through $\varphi$. \\
	A natural specification for $g$ is to enforce delta-hedging, penalizing deviations between the actual underlying inventory $q$ and the hedged position $i\partial_p b(t,e)$.  
	Similarly, the role of $h$ is to discourage purely passive quoting without generating trades, by penalizing intensities below a target level $\underline{\lambda}>0$.  
	Typical quadratic forms are therefore
	\begin{equation*}
		g(t,q,i,e) = (q - i\partial_p b(t,e))^2, \qquad
		h(t,q,i,e) := \Big( \max\{0, \underline{\lambda} - (\lambda^a(t,\alpha,\beta,e)+\lambda^b(t,\alpha,\beta,e)) \} \Big)^2,
	\end{equation*}
	which combine hedging discipline with a requirement to sustain a minimum level of trading activity. \\
	The market maker seeks to maximize her expected performance criterion: 
	\begin{equation*}
		\tilde{v}(t, x, q, i, e) := \underset{\gamma \in \Ac(p, s, q)}{\sup} J_\gamma(t, x, q, i, e)
	\end{equation*}
	
	Proposition~\ref{PROP:reduction_by_cash_additivity} shows that $x$ separates additively, so the effective value function is $v$, independent of initial cash.
	
	\begin{Proposition}[Reduction by cash additivity]
		\label{PROP:reduction_by_cash_additivity}
		For all $(t,x,q,i,e)$, the value function satisfies
		\begin{equation*}
			\tilde{v}\left(t, x, q, i, e\right) = x + v\left(t, q, i, e\right), 
		\end{equation*}
		where $v$ is given by
		\begin{align}
			\label{EQN:market_making_value_function}
			v\left(t, q, i, e\right) &= \underset{\gamma \in \Ac}{\sup} \mathbb{E} \Bigg[ \Int{t}{T} \left[ \alpha_u \lambda^a - \beta_u \lambda^b \right](u, \alpha_u, \beta_u, E_u) du \nonumber \\ 
			&\quad + \Sum{i=1}{+\infty} \Ind{\nu_i \in [t, T]} \left(P_B(P_{\nu_i} - S_{\nu_i}/2, \xi_i^-) - c \right) \nonumber \\
			&\quad + \Sum{i=1}{+\infty} \Ind{\tau_i \in [t, T]} \left(- P_A(P_{\tau_i} + S_{\tau_i}/2, \xi_i^+) - c \right) \nonumber \\
			&\quad - \Int{t}{T} (g + h)(u, Q_u, I_u, E_u) du + L(T, Q_T, I_T, P_T, S_T) \Bigg] 
		\end{align}
	\end{Proposition}
	
	\begin{proof} 
		Integrating the dynamics of the cash process over $[t,T]$ and using the compensated martingales for the option order arrival processes then taking the expectations yields the required result.
	\end{proof}
	
	\section{Model consistency and well-posedness}
	\label{SEC:model_consistency_well_posedness}
	The purpose of this section is to examine possible forms of arbitrage and 
	manipulation within our framework, in order to ensure the model’s internal 
	consistency and to identify whether manipulative strategies could arise as 
	optimal candidates.
	\subsection{Underlying market: absence of arbitrage and manipulation}
	In Lemma~\ref{LEM:market_making_no_instant_rt}, we first rule out arbitrage in the simplest case of an instantaneous round-trip. This result confirms that immediate buy–sell cycles are loss-making, 
	with the spread acting as a lower bound on trading frictions. 
	
	\begin{Lemma}[No instantaneous round-trip] 
		\label{LEM:market_making_no_instant_rt}
		For any market state $(p,s)$ with $\delta > 0$, any admissible order book satisfying 
		Assumptions~\ref{ASS:market_making_orderbook_integrability}–\ref{ASS:market_making_right_truncated_depth}, and any admissible trade size $q > 0$, the execution cost difference satisfies
		\begin{equation*}
			P_B(p - s/2, q) - P_A(p + s/2, q) - 2c \leq -\delta q - 2c < 0,
		\end{equation*}
		where $2c$ denotes the total fixed cost of the two transactions.
	\end{Lemma}
	
	\begin{proof}
		From the integral representations of $P_A$ and $P_B$ 
		(Equations~\ref{EQN:market_making_PA_rep}–\ref{EQN:market_making_PB_rep}), we have
		\begin{align*}
			P_B(p - s/2, q) - P_A(p + s/2, q) 
			&= (p-s/2)q - \int_{0}^{q} \Phi_B^{-1}(y)dy 
			- (p+s/2)q + \int_{0}^{q} \Phi_A^{-1}(y)dy \\
			&= \int_{0}^{q} \left[-s - \Phi_B^{-1}(y) - \Phi_A^{-1}(y)\right]dy .
		\end{align*}
		Since $s \ge \delta > 0$, the right-hand side is bounded above by $-\delta q$. Adding the fixed cost term $-2c \leq 0$ completes the proof.
	\end{proof}
	Beyond this elementary case, one must also rule out more elaborate manipulations over finite horizons. 
	In line with Huberman and Stanzl’s no–dynamic-arbitrage condition \cite{Huberman2004}, the next result shows that microstructure frictions in our model prevent any admissible strategy from generating a strictly positive pure execution profit.
	
	\begin{Proposition}[No round-trip arbitrage] 
		\label{PROP:market_making_no_dynamic_arbitrage_underlying}
		Let $(\nu_i,\xi_i^-)_{i \geq 1}$ and $(\tau_j,\xi_j^+)_{j \geq 1}$ denote the sequences of admissible trades executed by the market maker on the underlying up to a finite horizon $T$, where $\xi_i^-$ (resp.\ $\xi_j^+$) is the size of the $i$-th sale (resp.\ $j$-th purchase).  
		Define the total traded volume and the number of interventions by
		\begin{equation*}
			\mathcal V_T := \sum_{\nu_i \le T} \xi_i^- + \sum_{\tau_j \le T} \xi_j^+, \qquad \overline H_T := \sum_{\nu_i \le T} 1 + \sum_{\tau_j \le T} 1.
		\end{equation*}
		Then, for any round trip completed over $[0,T]$ such that $Q_T = Q_0$, the pure execution profit and loss $\Pi$ satisfies
		\begin{equation*}
			\Pi_T - \Pi_0 \leq - \frac{\delta}{2} \mathcal V_T - c \overline H_T \leq 0,
		\end{equation*}
		with strict inequalities whenever at least one trade is executed.
	\end{Proposition}
	
	\begin{proof}
		Let $C_T$ be the cumulative underlying cash flows up to $T$. Using \eqref{EQN:market_making_PB_rep}–\eqref{EQN:market_making_PA_rep} at pre-trade quotes,
		\begin{align*}
			C_T &= \Sum{i=1}{+\infty} \Ind{\nu_i \in [0, T]} \left(P_B(P_{\nu_i} - \tfrac12 S_{\nu_i}, \xi_i^-) - c \right) + \Sum{j=1}{+\infty} \Ind{\tau_j \in [0, T]} \left(- P_A(P_{\tau_j} + \tfrac12 S_{\tau_j}, \xi_j^+) - c \right) \\
			&= \Sum{i=1}{+\infty} \left( \Ind{\nu_i \in [0, T]} \left( P_{\nu_i} - \tfrac12 S_{\nu_i} \right) \xi_i^- - \Int{0}{\xi_i^-} \Phi_B^{-1}(y)dy \right) \\
			&\qquad - \Sum{j=1}{+\infty} \left( \Ind{\tau_j \in [0, T]} \left( P_{\tau_j} + \tfrac12 S_{\tau_j} \right) \xi_j^+ + \Int{0}{\xi_j^+} \Phi_A^{-1}(y)dy \right) - c \overline H_T \\
			&= \Sum{\nu_i \leq T}{} P_{\nu_i} \xi_i^- - \Sum{\tau_j \leq T}{} P_{\tau_j} \xi_j^+ - \frac{1}{2 }\Sum{\nu_i \leq T}{} S_{\nu_i} \xi_i^- - \frac{1}{2}\Sum{\tau_j \leq T}{} S_{\tau_j} \xi_j^+ - \Sum{\nu_i \leq T}{} \Int{0}{\xi_i^-} \Phi_B^{-1}(y)dy \\
			&\qquad - \Sum{\tau_j \leq T}{} \Int{0}{\xi_j^+} \Phi_A^{-1}(y)dy - c \overline H_T.
		\end{align*}
		We note that: 
		\begin{align*}
			\Sum{\nu_i \leq T}{} P_{\nu_i} \xi_i^- - \Sum{\tau_j \leq T}{} P_{\tau_j} \xi_j^+ &= - \Int{(0,T]}{} P_u^- dQ_u,
		\end{align*}
		and introduce the pure execution P\&L:
		\begin{align*}
			\Pi_T &= C_T + \Int{(0,T]}{} P_u^- dQ_u + Q_0 P_0 .
		\end{align*} 
		
		The additional term $\int_{(0,T]} P_u^- dQ_u$ accounts for the neutralisation of the variations of the mid-price.
		
		The total pure execution profit and loss can be expressed as
		\begin{align} 
			\label{EQN:market_making_pnl_mark_to_mid}
			\Pi_T &= - \frac{1}{2}\Sum{\nu_i \leq T}{} S_{\nu_i} \xi_i^- - \frac{1}{2}\Sum{\tau_j \leq T}{} S_{\tau_j} \xi_j^+ - \Sum{\nu_i \leq T}{} \Int{0}{\xi_i^-} \Phi_B^{-1}(y)dy - \Sum{\tau_j \leq T}{} \Int{0}{\xi_j^+} \Phi_A^{-1}(y)dy - c \overline H_T + Q_0 P_0.
		\end{align}
		Spread satisfy $S_t \ge \delta$ for all $t$,  $\Phi_B^{-1},\Phi_A^{-1}\ge 0$ by definition and $\Pi_0 = Q_0 P_0$.  
		Therefore, every term on the right-hand side of \eqref{EQN:market_making_pnl_mark_to_mid} is nonpositive, which yields
		\begin{equation*}
			\Pi_T - \Pi_0 \le - \frac{\delta}{2} \Bigg( \sum_{\nu_i \le T} \xi_i^- + \sum_{\tau_j \le T} \xi_j^+ \Bigg) - c \overline H_T
			= - \frac{\delta}{2} \mathcal V_T - c \overline H_T.
		\end{equation*}
		Since $\delta > 0$ and $c \geq 0$, the inequality is strict whenever at least one trade occurs, which completes the proof.
	\end{proof}
	
	Taken together, these results show that neither instantaneous nor multi-step round trips can be profitable once P\&L is evaluated on a pure execution basis, which isolates trading costs from inventory revaluations. 
	This condition parallels the classical notion of no price manipulation of Huberman and Stanzl~\cite{Huberman2004}, later extended to limit-order-book models \cite{GatheralSchied13, Alfonsi2014}.
	Still, one must also exclude transaction-triggered strategies, where preliminary trades are used to influence prices before unwinding a position. The next proposition shows that such manipulations are likewise ruled out in our framework.
	
	\begin{Proposition}[No transaction-triggered price manipulation] 
		\label{PROP:market_making_no_TTPM}
		Let $0 \leq \nu < \tau \leq T$. 
		Consider a two-step trading scheme consisting of a pre-trade followed by a target trade: 
		a pre-purchase of size $z \ge 0$ at time $\nu$, followed by a target sale of size $q \ge 0$ at time $\tau$. 
		Denote by $\Pi_T^{(z)}$ the pure execution profit and loss generated by this pair of trades, 
		and by $\Pi_T^{(0)}$ the corresponding P\&L when the pre-trade is omitted. 
		
		Then, for any admissible $z, q$, the incremental profit satisfies
		\begin{equation*}
			\Pi_T^{(z)} - \Pi_T^{(0)} \leq - \tfrac{\delta}{2} z - c < 0.
		\end{equation*}
		In particular, pre-trading to influence prices prior to execution of a target trade necessarily reduces the pure execution P\&L.
	\end{Proposition}
	
	\begin{proof}
		According to Equation~\ref{EQN:market_making_pnl_mark_to_mid}, with pre-trade, the pure execution P\&L is
		\begin{equation*}
			\Pi_T^{(z)} = -\frac{1}{2} S_{\nu} z - \Int{0}{z} \Phi_A^{-1}(y) dy - \frac{\eta}{2} z \Phi_A^{-1}(z) - \frac{1}{2} S_{\tau}^{(z)} q - \Int{0}{q} \Phi_B^{-1} (y) dy - \frac{\eta}{2} q \Phi_B^{-1}(q) - 2c.
		\end{equation*}
		In the absence of the pre-trade, the pure execution P\&L reduces to
		\begin{equation*}
			\Pi_T^{(0)} = - \frac{1}{2} S_{\tau}^{(0)} q - \Int{0}{q} \Phi_B^{-1} (y) dy - \frac{\eta}{2} q \Phi_B^{-1}(q) - c.
		\end{equation*}
		Subtracting the two expressions yields
		\begin{equation*}
			\Pi_T^{(z)} - \Pi_T^{(0)} = -\frac{1}{2} S_{\nu} z - \Int{0}{z} \Phi_A^{-1}(y) dy - \frac{\eta}{2} z \Phi_A^{-1}(z) - \frac{1}{2} q \left( S_{\tau}^{(z)} - S_\tau^{(0)} \right) - c.
		\end{equation*}
		Since $S_t \geq \delta$ for all $t$ and the inequality $S_\tau^{(z)} \geq S_\tau^{(0)}$ holds pathwise because the spread dynamics \eqref{EQN:spread_dynamics} are monotone in their initial condition. It follows that
		\begin{equation*}
			\Pi_T^{(z)} - \Pi_T^{(0)} \leq -\tfrac{\delta}{2}z - c < 0 .
		\end{equation*}
		The strict inequality holds because $\delta > 0$, which completes the proof.
	\end{proof}
	
	The framework excludes strategies based on preliminary trades to influence later executions: such manipulative tactics are necessarily loss-making, which ensures economic consistency by preventing price distortions through strategically timed trading and preserving the robustness of the market-making model.
	
	\begin{Corollary}[No arbitrage and no price manipulation on the underlying]
		\label{COR:no_arbitrage_no_manipulation}
		Under the assumptions of Lemma~\ref{LEM:market_making_no_instant_rt}, 
		Proposition~\ref{PROP:market_making_no_dynamic_arbitrage_underlying} and 
		Proposition~\ref{PROP:market_making_no_TTPM}, the proposed market model 
		precludes any form of arbitrage or price manipulation on the underlying. 
		In particular, every admissible trading strategy necessarily yields a 
		nonpositive pure execution P\&L.
	\end{Corollary}
	
	\subsection{Coupled markets and terminal manipulation}
	
	In the standalone option market, trades do not affect quotes and the spread ($\beta_t \le \alpha_t$) rules out profitable round-trips: any purchase at the ask must be unwound at the bid, yielding a nonpositive outcome. From the market maker’s perspective, attempts to manipulate the underlying to shift the reference price and option order intensities are also ineffective, as intensities are uniformly bounded by Assumption~\ref{ASS:market_making_option_intensities} and frictions in 
	the underlying make such schemes unprofitable.
	
	In coupled markets, identifying manipulative strategies is more intricate. A salient possibility in our setting is end-of-maturity manipulation: near $T$, a market maker holding a nonzero terminal option inventory $i$ may attempt to move the underlying to tilt the option payoff in her favor. Proposition~\ref{PROP:market_making_no_arb_coupled_cost_gain} quantifies this risk under a Lipschitz payoff and uniformly bounded order-book depth. From a control standpoint, however, this need not undermine well-posedness: the resulting gains are at most bounded at maturity, so the optimization problem and its value function remain viable. A fuller analysis of manipulation on coupled markets is an interesting direction for future research.
	
	\begin{Proposition}[Terminal no-arbitrage condition in the coupled market]
		\label{PROP:market_making_no_arb_coupled_cost_gain}
		Let $\overline H_T$ be the total number of interventions of the market maker on underlying market. Assume that there exists $m>0$ such that $f_A(u)\le m$ for all $u\ge0$ and $f_B(u)\le m$ for all $u\le0$, 
		and that the payoff function $\varphi$ is $L^{(\varphi)}$--Lipschitz in $p$. 
		Let $i\in\mathbb Z$ denote the terminal option inventory and $\Delta_T:=|P_T^{(H)}-P_T^{(0)}|$ the terminal price distortion generated by underlying impulses. 
		Then the terminal P\&L difference satisfies
		\begin{equation}\label{eq:terminal_noarb}
			\Delta \widehat \Pi_T
			:= \big(\Pi_T^{(H)} + i\varphi(P_T^{(H)})\big) 
			- \big(\Pi_T^{(0)} + i\varphi(P_T^{(0)})\big)
			\le -\big(\delta m - |i|L^{(\varphi)}\big)\Delta_T 
			- \frac{2m}{\overline H_T}\Delta_T^{2} 
			- c\overline H_T .
		\end{equation}
	\end{Proposition}
	
	\begin{proof}
		We compare two paths driven by the same exogenous order flow: one with the market maker’s underlying impulses (superscript $^{(H)}$) and one without (superscript $^{(0)}$). By \eqref{EQN:mid_price_dynamics}–\eqref{EQN:resilient_impact_dynamics}, a buy impulse of size $\xi$ at time $u$ shifts the mid price at time $T$ by
		\begin{equation*}
			\Big(\tfrac{\eta}{2}+\tfrac{1-\eta}{2}e^{-r(T-u)}\Big)\Phi_{A}^{-1}(\xi),
		\end{equation*}
		while a sell impulse of the same size shifts it by the negative of
		\begin{equation*}
			\Big(\tfrac{\eta}{2}+\tfrac{1-\eta}{2}e^{-r(T-u)}\Big)\Phi_{B}^{-1}(\xi).
		\end{equation*}
		Since $\tfrac{\eta}{2}+\tfrac{1-\eta}{2}e^{-r(T-u)}\le \tfrac12$, summing the contributions of all impulses and taking absolute values yields the elementary bound
		\begin{equation}\label{eq:deltaT_basic}
			\Delta_T:=\big|P_T^{(H)}-P_T^{(0)}\big|
			\le
			\frac12\Bigg(\sum_{\tau_j\le T}\Phi_A^{-1}(\xi_j^+)+\sum_{\nu_i\le T}\Phi_B^{-1}(\xi_i^-)\Bigg).
		\end{equation}
		We now turn to the cashflow side. Using the pure execution representation \eqref{EQN:market_making_pnl_mark_to_mid} and the bound $S_t\ge \delta$, the difference between the impacted and non-impacted pure execution P\&L satisfies
		\begin{equation*}
			\Pi_T^{(H)}-\Pi_T^{(0)} \le -\frac{\delta}{2}\sum_{\nu_i\le T}\xi_i^- -\frac{\delta}{2}\sum_{\tau_j\le T}\xi_j^+ -\sum_{\nu_i\le T}\int_0^{\xi_i^-}\!\Phi_B^{-1}(y)dy -\sum_{\tau_j\le T}\int_0^{\xi_j^+}\!\Phi_A^{-1}(y)dy - c\overline H_T.
		\end{equation*}
		Finally, since $\varphi$ is $L$-Lipschitz in $p$, we have
		\begin{equation*}
			i \big(\varphi(P_T^{(H)})-\varphi(P_T^{(0)})\big)\ \le\ |i|L^{(\varphi)}\Delta_T.
		\end{equation*}
		Adding this to the previous inequality gives the stated bound
		\begin{align}
			\label{EQN:market_making_no_arbitrage_coupled_market}
			\Delta \widehat \Pi_T &\le |i|L^{(\varphi)}\Delta_T - \frac{\delta}{2}\sum_{\nu_i\le T}\xi_i^- -\frac{\delta}{2}\sum_{\tau_j\le T}\xi_j^+ \nonumber \\
			&\qquad -\sum_{\nu_i\le T}\int_0^{\xi_i^-}\!\Phi_B^{-1}(y)dy -\sum_{\tau_j\le T}\int_0^{\xi_j^+}\!\Phi_A^{-1}(y)dy - c\overline H_T,
		\end{align}
		which holds pathwise. \\
		In particular, if $|i|L^{(\varphi)}\le \delta m$ and at least one impulse occurs so that $\overline H_T\ge1$, the right-hand side is strictly negative, and the claim follows.
		
		We have the local depth bound $0\le f_A(u)\le m$ (and symmetrically for $f_B$). Hence $\Phi_A(x)\le m x$ for every admissible quantity $x$, which implies $\Phi_A^{-1}(y)\ge y/m$ for $y\in[0,\Phi_A(x)]$, and analogously for $\Phi_B^{-1}$. 
		\begin{equation*}
			\Phi_A^{-1}(y)\ge \frac{y}{m},\qquad \Phi_B^{-1}(y)\ge \frac{y}{m}.
		\end{equation*}
		Substituting into \eqref{eq:deltaT_basic} gives
		\begin{equation}\label{eq:deltaT_sumxi}
			\Delta_T \le \frac{1}{2m}\Bigg(\sum_{\tau_j\le T}\xi_j^+ + \sum_{\nu_i\le T}\xi_i^- \Bigg).
		\end{equation}
		For any admissible $\xi \geq 0$, the integral inequality $\int_0^\xi \Phi^{-1}(y)dy\ge \xi^2/(2m)$. Applying it term by term and grouping the sums gives
		\begin{equation*}
			\Pi_T^{(H)}-\Pi_T^{(0)} \le -\frac{\delta}{2}\Bigg(\sum_{\nu_i\le T}\xi_i^-+\sum_{\tau_j\le T}\xi_j^+\Bigg) -\frac{1}{2m}\Bigg(\sum_{\nu_i\le T}(\xi_i^-)^2+\sum_{\tau_j\le T}(\xi_j^+)^2\Bigg) - c\overline H_T.
		\end{equation*}
		From Cauchy–-Schwarz inequality, $\sum (\xi)^2 \ge \frac{(\sum \xi)^2}{\overline H_T}$. Using this and \eqref{eq:deltaT_sumxi} yields
		\begin{equation*}
			\Pi_T^{(H)}-\Pi_T^{(0)} \le -\delta m\Delta_T -\frac{2m}{\overline H_T}\Delta_T^2 - c\overline H_T.
		\end{equation*}
		Hence Inequality~\ref{EQN:market_making_no_arbitrage_coupled_market} leads to
		\begin{equation*}
			\Delta \widehat \Pi_T \le -\big(\delta m - |i|L^{(\varphi)}\big)\Delta_T -\frac{2m}{\overline H_T}\Delta_T^{2} - c\overline H_T,
		\end{equation*}
		which holds pathwise. In particular, the inequality holds for all $\overline H_T \ge 1$. Hence, the right-hand side is strictly negative, and the claim follows.
	\end{proof}
	
	Proposition~\ref{PROP:market_making_no_arb_coupled_cost_gain} shows that an end-of-maturity manipulation channel does exist, yet under natural Lipschitz and depth bounds the associated gains remain bounded, so that the control problem retains its well-posedness, which will be the focus of the next section.
	
	\subsection{Well-posedness of the control problem}
	
	Before introducing the numerical method, we first need to ensure that the control problem is meaningful and non-trivial. 
	In particular, we must verify that the formulation given in \eqref{EQN:market_making_value_function} defines a well-posed optimization problem whose value function is finite. 
	Establishing this result guarantees that the model does not admit degenerate or unbounded solutions, and thus provides a solid foundation for the numerical analysis developed in the next section. 
	The following theorem formalizes these properties.
	
	\begin{Theorem}[Well-posedness and quadratic bounds]
		\label{TH:market_making_well_posedness_single}
		Under Assumptions~\ref{ASS:market_making_orderbook_integrability}, \ref{ASS:market_making_right_truncated_depth}, \ref{ASS:market_making_hawkes_stability}, \ref{ASS:market_making_ref_price_linear_growth}, \ref{ASS:market_making_option_intensities}  and \ref{ASS:market_making_penalties}, the value function $v$ in \eqref{EQN:market_making_value_function} is well defined and finite for all
		$(t,q,i,e)\in[0,T]\times\R\times\Z\times(\R^+)^5$.
		Moreover, there exist constants $C_T^{(-)},C_T^{(+)}\!>\!0$, depending only on \(T\) and model parameters, such that
		\begin{equation*}
			-C_T^{(-)}\bigl(1+q^2+i^2+\|e\|^2\bigr)
			\le v(t,q,i,e) \le
			C_T^{(+)}\bigl(1+|i|^2+\|e\|^2\bigr).
		\end{equation*}
	\end{Theorem}
	
	The argument relies on establishing stability and moment bounds for the model’s driving processes, which are then used to control the expected performance functional. 
	Lemma~\ref{LEM:market_making_inventory_moments} provides moment estimates for the option inventory, while Lemma~\ref{LEM:market_making_hawkes_bounds_with_intervention} ensures uniform bounds for the Hawkes intensities and their counting processes. 
	These results are propagated to the full market state in Proposition~\ref{PROP:market_making_state_moment_bounds}, which guarantees integrability and uniform moment control for all state variables on $[0,T]$. 
	Based on these estimates, a global lower bound for the value function is obtained in Proposition~\ref{PROP:market_making_lower_bound_value_function}, and an upper bound in Proposition~\ref{PROP:market_making_upper_bound_value_function}. 
	Together, these results establish the finiteness and quadratic growth of the value function, proving Theorem~\ref{TH:market_making_well_posedness_single}. 
	The detailed arguments are presented in Appendix~\ref{APPENDIX:proofs}.
	
	\section{Numerical methodology and experiments}
	\label{SEC:numerical_method}
	We propose a neural, simulation–based approach to approximate optimal quoting and hedging in our market–making model. Inspired by Deep Hedging~\cite{Buehler2019}, we use direct policy search: a network maps the current state to quotes and an inventory–normalized hedge, and its parameters are trained on a differentiable simulator to maximize the objective function.
	
	Although the model is formulated in continuous time, the numerical implementation developed in this section relies on a time-discretized simulator combined with neural-network-based policy optimization.
	While this approach does not provide guarantees of global optimality, it allows one to approximate solutions in high-dimensional stochastic control problems where classical grid-based methods suffer from the curse of dimensionality.
	To validate the numerical methodology, we first tested the learning procedure in a simplified setting with geometric diffusion dynamics, for which analytical benchmarks are available. 
	In this case, the learned policy successfully recovered Black--Scholes option prices together with the associated delta-hedging strategy. 
	This validation provided confidence in the ability of the method to approximate optimal solutions before extending it to the more complex dynamics considered in this paper.
	
	\subsection{Simulation and neural policy learning}
	
	\paragraph*{Discrete-time model and simulator} ~\\
	Recalling that $T > 0$ is the time horizon for the market making problem. We discretize the interval $[0, T]$ into $N > 0$ time steps. Let $\delta_t = T/N$ denote the time step size, and define the discrete time grid as
	\begin{equation*}
		\T = \{t_0 = 0, \dots, t_i = i \delta_t, \dots, t_N = T\}.
	\end{equation*}
	Let $\pi$ denote the agent’s discrete policy, which prescribes a decision at each point in the grid $\T$. For every $t_i \in \T$, the market maker chooses the controls $\left(\Gamma^\pi, \alpha^\pi, \beta^\pi\right)_{t_i}$. Inspired by the work of Buehler et al. \cite{Buehler2019}, we use an inventory-normalized hedge ratio $\Gamma^\pi_{t_i} \in [-1, 1]$. 
	
	Here, $I^\pi$ denotes the discretized option inventory process, which is controlled by the market maker’s option quotes $\beta^\pi_{t_i}$ and $\alpha^\pi_{t_i}$. These quotes influence the arrival intensities of the counting processes $N^a$ and $N^b$. Define the option–market counting increments
	\begin{equation}
		\label{EQN:market_making_option_arrivals_increments}
		\Delta N_{t_i}^a := N_{t_{i+1}}^a - N_{t_i}^a, \qquad
		\Delta N_{t_i}^b := N_{t_{i+1}}^b - N_{t_i}^b,
	\end{equation}
	then the option inventory position evolves as
	\begin{equation*}
		I_{t_{i+1}}^\pi = I_{t_i}^\pi + \Delta N_{t_i}^{a} - \Delta N_{t_i}^{b}, \\
	\end{equation*}
	and the hedged underlying
	\begin{equation*}
		Q_{t_{i+1}}^\pi = \Gamma_{t_i}^\pi I_{t_{i+1}}^\pi,
	\end{equation*}
	with initial conditions $I_0^\pi=i\in\mathbb{Z}$ and $Q_0^\pi=q\in\mathbb{R}$. In the following, we write $\Delta Q_{t_i}^\pi := Q_{t_{i+1}}^\pi - Q_{t_i}^\pi$.
	In our numerical implementation, the underlying inventory adjustment is clipped to remain within admissible trading volumes, according to:
	\begin{equation*}
		\Delta Q_{t_i}^\pi = 
		\begin{cases}
			\max \left\{ - \Phi_B(P_{t_i}^\pi - S_{t_i}^\pi/2), \Delta Q_{t_i}^\pi \right\} & \Delta Q_{t_i}^\pi < 0, \\
			\min \left\{ \Phi_A(+\infty), \Delta Q_{t_i}^\pi \right\} & \Delta Q_{t_i}^\pi > 0, \\
			0 & \text{otherwise}.
		\end{cases}
	\end{equation*}
	This ensures that the market maker does not attempt to trade beyond available liquidity. In practice, we use a smooth soft-clipping function, preserving the differentiability of the simulator required for gradient-based training. $\Phi_A$ and $\Phi_B$ are chosen such that the desired hedging action rarely reaches these bounds, reflecting realistic market behaviour. In configurations where this may not hold, the clipping rate is monitored throughout training as a diagnostic metric to ensure stability of the gradient estimates.
	
	We now express the evolution of the market maker's cash process $X^\pi$ along the grid. It is given by
	\begin{align}
		\label{EQN:discretized_cash}
		X_{t_{i+1}}^\pi = X_{t_i}^\pi &+ \alpha_{t_i}^\pi \Delta N_{t_i}^a - \beta_{t_i}^\pi \Delta N_{t_i}^b + P_B \left(P_{t_i}^\pi - \frac{1}{2} S_{t_i}^\pi, (\Delta Q_{t_i}^\pi)^- \right) \nonumber \\
		&\qquad - P_A \left(P_{t_i}^\pi + \frac{1}{2} S_{t_i}^\pi, (\Delta Q_{t_i}^\pi)^+ \right) - c \Ind{\Delta Q_{t_i}^\pi \neq 0} .
	\end{align}
	The inventory–normalized hedge ratio $\Gamma^\pi$ allows the policy to be specified in relative terms, ensuring scalability across inventory levels with fewer parameters and improved stability. Together with the discretized dynamics, this yields a self-contained simulator on the grid $\T$, directly amenable to Monte Carlo estimation and gradient-based training. All other state variables are discretized by standard schemes. Since their dynamics are driven by Hawkes arrivals, this introduces no additional numerical difficulty. We next specify the measurability of the discrete processes and the information available to the agent, completing the definition of admissible strategies in discrete time.
	
	\paragraph*{Measurability of processes} ~\\
	Let $t_i \in \T$ denote a discrete time point, we define the state of the system at that time as:
	\begin{equation*}
		Z_{t_i} := (P_{t_i}^\pi, D_{t_i}^\pi, S_{t_i}^\pi, \lambda_{t_i}^{\pi, -}, \lambda_{t_i}^{\pi, +}, I_{t_i}^\pi, Q_{t_i}^\pi, N_{t_i}^a, N_{t_i}^b, N_{t_i}^-, N_{t_i}^+).
	\end{equation*}
	We introduce the filtration $(\Fc_{t_i})_{i \in \N}$ describing the information available up to time $t_i$:
	\begin{equation*}
		\Fc_{t_i} := \sigma(Z_{t_0}, \cdots, Z_{t_i}).
	\end{equation*}
	At each time $t_i$, the market state variables $(P_{t_i}^\pi, D_{t_i}^\pi, S_{t_i}^\pi, \lambda_{t_i}^{-,\pi}, \lambda_{t_i}^{+,\pi})$ are $\Fc_{t_i}$–measurable, as they are entirely determined by past realizations.  
	Similarly, the agent’s controls $(\alpha_{t_i}^\pi,\beta_{t_i}^\pi,\Gamma_{t_i}^\pi)$ are chosen on the basis of $\Fc_{t_i}$, and are therefore $\Fc_{t_i}$–measurable.  
	By contrast, the order-flow increments $(\Delta N_{t_i}^a, \Delta N_{t_i}^b, \Delta N_{t_i}^-, \Delta N_{t_i}^+)$ are only revealed over the interval $[t_i,t_{i+1})$ and hence belong to $\Fc_{t_{i+1}}$.  
	These properties ensure the internal consistency of the discrete-time model and clarify the informational structure available to the agent at each step of the simulation.
	
	\paragraph*{Objective function} ~\\
	Within the discrete-time setting and the measurability framework established above, we now introduce a tractable version of the objective originally defined in \eqref{EQN:objective_function}. For a given policy $\pi$, the corresponding criterion reads
	\begin{equation}
		\label{EQN:discretized_objective_function}
		J_\pi(x, q, i, e) = \Expect{X_T^{\pi} - \delta_t \Sum{k=0}{N - 1} (g+h)(t_{k + 1}, Q_{t_{k + 1}}^\pi, I_{t_{k + 1}}^\pi, E_{t_{k + 1}}^\pi) + L(T, Q_T^\pi, I_T^\pi, P_T^\pi, S_T^\pi)} .
	\end{equation}
	
	The function $L$ denotes the terminal liquidation function introduced in \eqref{EQN:liquidation_function}, while $g$ and $h$ collect the running penalty terms reflecting hedging frictions and quoting incentives. The vector $e = (p, d, s, \lambda_0^-, \lambda_0^+)$ specifies the initial exogenous market configuration.
	
	Using the same arguments than in Proposition~\ref{PROP:reduction_by_cash_additivity}, the objective function in \eqref{EQN:discretized_objective_function} admits the following decomposition:
	\begin{align*}
		J_\pi(x, q, i, e) &= x + \mathbb{E}\Bigg[\delta_t \Sum{k=0}{N - 1} \left( \alpha_{t_k}^\pi \lambda^a - \beta_{t_k}^\pi \lambda^b \right)(t_k, \alpha_{t_k}^\pi, \beta_{t_k}^\pi, E_{t_k}^\pi) \\
		&\quad + \Sum{k=0}{N - 1}  \Bigg(P_B\left(P_{t_k}^\pi - \frac{1}{2} S_{t_k}^\pi, \left(\Delta Q_{t_k}^\pi \right)^-\right) \\
		&\quad - P_A\left(P_{t_k}^\pi + \frac{1}{2} S_{t_k}^\pi, \left(\Delta Q_{t_k}^\pi \right)^+\right) - c \Ind{\Delta Q_{t_k}^\pi \neq 0} \Bigg) \\
		&\quad - \delta_t \Sum{k=0}{N - 1} (g+h)(t_{k + 1}, Q_{t_{k + 1}}^\pi, I_{t_{k + 1}}^\pi, E_{t_{k + 1}}^\pi) + L(T, Q_T^\pi, I_T^\pi, P_T^\pi, S_T^\pi) \Bigg]					
	\end{align*}
	The problem faced by the agent is to select an admissible policy $\pi^\star$ that maximizes the expected objective, namely
	\begin{equation*}
		\pi^\star = \underset{\pi \in \Pi}{\arg \max} J_\pi(x, q, i, e) .
	\end{equation*}
	This optimization naturally leads to the definition of the discrete-time value function,
	\begin{equation*}
		\hat{v}(t, q, i, e) = \underset{\pi \in \Pi}{\sup} J_\pi(x, q, i, e)
	\end{equation*}
	which represents the maximal attainable performance given the current state. 
	
	\paragraph*{Training by Monte Carlo and autodifferentiation} ~\\
	The policy $\pi_\theta$ is parameterized by a neural network with weights $\theta$. 
	Given $\theta$, we simulate $M$ independent market trajectories under the discrete-time dynamics. 
	For each trajectory, the agent sequentially selects at each step a bid quote, a spread (with the ask quote set to bid plus spread, ensuring bid $\leq$ ask), and a normalized hedge ratio $\Gamma^\pi$. 
	The resulting controls generate order arrivals and inventory updates, and the simulator propagates the impacts so that all state variables evolve consistently with the model. 
	The realized objective values $J_\pi^{(m)}(\theta)$ are then collected, and their empirical average
	\begin{equation*}
		\widehat{J}_\pi(\theta) := \frac{1}{M}\sum_{m=1}^M J_\pi^{(m)}(\theta)
	\end{equation*}
	provides a Monte Carlo estimate of the expected objective.
	
	To render the simulator almost everywhere differentiable, we assume that the bid and ask intensity functions are differentiable almost everywhere. 
	Under this assumption, automatic differentiation yields unbiased gradients $\nabla_\theta \widehat{J}_\pi(\theta)$, which are used to update the network parameters via stochastic gradient ascent. 
	In practice, we employ a standard multilayer perceptron architecture, together with feature transformations that enhance training stability and improve sample efficiency. 
	Potential pointwise non-differentiabilities are handled by standard machine-learning techniques, ensuring that gradient-based optimization remains effective in practice.
	
	\subsection{Experimental results}
	\subsubsection{Parameters}				
	\paragraph*{Underlying market} ~\\
	While our model could in principle be calibrated to historical data, the parameter values used in our experiments were chosen to produce plausible market dynamics and consistent orders of magnitude for both the underlying and the option markets.
	We consider a trading horizon of five trading weeks, corresponding to 
	\begin{equation*}
		T = \tfrac{25}{252} \text{years}.
	\end{equation*}
	In our numerical experiments we assume that the agent has ten trading opportunities per day, so that the interval $[0,T]$ is discretized into $N=250$ steps.
	To describe the underlying market, we adopt a linear order book representation with constant depth densities on both sides up to finite cutoffs.  
	On the bid side, liquidity is specified as
	\begin{equation*}
		f_B(u) = c_B \mathbf{1}_{\{-U_B \le u \le 0\}},
	\end{equation*}
	yielding the cumulative depth $\Phi_B(x)=c_Bx$ for $0 \le x \le U_B$, with generalized inverse $\Phi_B^{-1}(y)=y/c_B$ for $y \in [0,c_B U_B]$.  
	On the ask side, we similarly set
	\begin{equation*}
		f_A(u) = c_A \mathbf{1}_{\{0 \le u \le U_A\}},
	\end{equation*}
	so that $\Phi_A(x)=c_Ax$ for $0 \le x \le U_A$, saturating at $\Phi_A(+\infty)=c_AU_A$, with inverse $\Phi_A^{-1}(y)=y/c_A$ for $y \in [0,c_A U_A]$.
	These explicit forms make execution costs analytically tractable.  
	If $b>0$ denotes the best bid and $a>0$ the best ask, then selling a quantity $q^- \in [0,c_B U_B]$ generates the revenue
	\begin{equation*}
		P_B(b,q^-) = \int_{-q^-/c_B}^{0} (b+u)c_Bdu
		= bq^- - \frac{(q^-)^2}{2c_B},
	\end{equation*}
	while purchasing a quantity $q^+ \in [0,c_A U_A]$ entails the cost
	\begin{equation*}
		P_A(a,q^+) = \int_{0}^{q^+/c_A} (a+u)c_Adu
		= aq^+ + \frac{(q^+)^2}{2c_A}.
	\end{equation*}
	
	We fix linear book densities at $(c_A,c_B)=(100,100)$ and finite depths $(U_A,U_B)=(0.5,0.5)$, implying a total liquidity of $50$ units on each side.  
	At the start of the experiment, the underlying mid-price is set to $P_0 = 100$, with an initial bid–ask spread $S_0 = 0.10$ and a minimal admissible spread $\delta = 0.02$.  
	Furthermore, we assume that $D_0 = 0$.
	
	For simplicity, we model exogenous buy and sell orders arrive at the best quotes according to independent homogeneous Poisson processes with intensities $\lambda^- > 0$ and $\lambda^+ > 0$. This tractable specification is particularly convenient for the construction of the option reference price, since it yields explicit expressions for the variance of the mid-price dynamics. Each order is associated with a random mark $M_t^\pm \in [0,1]$ representing the fraction of available depth consumed.  
	We assume $M^- \sim \mathrm{Beta}(a_-,b_-)$ and $M^+ \sim \mathrm{Beta}(a_+,b_+)$, independent of the Poisson clocks and of the past.  
	The Beta distribution provides a convenient way to generate bounded marks in $[0,1]$, with the parameters controlling the typical trade size.  
	In our experiments, we set $(a_\pm, b_\pm)=(2,5)$, producing a majority of small trades. 
	
	Exogenous order flow is set to produce, on average, 30 events per trading day on each side. Annualized with 252 trading days, this corresponds to $\lambda^+ = \lambda^- = 30 \times 252 = 7560$ arrivals per year.  
	We assume that 30\% of price impact is permanent, while the remaining 70\% mean-reverts at resilience $r=60$ events per day (that is, $r=60 \times 252 = 15{,}120$ annually).  
	The spread reverts to its lower bound at rate $\rho=200$ per day, or $\rho=200 \times 252 = 50{,}400$ annually.  
	For the baseline parameter set, we normalize the fixed cost of each impulse trade in the underlying to zero, i.e., $c=0$. All parameter values are collected in Table~\ref{tab:baseline-params}.
	
	\begin{table}[H]
		\centering
		\begin{tabular}{cccccccccccc}
			\toprule
			$P_0$ & $D_0/Q_0$ & $S_0$ & $\delta$ & $U_A,U_B$ &  $c_A,c_B$ & $\lambda^{+/-}$ & $(a_\pm, b_\pm)$ & $\eta$ & $r$ & $\rho$ & $T$ \\
			\midrule
			$100$ & $0$ & $0.10$ & $0.02$ & $0.5$ & $100$ & $7560$ & $(2,5)$ & $0.3$ & $15{,}120$ & $50{,}400$ & $25/252$ \\
			\bottomrule
		\end{tabular}
		\caption{Baseline parameters of the underlying market.}
		\label{tab:baseline-params}
	\end{table}
	Since $D_0 = 0$ and $B_0 = P_0 - S_0/2 > 0$, the initial conditions of Lemma~\ref{LEM:market_making_positivity_quotes} are satisfied. 
	Therefore, bid and ask quotes remain nonnegative throughout all simulations for any admissible strategy.
	
	\paragraph*{Option market} ~\\
	The market maker is assumed to provide liquidity on a European call option with strike $K=98$ and payoff 
	\begin{equation*}
		(x-K)_+ = \max(x-K,0).
	\end{equation*}
	To model the option order arrival rates, we specify the following functional form for the bid and ask intensities, which allows for flexible specification through scale and shift parameters $\mu_a,\mu_b \in \R$. 
	Specifically, for $(t,\alpha,\beta,e) \in [0,T]\times\R^7$, we set
	\begin{equation}
		\label{EQN:market_making_option_intensities}
		\lambda^b(t,\alpha,\beta,e) = \overline{\lambda}^b  \sigma\!\left(-k_b (b(t,e)-\beta)+\mu_b\right), 
		\qquad
		\lambda^a(t,\alpha,\beta,e) = \overline{\lambda}^a  \sigma\!\left(k_a (b(t,e)-\alpha)+\mu_a\right),
	\end{equation}
	where $\sigma(x)=(1+e^{-x})^{-1}$ is the logistic sigmoid, $\overline{\lambda}^a,\overline{\lambda}^b>0$ and $k_a,k_b>0$. As before, $b$ denotes the option reference price. Its construction is detailed below.
	
	On a daily scale, we set $\overline\lambda^a=\overline\lambda^b=200$ arrivals at competitive quotes.  
	Annualized, this corresponds to $\overline\lambda^a=\overline\lambda^b=200\times 252 = 50{,}400$ arrivals per year.  
	The exponential slopes are fixed at $k_a=k_b=50$, so that intensities decay rapidly once quotes deviate from the efficient level by more than a few ticks. We set the initial option inventory to $I_0=0$, and summarize the full set of parameters of the option order flow model in Table~\ref{tab:market_making_option_params}.
	
	\begin{table}[H]
		\centering
		\begin{tabular}{ccccc}
			\toprule
			$I_0$ & $\overline\lambda^a$ & $\overline\lambda^b$ & $k_a,k_b$ & $K$  \\
			\midrule
			$0$ &$50{,}400$ & $50{,}400$ & $50$ & $98$ \\
			\bottomrule
		\end{tabular}
		\caption{Baseline parameters of the option order flow model.}
		\label{tab:market_making_option_params}
	\end{table}
	
	\paragraph*{Reference price and penalties} ~\\
	The option reference price $b$ is defined as a Black–Scholes value computed with an effective volatility extracted from the order-driven mid-price dynamics. 
	The idea is to match the instantaneous variance of the compound-Poisson mid-price process with that of a diffusion, and then to value the option under the corresponding proxy diffusion.
	With independent Poisson flows on the bid and ask at intensities $(\lambda^-,\lambda^+)$ and marks $M_t^\pm\in[0,1]$, a jump on the bid shifts the mid by
	$-\tfrac12\Phi_B^{-1}\!\big(M_t^-\Phi_B(B_{t^-})\big)$ and a jump on the ask by $+\tfrac12\Phi_A^{-1}\!\big(M_t^+\Phi_A(+\infty)\big)$.
	In the linear book,
	\begin{equation*}
		\Phi_A(+\infty)=c_AU_A,\quad \Phi_A^{-1}(y)=\tfrac{y}{c_A}\ \Rightarrow\ 
		\Phi_A^{-1}\!\big(M^+\Phi_A(+\infty)\big)=M^+U_A,
	\end{equation*}
	and similarly $\Phi_B^{-1}\!\big(M^-\Phi_B(\cdot)\big)=M^-U_B$ (with $U_B$ small in price units, $B_{t^-}\ge U_B$ is natural in practice).
	Over $[0,t]$, the quadratic variation is the sum of squared jumps. Taking expectations and using that the numbers of bid/ask arrivals are Poisson with means $\lambda^\pm t$, while the marks are i.i.d. and independent of the counts, the expected quadratic variation is the mean count times the mean squared jump size on each side. Hence
	\begin{equation*}
		\Expect{[P]_t} = \tfrac14\Big(\lambda^+\mathbb{E}[(M^+U_A)^2]+\lambda^-\mathbb{E}[(M^-U_B)^2]\Big)t,
	\end{equation*}
	so the variance rate of the mid is
	\begin{equation*}
		\nu = \frac{1}{t}\Expect{[P]_t} = \tfrac14\Big(\lambda^+U_A^2\mathbb{E}[(M^+)^2]+\lambda^-U_B^2\mathbb{E}[(M^-)^2]\Big).
	\end{equation*}
	As $M^\pm\sim\mathrm{Beta}(a_\pm,b_\pm)$, we have
	\begin{equation*}
		\Expect{(M^\pm)^2} = \frac{a_\pm(a_\pm+1)}{(a_\pm+b_\pm)(a_\pm+b_\pm+1)}.
	\end{equation*}
	The effective volatility is then defined as
	\begin{equation}\label{eq:sigma_eff}
		\sigma_{\mathrm{eff}}(t,e)=\frac{\sqrt{\nu}}{P_t}
		=\frac{1}{2P_t}\sqrt{\lambda^+U_A^2\E[(M^+)^2]+\lambda^-U_B^2\E[(M^-)^2]}.
	\end{equation}
	Given time-to-maturity $T-t$ and strike $K$, the option reference price is defined as
	\begin{equation*}
		b(t,e) = \mathrm{BS}\!\big(P_t, K, T-t, r, \sigma_{\mathrm{eff}}(t,e)\big),
	\end{equation*}
	where $\mathrm{BS}$ denotes the standard Black-Scholes formula, evaluated at the effective volatility $\sigma_{\mathrm{eff}}$ obtained from the mid-price dynamics.  
	In our numerical implementation we set the risk-free rate $r=0$, which is consistent with the symmetric order book assumption.
	
	We assume that market participants observe the underlying and estimate the option value using the Black–Scholes formula with effective volatility $\sigma_{\mathrm{eff}}$. 
	This choice serves as a practical specification for the reference price, which in turn shapes the intensities of buy and sell orders in the option market. 
	Alternative specifications of the reference price could be considered, allowing the framework to accommodate different perceptions of option value by the market. 
	The reference price is used solely within the intensity and penalty functions and does not form part of the information directly available to the market maker.
	
	To enforce hedging discipline, we introduce a quadratic penalty anchored to the Black-Scholes delta,
	\begin{equation*}
		g(t,q,i,e) = 
		\kappa_{\mathrm{hedge}} \Big(q+i\partial_p b(t,e)\Big)^2,
	\end{equation*}
	where $\partial_p b$ is the option delta under $\sigma_{\mathrm{eff}}$.  
	To complement the hedging penalty, we introduce an incentive term that only penalizes under-provision of liquidity.  
	Let $\bar\lambda^a,\bar\lambda^b$ denote the maximal ask and bid intensities, and set the benchmark
	\begin{equation*}
		\Lambda := \theta_{\mathrm{flow}}\big(\bar\lambda^a + \bar\lambda^b\big),
	\end{equation*}
	with $\theta \in (0,1)$ a fixed proportion of the maximal total flow.  
	Given quotes $(\alpha,\beta)$ generating order flow intensities $(\lambda^a,\lambda^b)$, the activity penalty is then defined as
	\begin{equation}
		\label{EQN:incentive_pen}
		h(t,\alpha,\beta,e) = 
		\kappa_{\mathrm{act}} \Big(\Lambda - \big(\lambda^a(t,\alpha,\beta,e)+\lambda^b(t,\alpha,\beta,e)\big)\Big)_+^2,
	\end{equation}
	with $\kappa_{\mathrm{act}}>0$ a penalty coefficient and $(x)_+=\max(x,0)$.  
	This formulation ensures that the penalty vanishes whenever the total captured flow exceeds the benchmark $\Lambda$, while increasing quadratically otherwise.  
	Unless specified otherwise, all subsequent experiments use these parameter values; they are consistent with Assumption~\ref{ASS:market_making_penalties} and preserve the qualitative behavior described in Section~\ref{SUBSEC:market_making_objective_market_maker}.
	In our numerical applications, we adopt the following parameter values:
	\begin{equation*}
		\kappa_{\mathrm{hedge}} = 4, 
		\qquad \theta_{\mathrm{flow}} = 5\%, 
		\qquad \kappa_{\mathrm{act}} = 0.1.
	\end{equation*}
	These choices provide a sufficiently strong incentive for hedging discipline while ensuring a minimal level of liquidity provision without excessively penalizing the agent.
	
	\subsubsection{Naive benchmark agent}
	Throughout the experiments, the learned neural policy is systematically compared against a naive benchmark agent that ignores hedging-induced market impact.
	The benchmark agent follows a simple parametric quoting rule: it posts constant bid and ask spreads around the Black--Scholes reference price $b$, namely
	\begin{equation*}
		\beta_t^{\mathrm{naive}}(t, e) = b(t,e) - \delta_{\mathrm{bid}}^{\mathrm{naive}}, 
		\qquad 
		\alpha_t^{\mathrm{naive}}(t, e) = b(t,e) + \delta_{\mathrm{ask}}^{\mathrm{naive}},
	\end{equation*}
	where $\delta_{\mathrm{bid}}^{\mathrm{naive}}, \delta_{\mathrm{ask}}^{\mathrm{naive}} > 0$ are constant half-spreads, potentially asymmetric between the bid and ask sides.
	The agent hedges its option inventory using the Black--Scholes delta, targeting an underlying position given by
	\begin{equation*}
		Q_t^{\mathrm{naive}} = -I_t \cdot \partial_p b(t, e).
	\end{equation*}
	
	To ensure a fair comparison within this restricted class of policies, the parameters $(\delta_{\mathrm{bid}}^{\mathrm{naive}}, \delta_{\mathrm{ask}}^{\mathrm{naive}})$ are optimized independently for each experimental configuration. The optimization is performed using a derivative-free Nelder--Mead procedure based on Monte Carlo simulations, with the objective of maximizing \eqref{EQN:discretized_objective_function}.
	In each experimental setting, the performance of the naive benchmark is reported alongside that of the learned policy, enabling a direct assessment of the value added by impact-aware and adaptive strategies.
	
	\subsubsection{Impact of initial inventory}
	
	Training is conducted on batches of $10{,}000$ simulated paths over $500$ epochs, with a learning rate of $10^{-4}$.  
	The policy network is implemented as a standard multilayer perceptron with ReLU activations and input normalization. 
	
	\paragraph*{Illustration in a strongly imbalanced inventory setting} ~\\
	We start by investigating the impact of asymmetric initial conditions by considering non-zero initial option inventories $I_0 \neq 0$. Non-zero inventories introduce a structural imbalance in the market maker’s portfolio, which must be dynamically managed through a combination of hedging and order flow control. This setting gives rise to inherently asymmetric optimal policies.
	We start by analyzing an extreme configuration with a large negative initial inventory, $I_0 = -100$. This stressed scenario allows us to highlight the qualitative mechanisms learned by the agent.
	
\begin{figure}[H]
		\centering
		\includegraphics[width=0.75\textwidth]{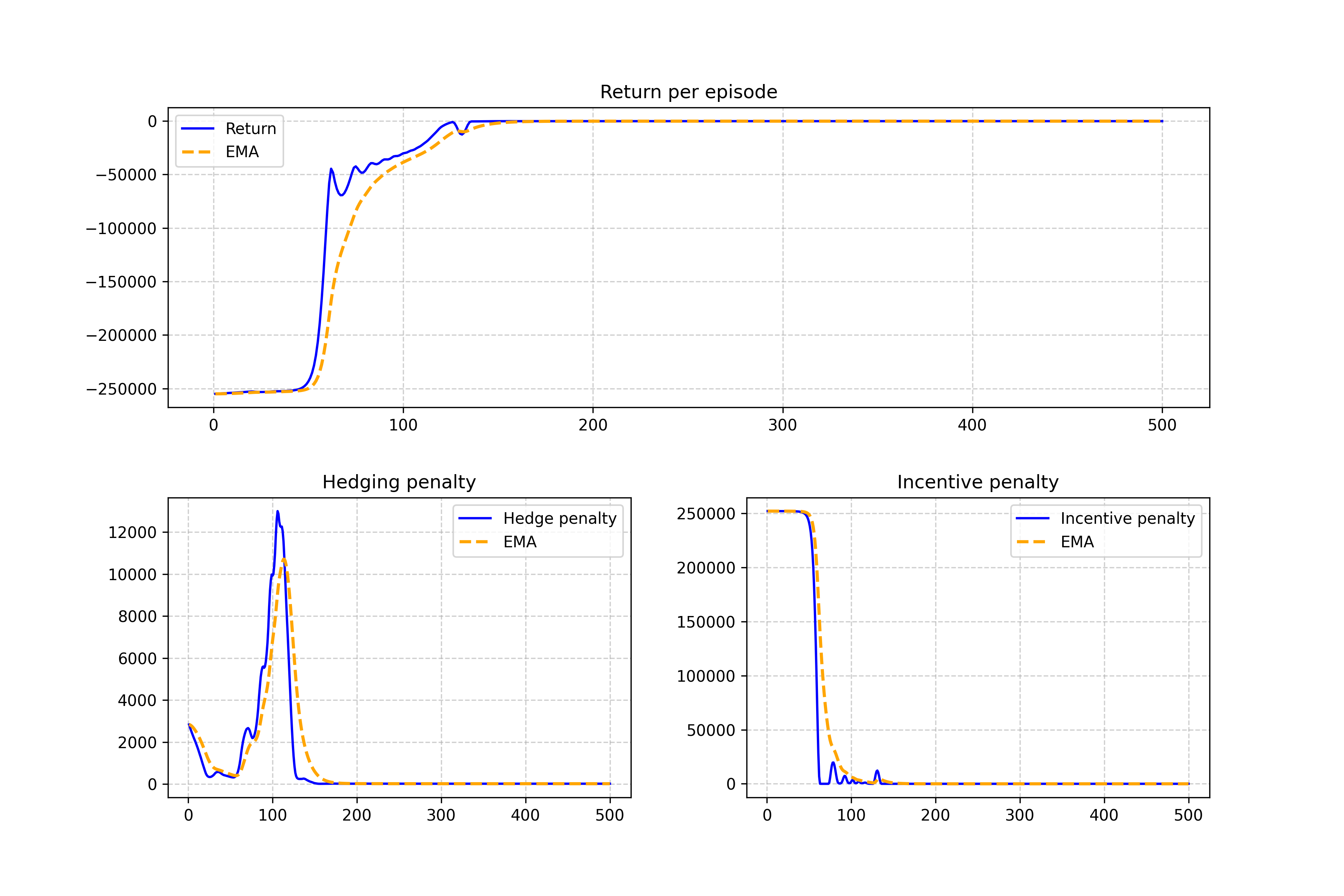}
		\caption{Learning metrics for linear order book with $I_0 = -100$.}
		\label{FIG:market_making_linear_impact_learning_metrics_neg_inv}
	\end{figure}

\begin{figure}[H]
		\centering
		\begin{subfigure}[t]{0.48\textwidth}
			\centering
			\includegraphics[width=\linewidth]{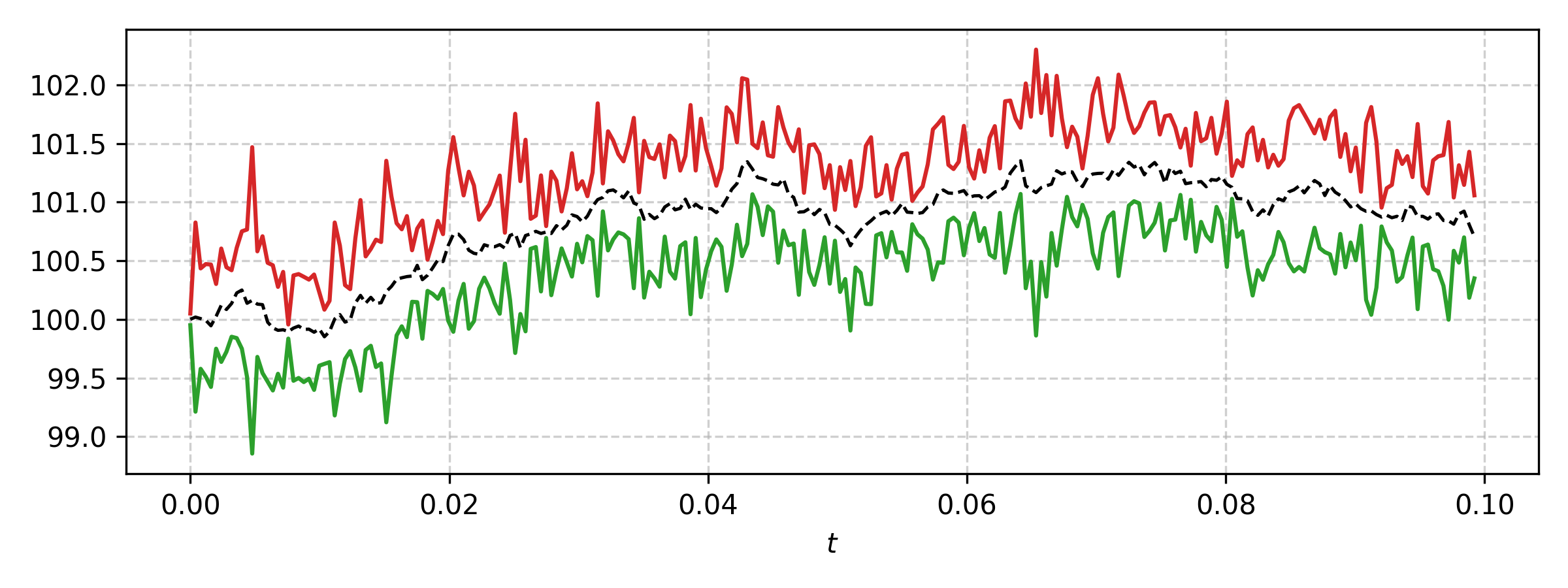}
			\caption{Underlying mid-price path ($I_0=-100$).}
			\label{FIG:market_making_linear_impact_symetric_case_underlying_evolution_idx0}
		\end{subfigure}
		\hfill
		\begin{subfigure}[t]{0.48\textwidth}
			\centering
			\includegraphics[width=\linewidth]{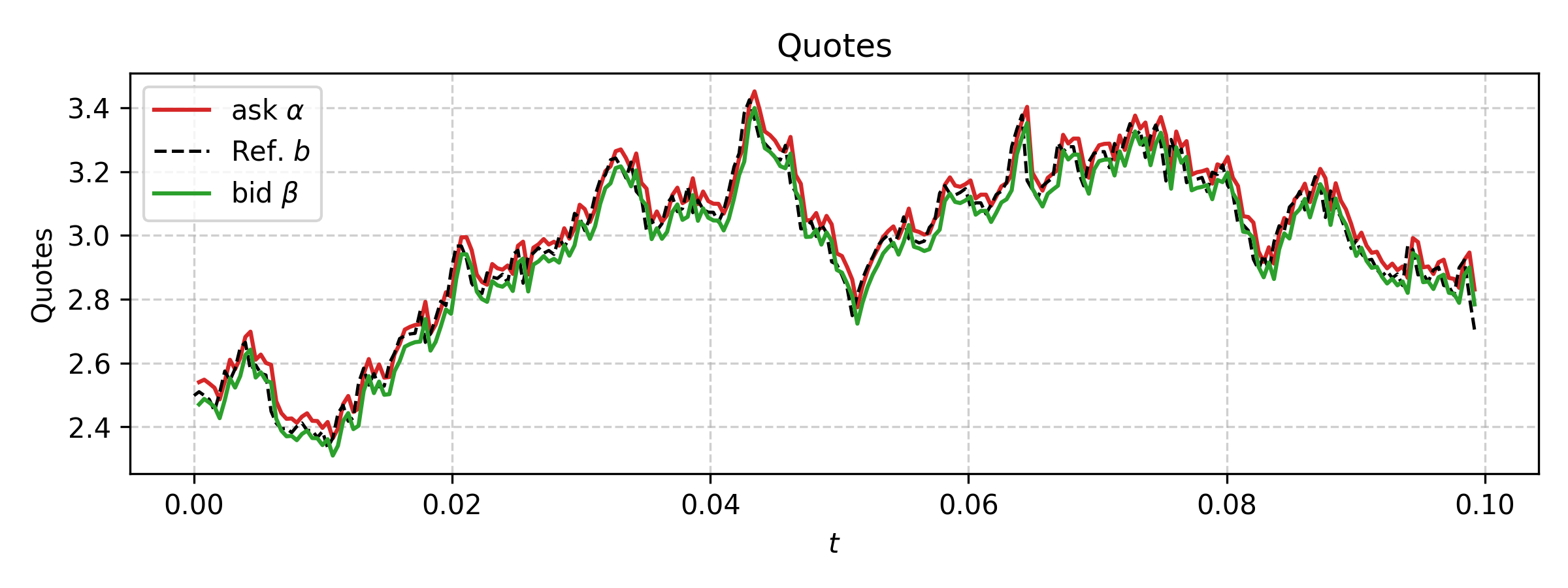}
			\caption{Bid and ask quotes together with the reference price ($I_0=-100$).}
			\label{FIG:market_making_linear_impact_symetric_case_quotes_idx0}
		\end{subfigure}
		\caption{Sample path: underlying price and corresponding quoting behavior of the agent.}
		\label{FIG:market_making_linear_impact_symetric_case_underlying_quotes}
	\end{figure}

	Figure~\ref{FIG:market_making_linear_impact_learning_metrics_neg_inv} documents the training dynamics across episodes: the return rises steadily and then plateaus, indicating that the policy discovers a stable quoting and hedging regime.  
	The hedging penalty declines markedly as the agent learns to align its underlying position with the option exposure, thereby reducing costly discrepancies.  
	The incentive penalty quickly drops toward low levels and remains stable, which shows that the agent consistently meets the activity requirement without being punished for supplying additional liquidity.  
	To gain intuition on the learned strategy, we first examine a representative sample path. 
	Figure~\ref{FIG:market_making_linear_impact_symetric_case_underlying_quotes} shows the joint evolution of the underlying price and the corresponding bid--ask quotes. 
	This pathwise view highlights how the agent dynamically adapts its quoting behavior to both price movements and inventory imbalances.
	To better understand the policy learned in this setting, we now examine the average evolution of both the strategy and the environment across $10{,}000$ simulated paths.
	
\begin{figure}[H]
		\centering
		\begin{subfigure}[t]{0.49\textwidth}
			\centering
			\includegraphics[width=1\textwidth]{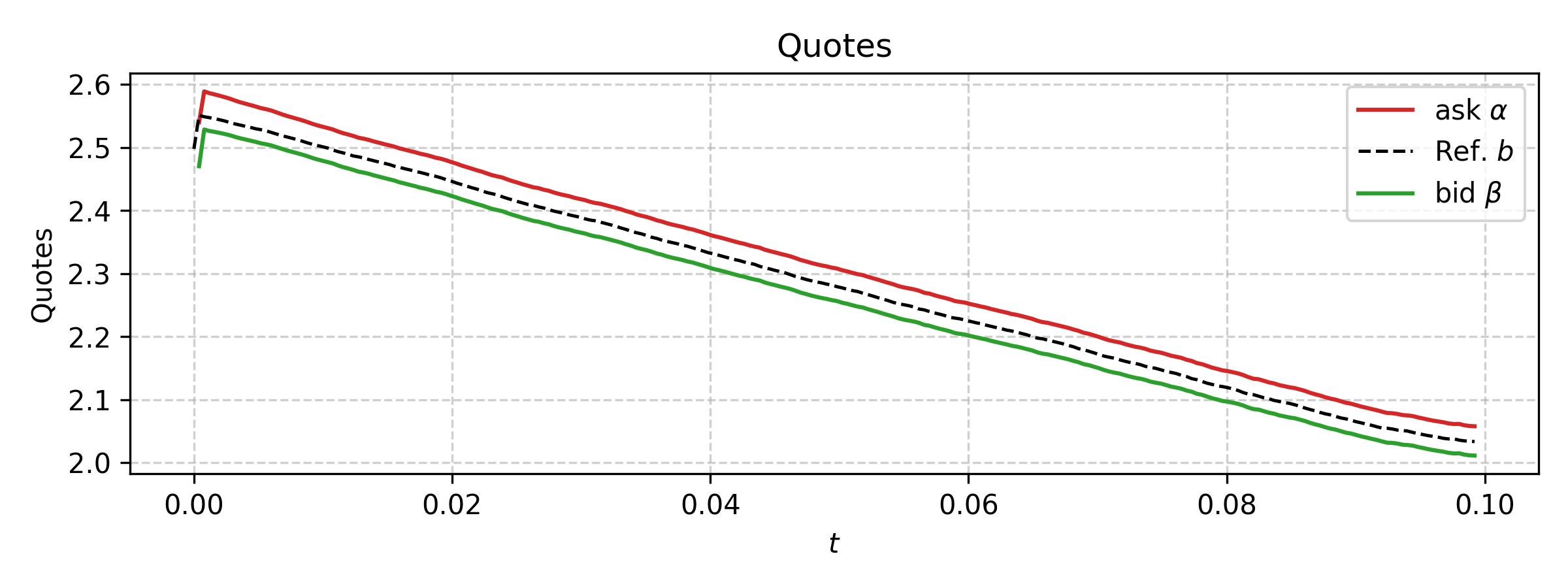} 
			\caption{Average quoting strategy ($I_0 = -100$).}
			\label{FIG:market_making_linear_impact_symetric_case_quotes_neg_inv}
		\end{subfigure}
		\hfill
		\begin{subfigure}[t]{0.49\textwidth}
			\centering
			\includegraphics[width=1\textwidth]{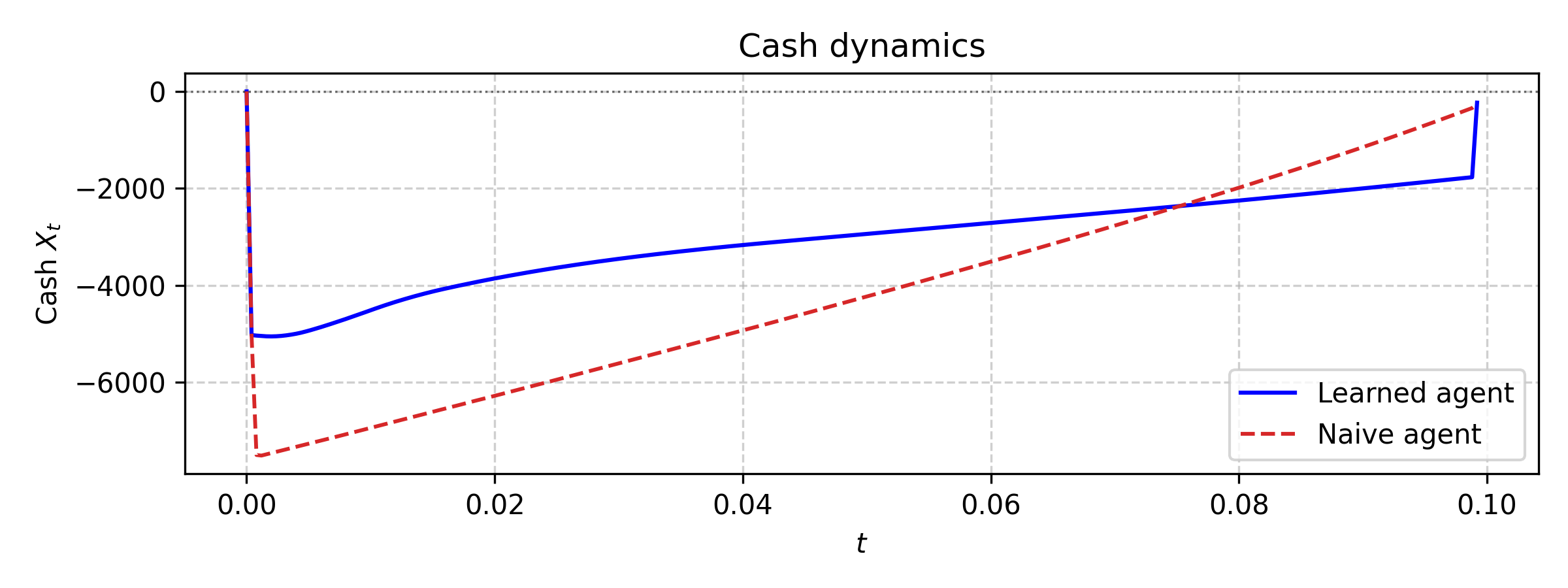} 
			\caption{Average cash trajectory of the market maker ($I_0 = -100$).}
			\label{FIG:market_making_linear_impact_symetric_case_pnl_neg_inv}
		\end{subfigure}
		\caption{Quoting strategy and cash dynamics under a large initial short position.}
	\end{figure}

	Turning first to the quoting behavior, Figure~\ref{FIG:market_making_linear_impact_symetric_case_quotes_neg_inv} shows that the agent adopts an asymmetric quoting policy, with wider spreads on the ask side than on the bid side. This asymmetry reflects a deliberate control of order flow: the agent makes its bid relatively more aggressive to attract sell orders and reduce its short option position, while discouraging trades that would further increase it.
	
\begin{figure}[H]
		\centering
		\begin{subfigure}[t]{0.49\textwidth}
			\centering
			\includegraphics[width=1\textwidth]{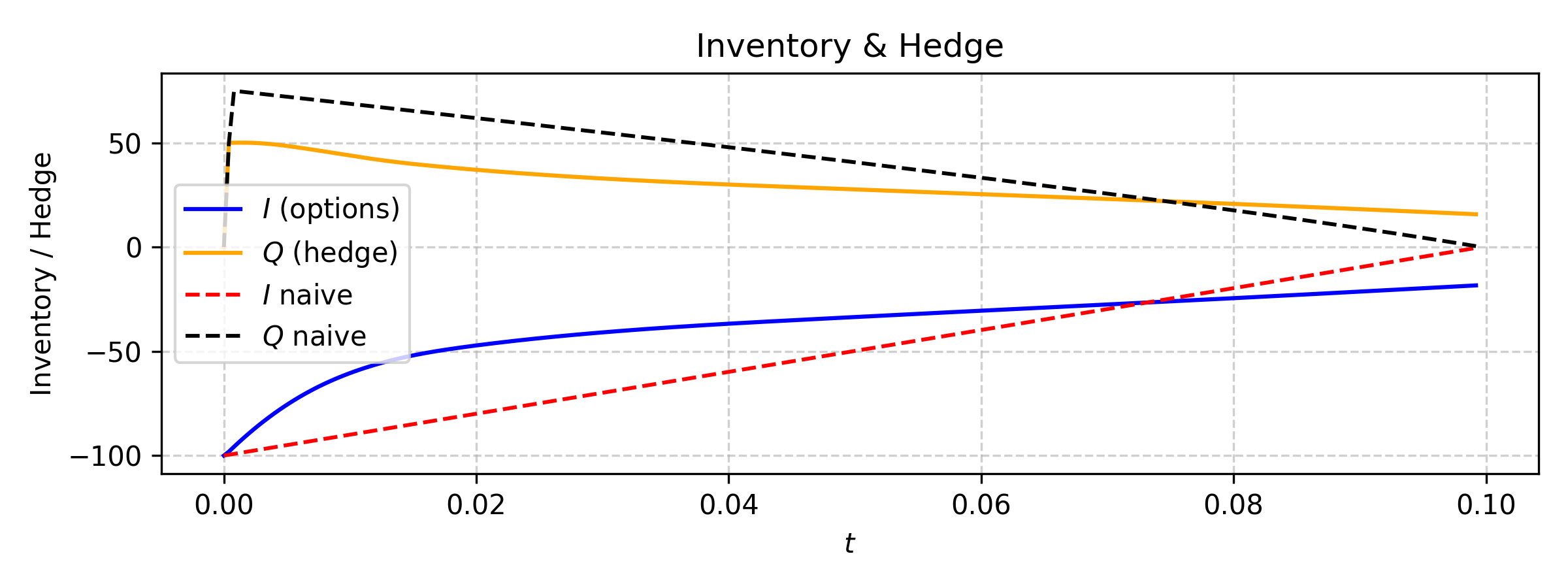} 
			\caption{Average option inventory and hedging position over time ($I_0 = -100$).}
			\label{FIG:market_making_linear_impact_symetric_case_inventories_neg_inv}
		\end{subfigure}
		\hfill
		\begin{subfigure}[t]{0.49\textwidth}
			\centering
			\includegraphics[width=1\textwidth]{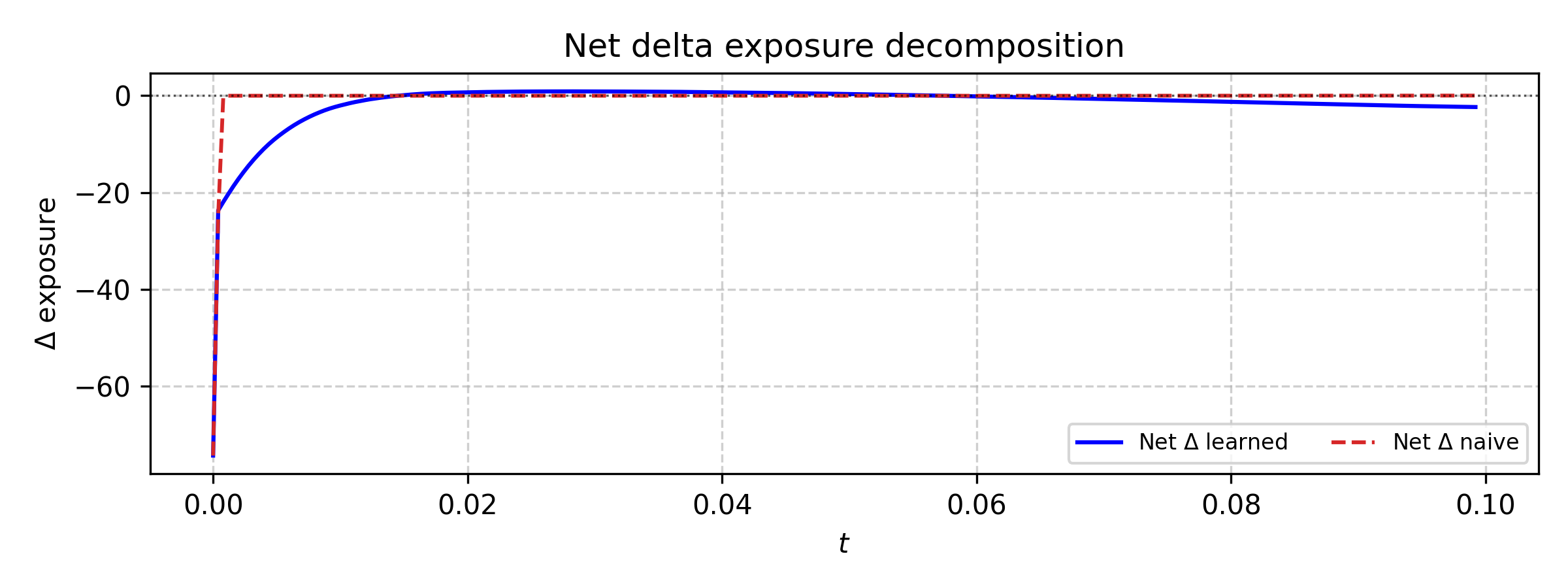} 
			\caption{Average net delta decomposition over time ($I_0 = -100$).}
			\label{FIG:market_making_linear_impact_symetric_case__inventories_neg_net_delta}
		\end{subfigure}
		\caption{Inventory and net delta dynamics under a large initial short position.}
	\end{figure}

	Figure~\ref{FIG:market_making_linear_impact_symetric_case_pnl_neg_inv} reports the resulting cash trajectory. The learned policy exhibits an initial drawdown due to the cost of establishing a partial hedge, but this is significantly mitigated compared to the naive strategy. As the agent progressively unwinds both its option and hedge positions, the cash recovers through spread capture and controlled execution. Although the terminal P\&L remains negative in this extreme configuration, the loss is substantially reduced relative to the naive benchmark, as confirmed by the aggregated results in Table~\ref{tab:perf_asymmetry_inventory_all}.
	The inventory dynamics reported in Figure~\ref{FIG:market_making_linear_impact_symetric_case_inventories_neg_inv} reveal the mechanism behind this behavior. Rather than fully hedging the initial short option exposure, the agent builds only a partial position in the underlying. A complete delta-neutral hedge would require purchasing approximately $75$ units, but the agent limits its initial hedge to roughly $50$, as the full cost of immediate neutralization through market impact would outweigh the benefit. Instead, it simultaneously reduces the option inventory through order flow control, which progressively lowers the required hedge and leads the portfolio toward a near-neutral configuration before maturity.
	
	Figure~\ref{FIG:market_making_linear_impact_symetric_case__inventories_neg_net_delta} confirms this through the net delta decomposition. The learned agent reduces its net exposure rapidly in the first few time steps through the partial hedge, then maintains it at a controlled level as the option inventory is unwound. The naive benchmark, by contrast, attempts full delta neutralization immediately through aggressive underlying purchases, incurring larger impact costs. This contrast illustrates the central trade-off exploited by the learned policy: coordinating partial hedging with order flow control to limit market impact, rather than treating hedging and inventory liquidation as independent decisions.
	
	\paragraph*{Performance across option inventory levels} ~\\
	The case $I_0=-100$ highlights the main qualitative mechanisms induced by a large inventory imbalance, namely asymmetric quoting, gradual inventory unwinding, and impact-aware hedging. We now examine whether these gains persist more systematically across the full range of initial inventories.
	Table~\ref{tab:perf_asymmetry_inventory_all} reports the distribution of terminal P\&L for both the naive and learned policies across different initial inventory levels.
	
	In the balanced case $I_0=0$, both policies achieve nearly identical performance, indicating that the learned agent effectively recovers the benchmark behavior in the absence of inventory imbalance. The slightly larger dispersion observed for the learned policy is consistent with approximation error induced by the neural network parametrization.
	As soon as the initial inventory departs from zero, a clear performance gap emerges. 
	\begin{table}[H]
		\centering
		\small
		\begin{tabular}{lrrrrrr}
			\toprule
			& \multicolumn{3}{c}{Naive} & \multicolumn{3}{c}{Learned} \\
			\cmidrule(lr){2-4} \cmidrule(lr){5-7}
			$I_0$ & Mean & Std & Median & Mean & Std & Median \\
			\midrule
			$-100$ & $-263.43$ & $6.38$  & $-262.94$ & $-232.58$ & $8.30$  & $-231.67$ \\
			$-75$  & $-173.19$ & $2.84$  & $-173.15$ & $-157.57$ & $10.19$ & $-155.09$ \\
			$-50$  & $-91.45$  & $1.67$  & $-91.47$  & $-81.62$  & $7.78$  & $-79.79$ \\
			$-25$  & $-15.56$  & $0.83$  & $-15.56$  & $-10.18$  & $4.96$  & $-9.60$ \\
			$0$    & $55.68$   & $0.00$  & $55.68$   & $55.43$   & $0.17$  & $55.42$ \\
			$+25$  & $96.72$   & $1.70$  & $96.81$   & $116.52$  & $1.68$  & $116.51$ \\
			$+50$  & $133.19$  & $3.64$  & $133.42$  & $173.85$  & $2.86$  & $173.83$ \\
			$+75$  & $172.30$  & $6.58$  & $172.30$  & $196.56$  & $1.57$  & $196.57$ \\
			$+100$ & $181.35$  & $8.86$  & $181.65$  & $252.13$  & $2.61$  & $252.10$ \\
			\bottomrule
		\end{tabular}
		\caption{Summary statistics of terminal cash for the naive and learned
			policies across initial inventory levels.}
		\label{tab:perf_asymmetry_inventory_all}
	\end{table}
	For negative initial inventories, both policies incur losses, but the learned strategy consistently mitigates these losses relative to the naive benchmark. For positive inventories, both policies remain profitable, yet the learned agent achieves significantly higher terminal cash. 
	The magnitude of this improvement increases with the size of the imbalance. When $|I_0|$ is large, liquidation and hedging decisions become strongly coupled with market impact, and the limitations of the naive strategy become more pronounced. Its static quoting rule and frictionless hedging assumption lead to inefficient execution, whereas the learned policy adapts dynamically its quotes and hedging intensity to the prevailing inventory pressure.
	
\begin{figure}[H]
		\centering
		
		\begin{subfigure}[t]{0.48\textwidth}
			\centering
			\includegraphics[width=\linewidth]{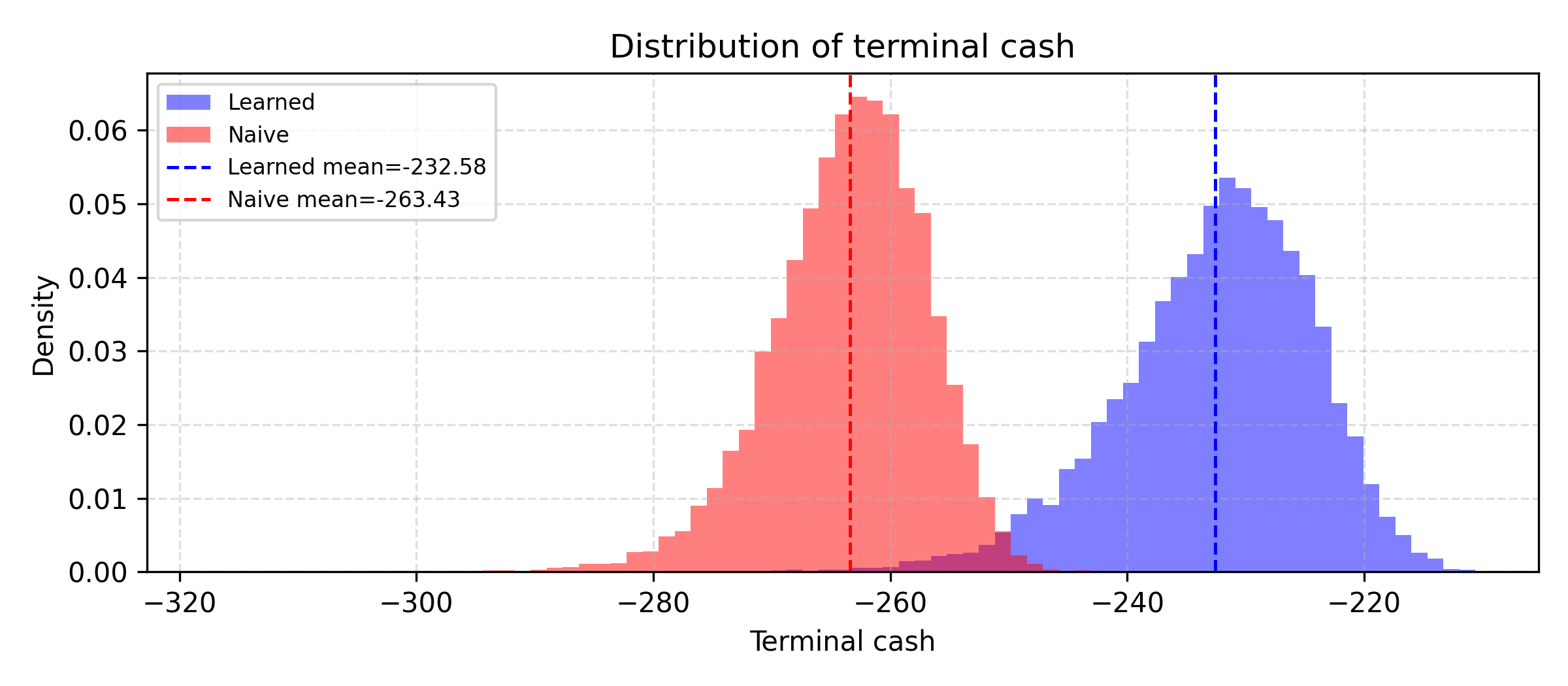}
			\caption{$I_0 = -100$}
		\end{subfigure}
		\hfill
		\begin{subfigure}[t]{0.48\textwidth}
			\centering
			\includegraphics[width=\linewidth]{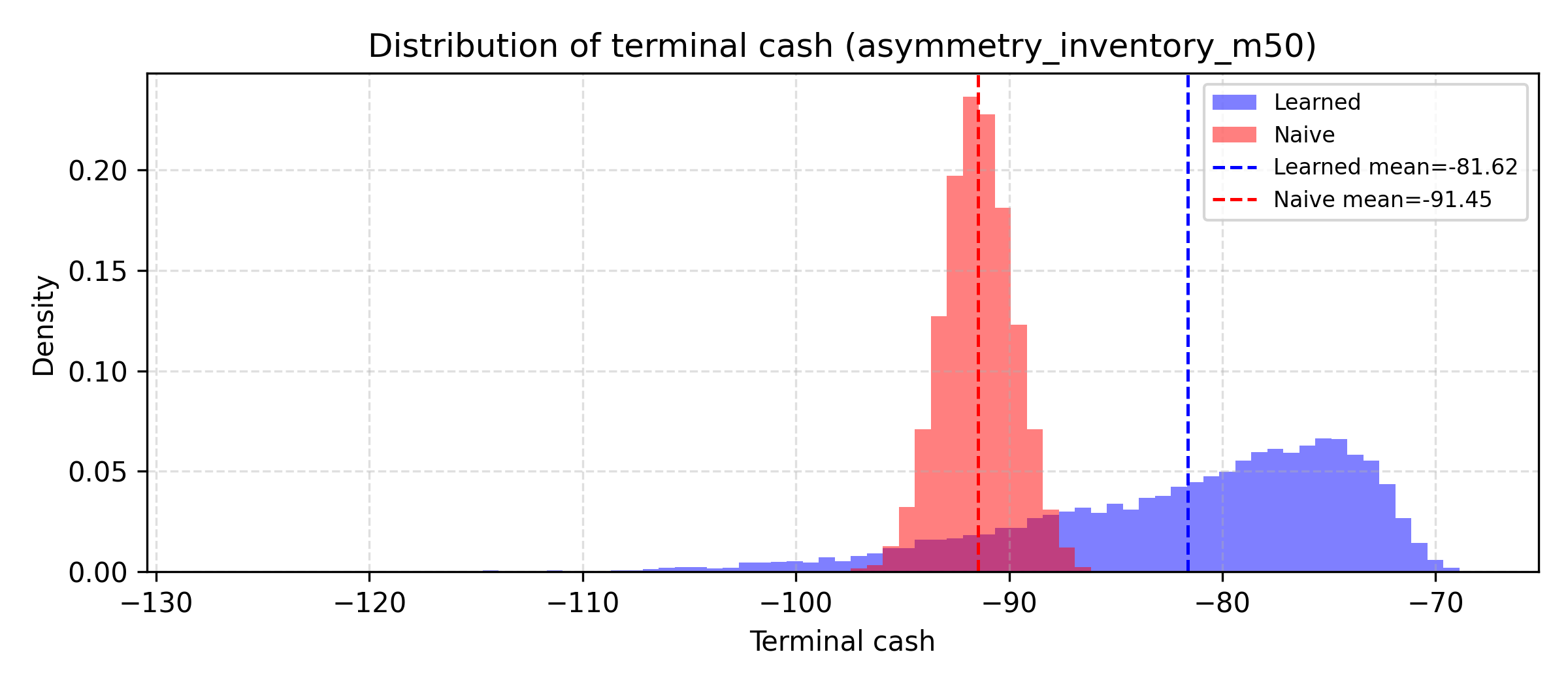}
			\caption{$I_0 = -50$}
		\end{subfigure}
		
		\vspace{0.4cm}
		
		\begin{subfigure}[t]{0.48\textwidth}
			\centering
			\includegraphics[width=\linewidth]{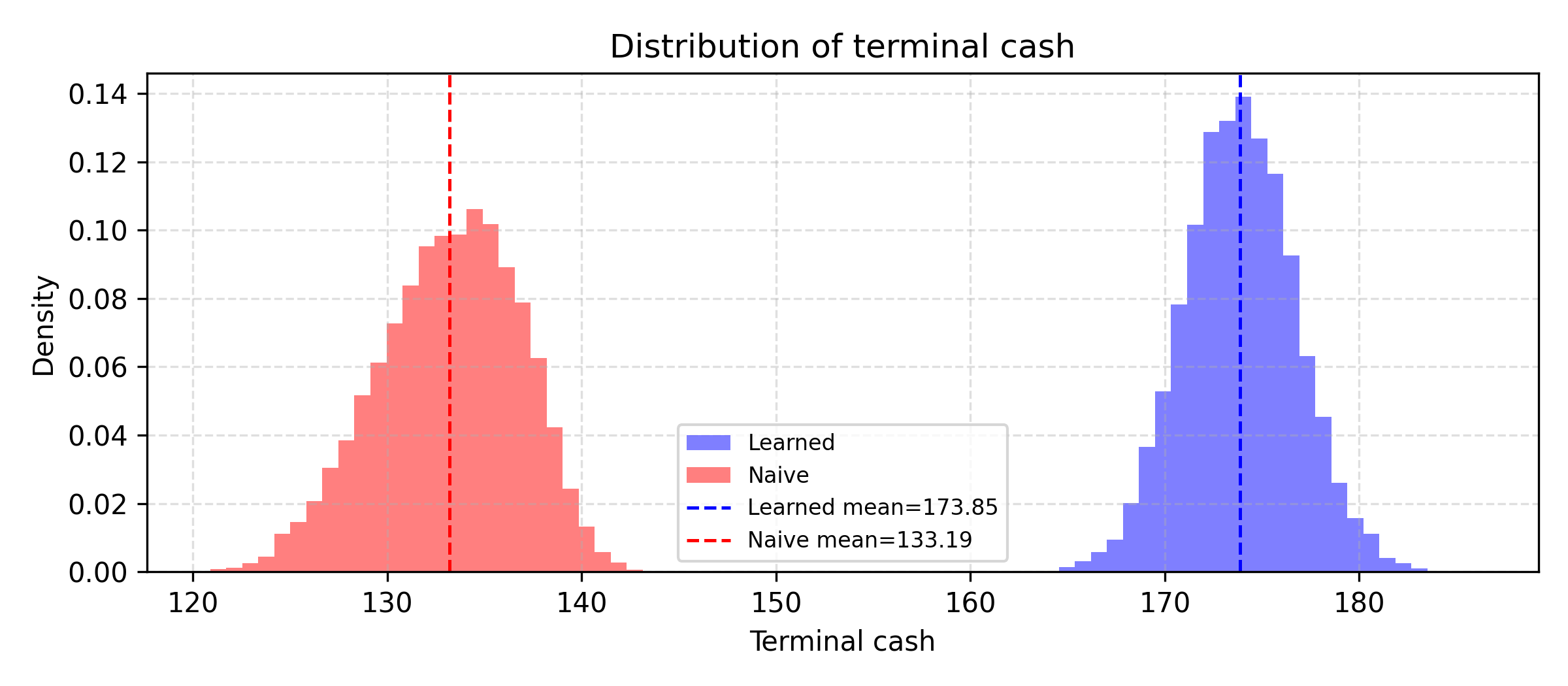}
			\caption{$I_0 = 50$}
		\end{subfigure}
		\hfill
		\begin{subfigure}[t]{0.48\textwidth}
			\centering
			\includegraphics[width=\linewidth]{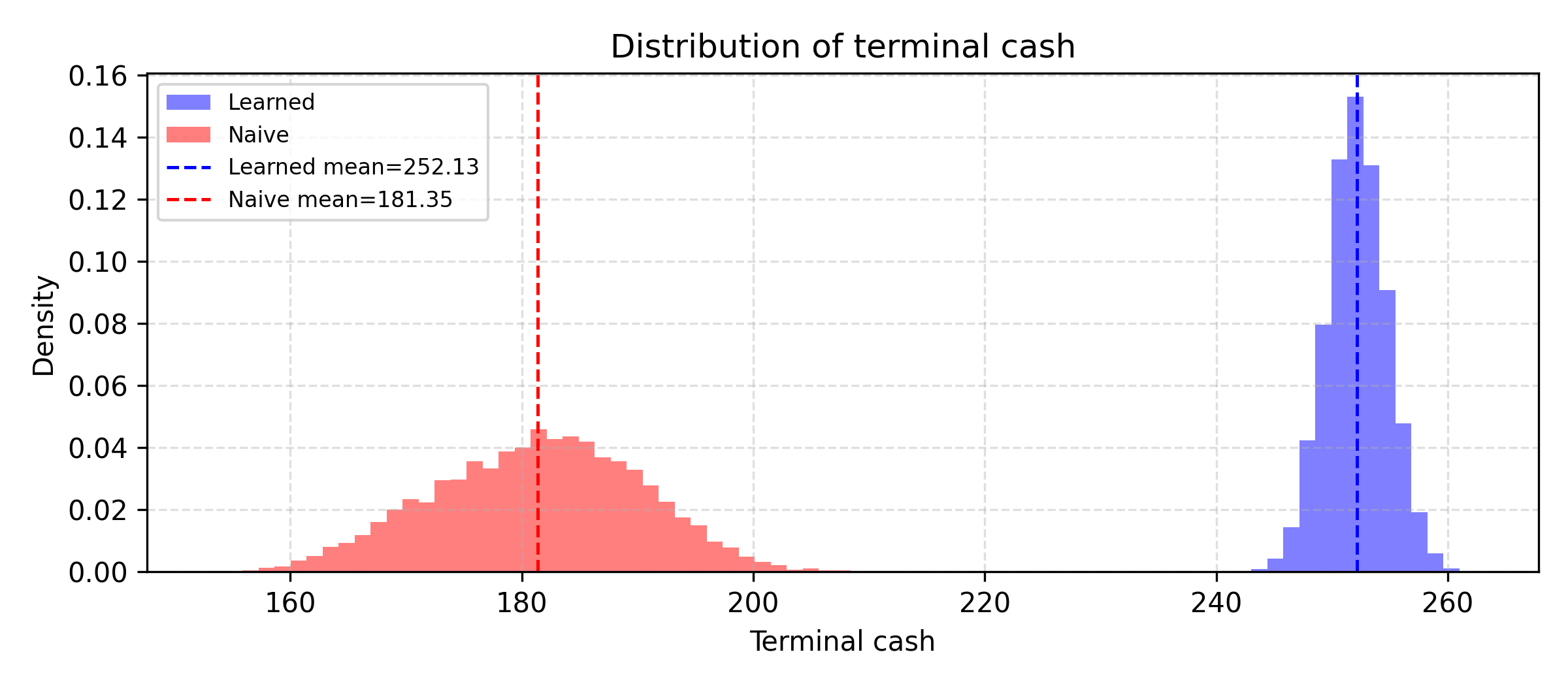}
			\caption{$I_0 = 100$}
		\end{subfigure}
		
		\caption{Empirical distributions of terminal P\&L for the naive and learned policies across selected initial inventory levels.}
		\label{fig:pnl_distributions_inventory}
	\end{figure}

\begin{figure}[H]
		\centering
		\begin{subfigure}[t]{0.48\textwidth}
			\centering
			\includegraphics[width=\linewidth]{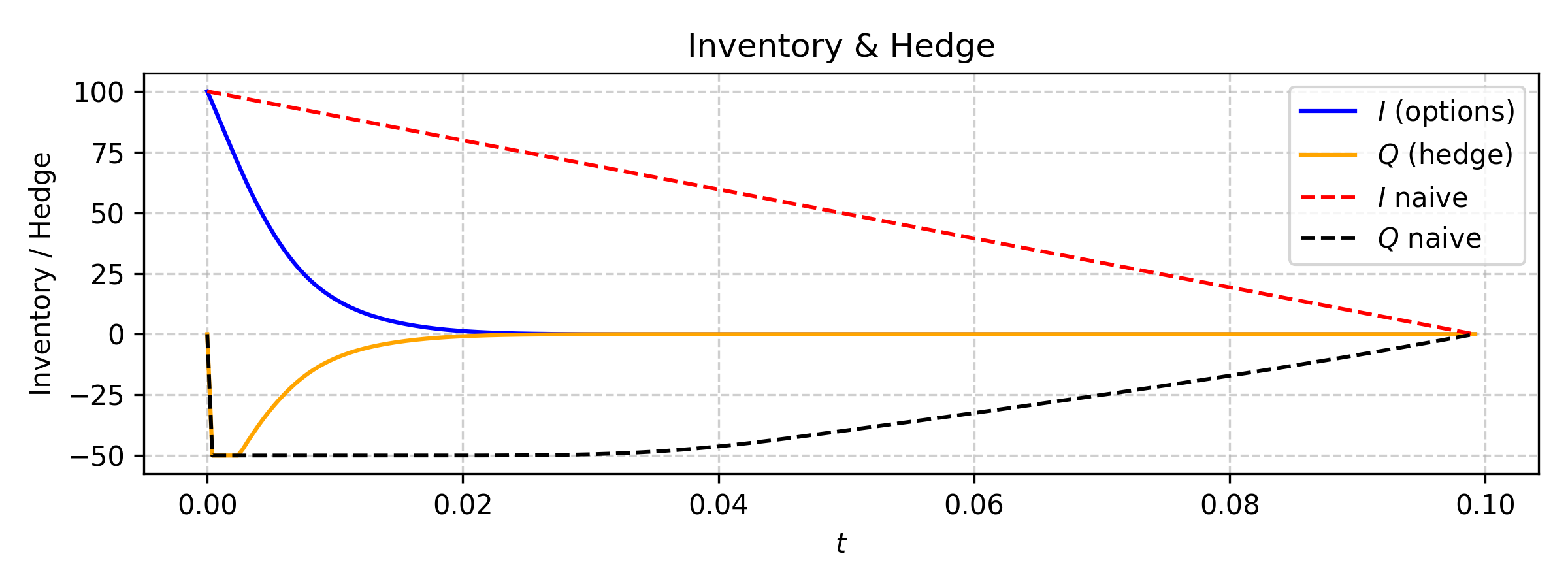}
			\caption{Inventory dynamics, $I_0 = 100$.}
			\label{FIG:inv_dyn_p100}
		\end{subfigure}
		\hfill
		\begin{subfigure}[t]{0.48\textwidth}
			\centering
			\includegraphics[width=\linewidth]{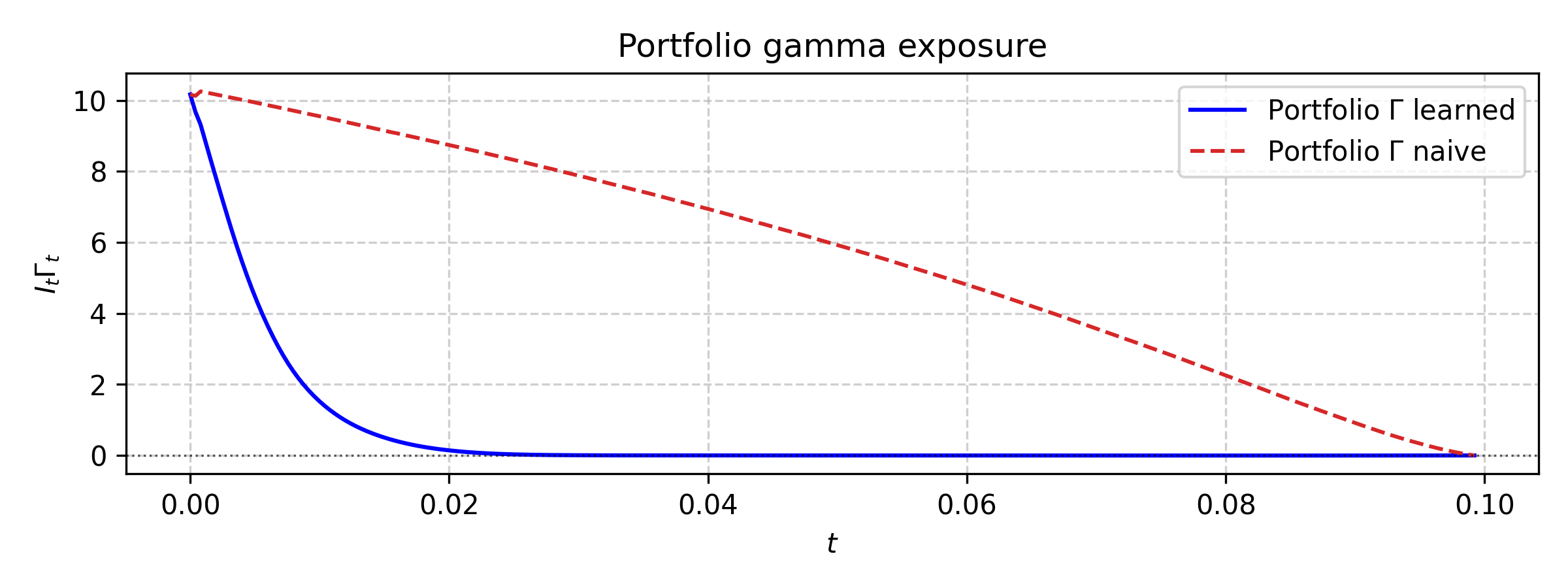}
			\caption{Portfolio gamma, $I_0 = 100$.}
			\label{FIG:port_gamma_p100}
		\end{subfigure}
		
		\vspace{0.4cm}
		
		\begin{subfigure}[t]{0.48\textwidth}
			\centering
			\includegraphics[width=\linewidth]{imgs/asymmetry_inventory_m100_decisions_inventory_hedge_avg.png}
			\caption{Inventory dynamics, $I_0 = -100$.}
			\label{FIG:inv_dyn_m100}
		\end{subfigure}
		\hfill
		\begin{subfigure}[t]{0.48\textwidth}
			\centering
			\includegraphics[width=\linewidth]{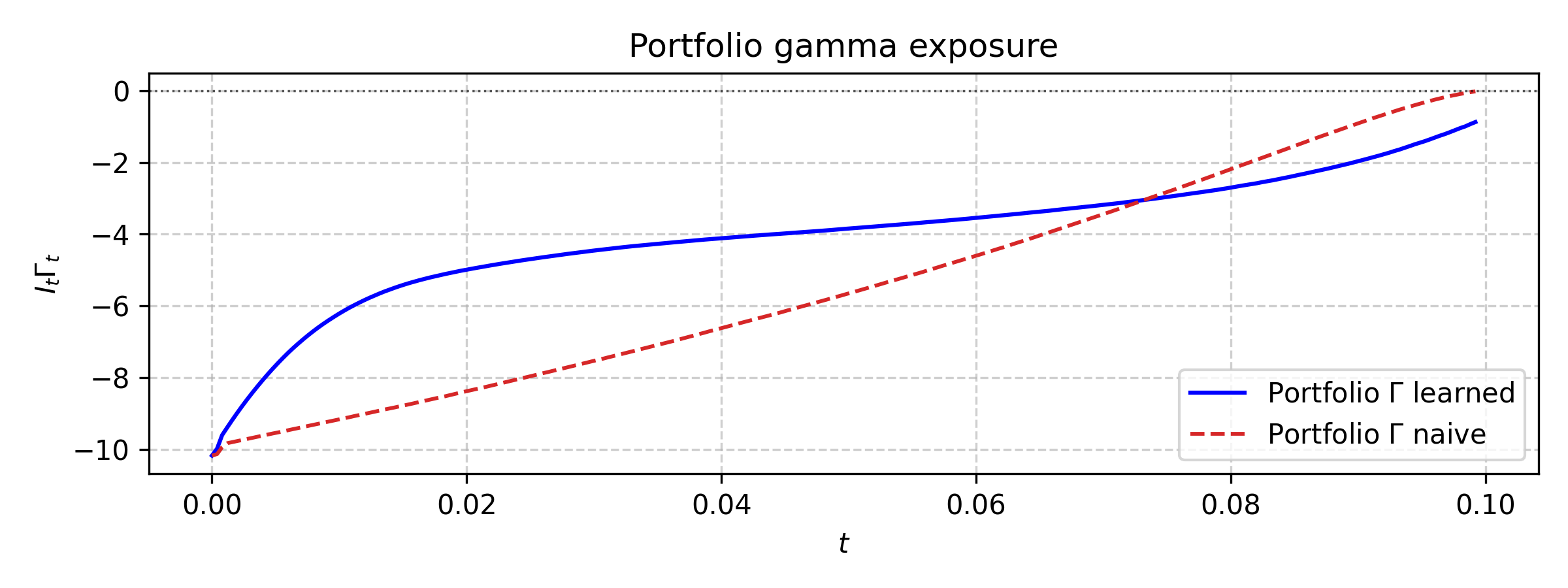}
			\caption{Portfolio gamma, $I_0 = -100$.}
			\label{FIG:port_gamma_m100}
		\end{subfigure}
		
		\caption{Asymmetry in liquidation and gamma management.}
		\label{fig:asymmetry_inventory_gamma}
	\end{figure}

	Figure~\ref{fig:pnl_distributions_inventory} further illustrates this comparison by displaying the empirical distributions of terminal P\&L for selected inventory levels. For $I_0 = -100$ and $I_0 = -50$, the learned policy shifts the entire distribution to the right, reducing tail losses. For $I_0 = 50$ and $I_0 = 100$, the distributions are more concentrated and centered at higher values, confirming that the learned agent captures a larger share of the available spread revenues while maintaining tighter risk control.
	The relative improvement of the learned policy over the naive benchmark is notably larger for positive initial inventories. This asymmetry results from the interaction between the gamma-theta trade-off and the cost structure of inventory liquidation.
	When $I_0 > 0$, the market maker holds a long option position with positive gamma and negative theta. Time decay erodes the value of the position at each time step, creating a strong incentive to liquidate quickly. The learned agent exploits this by aggressively tightening its ask quotes to attract client buy orders, eliminating its option inventory within the first few time steps. This rapid liquidation simultaneously removes the need for a costly underlying hedge: the agent bypasses hedging entirely for moderate inventories ($I_0 = 25, 50$) and only partially hedges for larger ones, as illustrated in Figure~\ref{FIG:inv_dyn_p100}. The portfolio gamma exposure, displayed in Figure~\ref{FIG:port_gamma_p100}, confirms that the learned agent eliminates its positive convexity within a few time steps, forgoing the associated benefit in order to avoid hedging costs. The combined savings, namely avoided hedging impact, reduced hedge penalty, and captured spread on each option sale, account for the large performance gap relative to the naive benchmark, which liquidates linearly over the full horizon while maintaining a costly hedge throughout.
	
	When $I_0 < 0$, the situation is reversed. The short option position carries negative gamma but positive theta: time decay works in the agent's favor, as the options it must repurchase become cheaper over time. This creates an incentive to delay liquidation. At the same time, the negative gamma exposure means that each price movement generates a loss, which compels the agent to hedge immediately. As illustrated in Figure~\ref{FIG:inv_dyn_m100}, the learned policy resolves this tension by hedging first and liquidating gradually. It accepts the cost of carrying a hedge over a longer period in exchange for more favorable repurchase prices. Figure~\ref{FIG:port_gamma_m100} shows that the negative gamma exposure is accordingly carried over a longer period before converging to zero. The margin for improvement over the naive benchmark is correspondingly smaller, since both strategies must bear significant hedging costs throughout the horizon.
	Overall, these results indicate that the learned policy remains competitive in near-symmetric settings and delivers substantial gains in imbalanced regimes, where impact-aware and state-dependent decisions are critical.
	
	\paragraph*{Impact of hedging and execution constraints on the learned policy} ~\\
	The coefficients $\kappa_{\mathrm{hedge}}$ and $\kappa_{\mathrm{act}}$ govern the relative weight assigned to hedging and execution considerations in the agent’s objective. 
	The hedging penalty promotes risk control by penalizing deviations from delta neutrality, while the incentive penalty discourages insufficient trading activity relative to a target transaction level. 
	Varying these coefficients therefore provides insight into how the learned policy adjusts to different operational priorities. 
	
\begin{figure}[H]
		\centering
		\begin{subfigure}[t]{0.49\textwidth}
			\centering
			\includegraphics[width=1\textwidth]{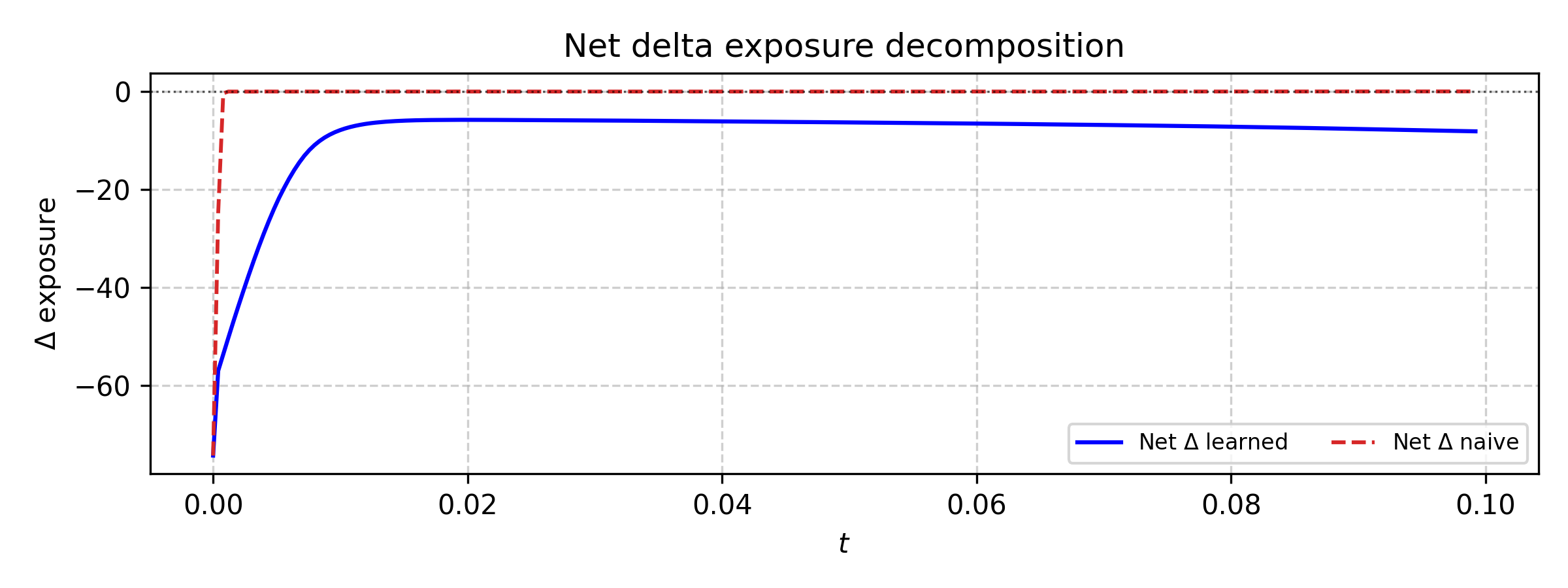} 
			\caption{Average net delta dynamics for $\kappa_{\mathrm{hedge}} = 0.5$.}
			\label{FIG:net_delta_kappa_hedge_05}
		\end{subfigure}
		\hfill
		\begin{subfigure}[t]{0.49\textwidth}
			\centering
			\includegraphics[width=1\textwidth]{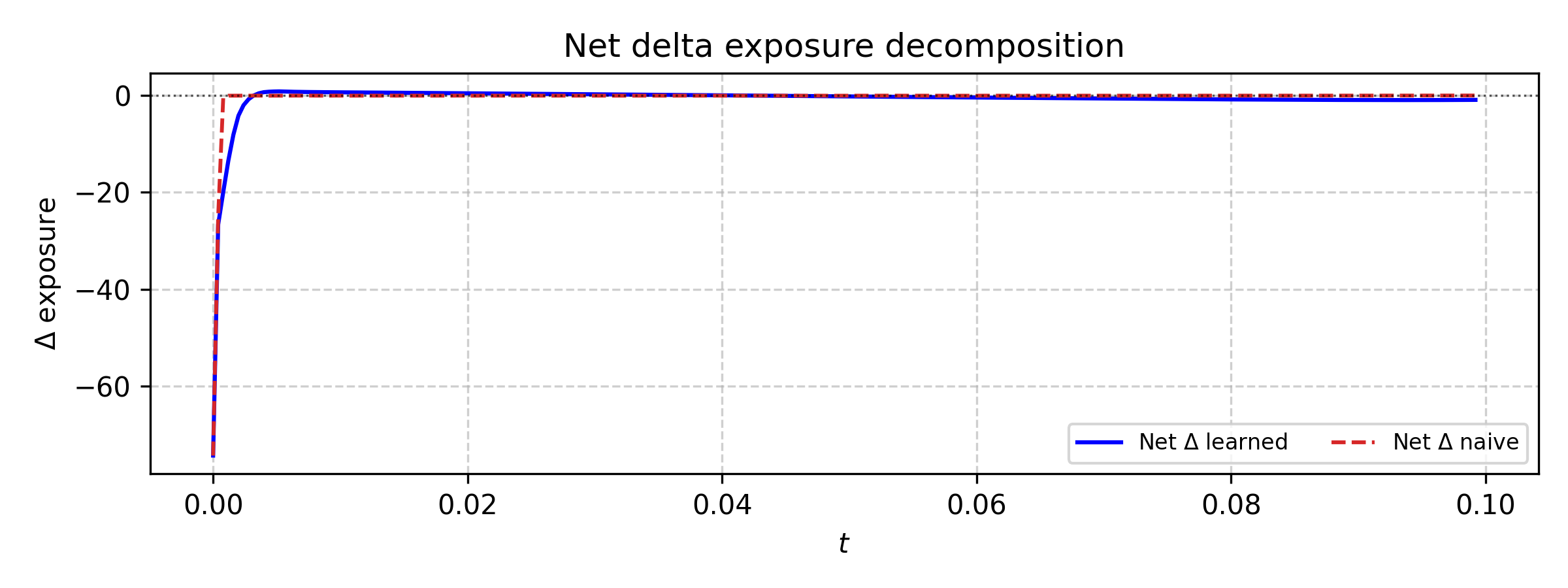} 
			\caption{Average net delta dynamics for $\kappa_{\mathrm{hedge}} = 16$.}
			\label{FIG:net_delta_kappa_hedge_16}
		\end{subfigure}
		\caption{Impact of the hedging penalty $\kappa_{\mathrm{hedge}}$ on the net delta dynamics.}
	\end{figure}

	To assess the impact of these coefficients on the learned policy, we conducted a sensitivity analysis. The resulting performance metrics are reported in Table~\ref{tab:sensitivity_penalties_clean}, and further insights are provided through the net delta dynamics displayed in Figures~\ref{FIG:net_delta_kappa_hedge_05} and~\ref{FIG:net_delta_kappa_hedge_16}.
	\begin{table}[H]
		\centering
		\begin{tabular}{c|cc|c}
			\toprule
			$\kappa_{\mathrm{hedge}}$ & Avg.\ P\&L & Std.\ P\&L & Norm.\ hedge pen. \\
			\midrule
			0.50 & -199.72 & 14.64 & 15.48 \\
			2.00 & -223.93 & 10.70 & 5.09  \\
			4.00 & -232.62 & 8.30 & 4.28  \\
			8.00 & -245.89 & 6.69 & 3.60   \\
			16.00 & -263.02 & 5.07 & 2.93  \\
			\bottomrule
		\end{tabular}
		\caption{Sensitivity of the learned policy to hedging penalty coefficient.}
		\label{tab:sensitivity_penalties_clean}
	\end{table}
	For low values of $\kappa_{\mathrm{hedge}}$ (e.g., $\kappa_{\mathrm{hedge}} \approx 0.5$), the agent adopts a risk-taking behavior. As shown in Figure~\ref{FIG:net_delta_kappa_hedge_05}, the net delta is only partially corrected over time and remains significantly different from zero, indicating that residual exposure is deliberately maintained. This is consistent with the relatively high P\&L reported in Table~\ref{tab:sensitivity_penalties_clean}, as well as the large normalized hedge penalty, which reflects that hedging is used sparingly. In this regime, the option inventory is allowed to evolve freely, and hedging is performed gradually, as the cost of rebalancing dominates the associated penalty.
	For intermediate values ($\kappa_{\mathrm{hedge}} \approx 2$--$4$), the strategy enters a balanced regime where risk control and profitability are jointly optimized. Although not shown explicitly in the figure, the net delta converges rapidly toward zero and remains tightly controlled thereafter. This behavior corresponds to a significant reduction in the normalized hedge penalty in Table~\ref{tab:sensitivity_penalties_clean}, while maintaining a relatively high level of P\&L. The agent performs an initial hedge followed by continuous adjustments, leading to a near delta-neutral strategy consistent with classical optimal hedging under transaction costs.
	For large values of $\kappa_{\mathrm{hedge}}$ (e.g., $\kappa_{\mathrm{hedge}} \approx 8$--$16$), the policy becomes effectively constrained. As illustrated in Figure~\ref{FIG:net_delta_kappa_hedge_16}, the net delta is brought extremely close to zero almost immediately and remains tightly centered around zero throughout the episode. This reflects a regime of quasi-continuous rebalancing, where hedging adjustments are frequent and precise. Consistently, Table~\ref{tab:sensitivity_penalties_clean} shows a further decrease in the normalized hedge penalty, accompanied by a marked deterioration in P\&L. In parallel, the agent tends to reduce its option inventory more aggressively over time, limiting the build-up of exposure. Overall, the penalty dominates the objective, effectively enforcing delta neutrality and leading to over-hedging and reduced economic efficiency.
	The activity penalty defined in~\eqref{EQN:incentive_pen} enforces a minimal level of transaction intensity. 
	While it is initially active during training (see Figure~\ref{FIG:market_making_linear_impact_learning_metrics_neg_inv}), it quickly vanishes as the policy adapts its quotes to satisfy the constraint. 
	As a result, $\kappa_{\mathrm{act}}$ mainly acts as a regularization device during learning and does not bind at optimality, which explains the negligible activity penalties reported in Table~\ref{tab:sensitivity_penalties_clean}.
	
	\subsubsection{Shifted order intensities}
	
\begin{figure}[H]
		\centering
		\includegraphics[width=0.6\textwidth]{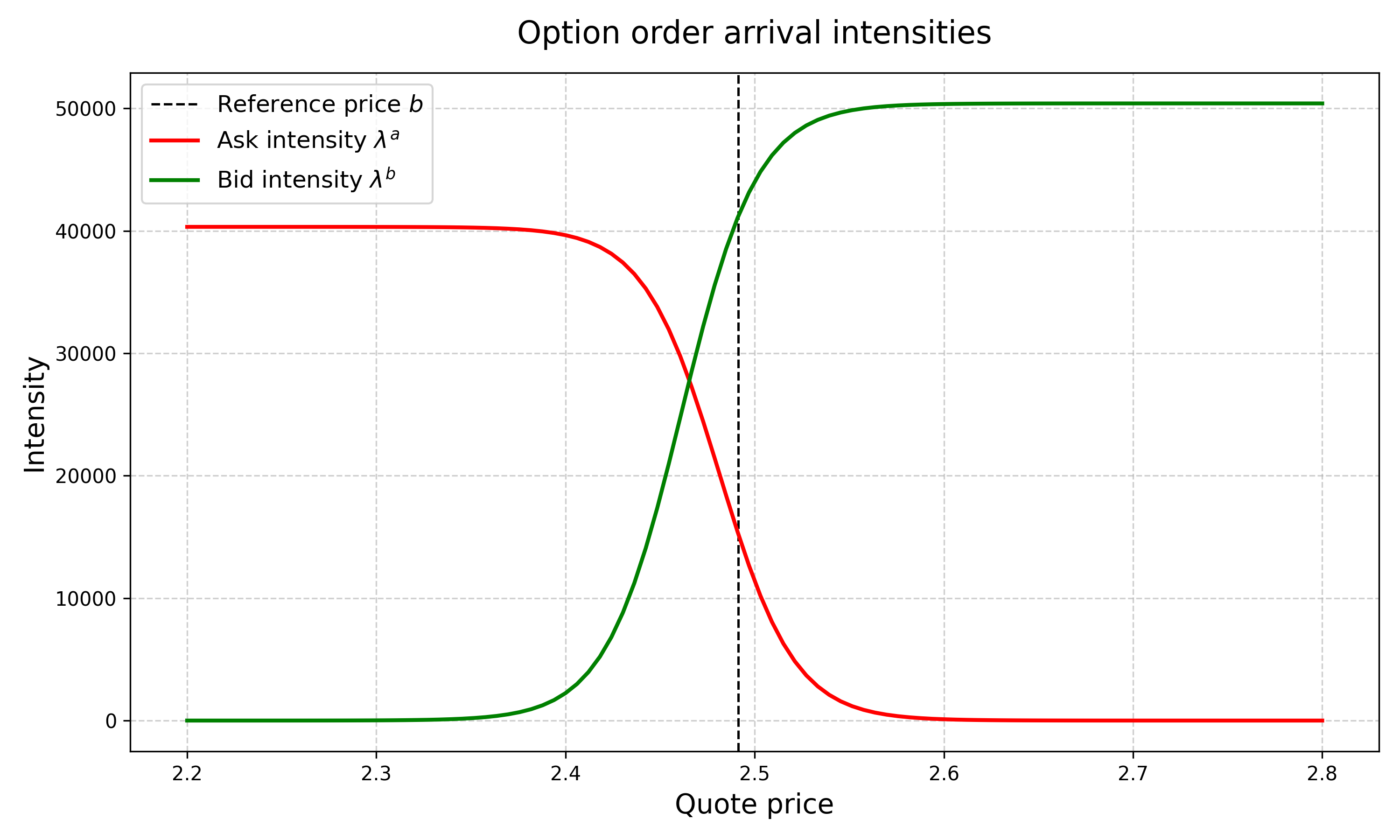} 
		\caption{Shape of asymmetric option order flow intensities.}
		\label{FIG:market_making_linear_impact_option_intensities_ass_intensities}
	\end{figure}

	In this second asymmetric configuration, we do not impose a negative initial option inventory. 
	Instead, we distort the option market itself by shifting the order arrival intensities away from symmetry. 
	Specifically, we use the specification of Equation~\eqref{EQN:market_making_option_intensities} with $\mu_b = 3/2$ and $\mu_a = -1/2$, and reduce the maximal ask intensity to $80\%$ of the reference value. 
	As shown in Figure~\ref{FIG:market_making_linear_impact_option_intensities_ass_intensities}, this parametrization makes bid intensities systematically higher than ask intensities near the reference price, with the two curves intersecting only at quotes placed well below $b$. 
	As a consequence, client buy orders (hitting the bid) occur more frequently than sell orders, so the market maker is structurally pushed toward a negative option inventory. 
	This imbalance in option flows then translates into a persistent exposure in the underlying market.
	
\begin{figure}[H]
		\centering
		\begin{subfigure}[t]{0.49\textwidth}
			\centering
			\includegraphics[width=1\textwidth]{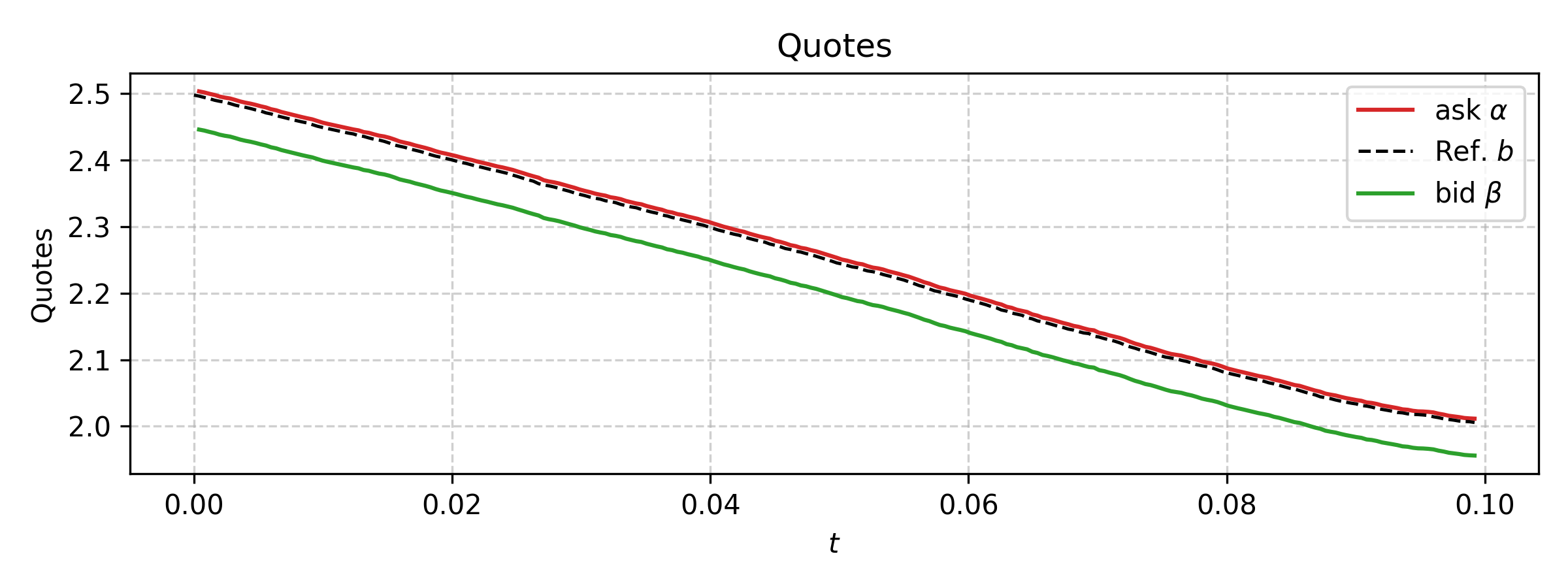} 
			\caption{Average quoting strategy ($I_0 = 0$ and asymmetric intensities).}
			\label{FIG:market_making_linear_quotes_ass_intensities}
		\end{subfigure}
		\hfill
		\begin{subfigure}[t]{0.49\textwidth}
			\centering
			\includegraphics[width=1\textwidth]{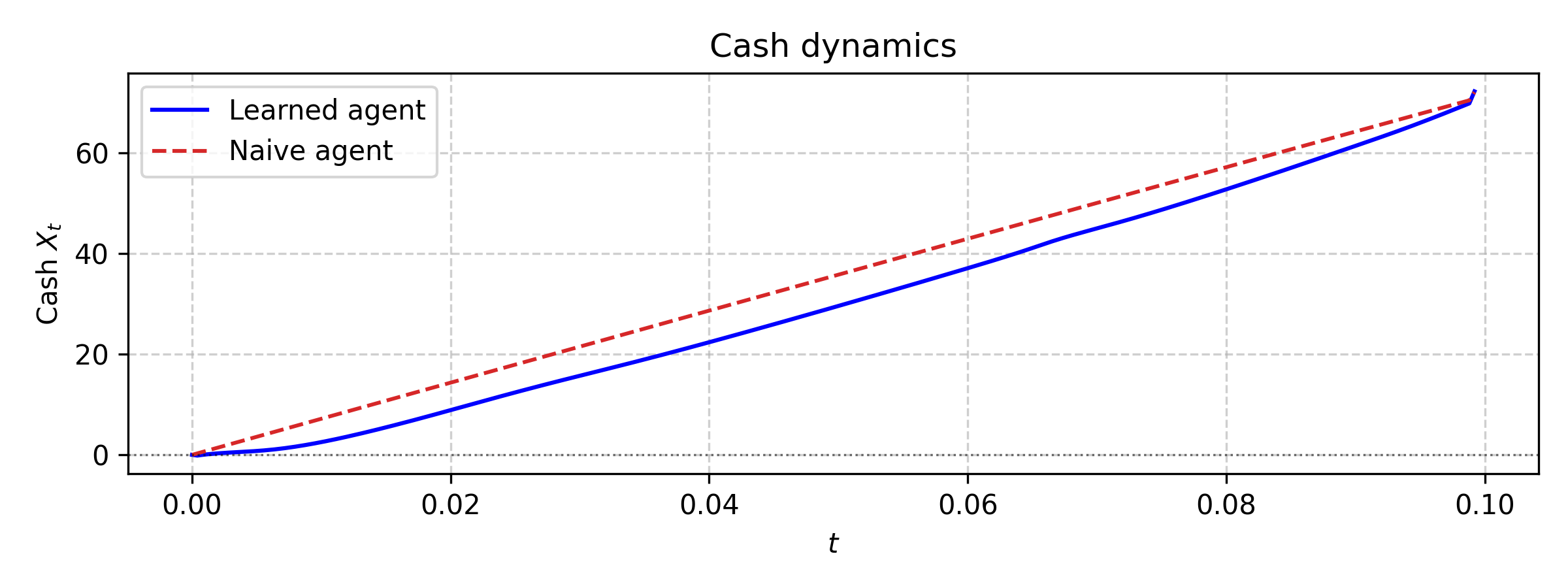} 
			\caption{Average cash trajectory of the market maker ($I_0 = 0$ and asymmetric intensities).}
			\label{FIG:market_making_linear_pnl_ass_intensities}
		\end{subfigure}
		\caption{Quoting behavior and cash dynamics under asymmetric order flow.}
	\end{figure}

\begin{figure}[H]
		\centering
		\begin{subfigure}[t]{0.49\textwidth}
			\centering
			\includegraphics[width=1\textwidth]{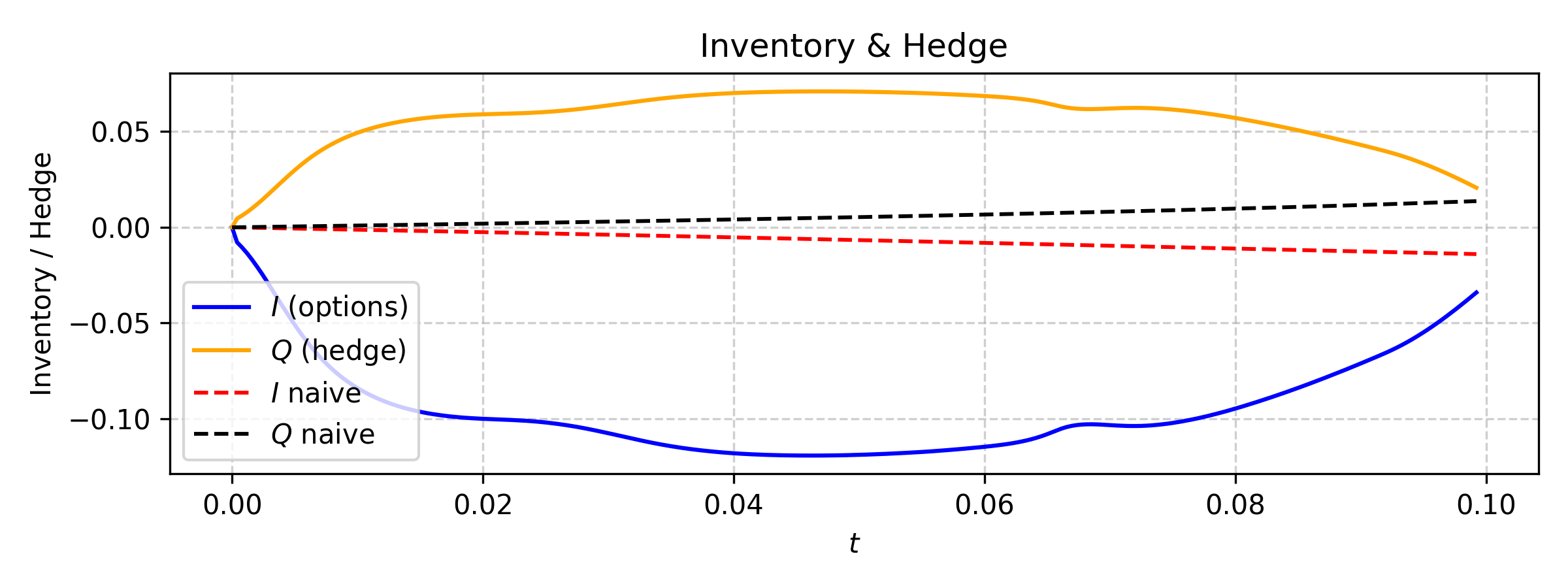} 
			\caption{Average option inventory and hedging position over time ($I_0 = 0$ and asymmetric intensities).}
			\label{FIG:market_making_inventories_ass_intensities}
		\end{subfigure}
		\hfill
		\begin{subfigure}[t]{0.49\textwidth}
			\centering
			\includegraphics[width=1\textwidth]{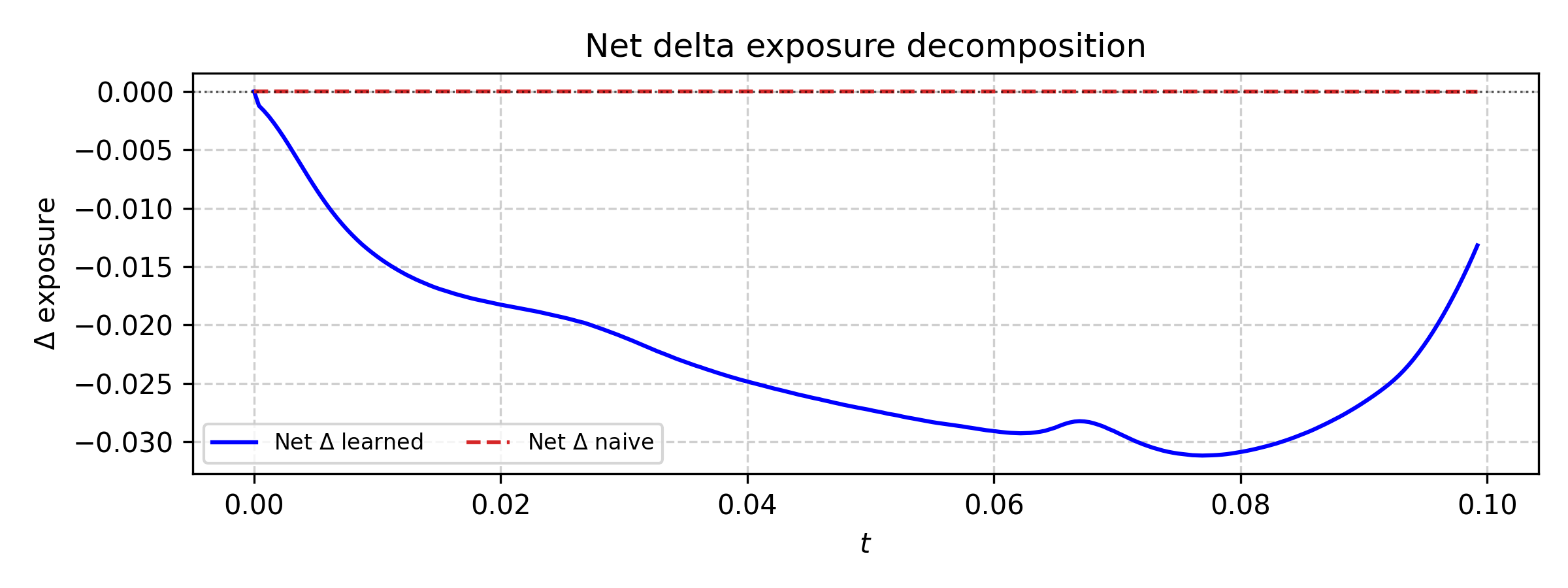} 
			\caption{Average net delta decomposition over time ($I_0 = 0$ and asymmetric intensities).}
			\label{FIG:market_making_linear_impact_ass_intensities_net_delta}
		\end{subfigure}
		\caption{Inventory and net delta dynamics under asymmetric order flow.}
	\end{figure}

	The average quoting strategy is shown in Figure~\ref{FIG:market_making_linear_quotes_ass_intensities}. 
	The agent systematically quotes with a wider spread on the bid side than on the ask side. 
	This asymmetry reflects an attempt to discourage client buy orders, which are structurally more frequent due to the shifted intensities, while still keeping competitive ask quotes to facilitate inventory reduction.  
	In other words, the market maker uses its quotes to locally rebalance order flow, preventing the emergence of persistent inventory imbalances.
	In contrast with the previous configuration, the learned policy does not passively absorb the imbalance induced by asymmetric order flow. 
	Instead, it actively controls its inventory so as to remain close to a delta-neutral position throughout the horizon.
	As shown in Figure~\ref{FIG:market_making_inventories_ass_intensities}, the option inventory remains centered around zero despite the structural asymmetry in order arrivals. 
	Rather than allowing inventory to drift and correcting it ex post, the agent continuously adjusts its quotes to balance incoming buy and sell orders at the flow level. 
	As a result, the inventory process remains tightly controlled around zero, reducing the need for large hedging adjustments.
	
	This mechanism is further confirmed by the net delta decomposition in Figure~\ref{FIG:market_making_linear_impact_ass_intensities_net_delta}. 
	The learned policy maintains a near-zero net delta exposure over time, coordinating option market making and underlying hedging in a way that prevents the accumulation of directional risk. 
	While the naive strategy also aims to maintain a near-zero delta exposure, it does so in a less adaptive manner and therefore fails to optimally control transient imbalances induced by the stochastic order flow.
	
	Turning to performance, Table~\ref{tab:perf_asymmetry_intensities} show that the learned policy yields a systematic, albeit moderate, improvement over the naive benchmark.
	While mean P\&L differences remain limited, the learned strategy improves the distribution of terminal outcomes, with a rightward shift and a more favorable lower tail. 
	
	This gain is not reflected in average inventory or hedging trajectories, which are similar across both strategies. 
	Instead, it arises from a more precise control of order flow, allowing the agent to better manage transient imbalances and execution costs. 
	These second-order effects do not impact mean trajectories but translate into improved terminal P\&L distributions.
	\begin{table}[H]
		\centering
		\begin{tabular}{l|ccccc}
			\toprule
			Agent & Mean & Std & Median & Q5 & Q95 \\
			\midrule
			Learned & $72.16$ & $0.05$ & $72.18$ & $72.06$ & $72.20$ \\
			Naive   & $72.00$ & $0.00$ & $72.00$ & $72.00$ & $72.01$ \\
			\bottomrule
		\end{tabular}
		\caption{Terminal cash distribution for the asymmetry intensities configuration.}
		\label{tab:perf_asymmetry_intensities}
	\end{table}
	Overall, this configuration highlights a different mechanism from the previous case: rather than smoothing the liquidation of an initial imbalance, the learned policy prevents the imbalance from emerging in the first place. 
	From a control perspective, this corresponds to maintaining the system close to its optimal manifold (delta neutrality) at the level of order flow, thereby minimizing both impact costs and hedging needs.
	
	\subsubsection{Inventory-risk trade-off}
	
	In this final experiment, we initialize the market maker with a negative option inventory $I_0 = -100$ and reduce the available liquidity.  
	Specifically, we set $c_A = c_B = 2.5$ and $U_A = U_B = 2$.  
	Increasing the depth parameters $U_A$ and $U_B$ amplifies the volatility of the underlying asset, since individual order arrivals trigger larger price movements.  
	Under the reference parameterization, the effective volatility was around $10\%$, whereas under the present low-liquidity configuration it rises to approximately $40\%$.  
	
	The combination of a negative initial option inventory and reduced liquidity induces a markedly different operating regime.
	At the start of the horizon, the learned agent does not attempt to immediately neutralize its risk through aggressive hedging. 
	Instead, it adapts its quoting strategy to the liquidity constraints by introducing a marked asymmetry, favoring option purchases in order to progressively unwind its initial short inventory.
	This behavior reflects an important trade-off: due to limited market depth, rapidly hedging the option exposure through the underlying would induce large impact costs. 
	The agent therefore prioritizes inventory reduction via the option market, while only partially hedging its residual exposure.
	\begin{figure}[H]
		\centering
		\begin{subfigure}[t]{0.49\textwidth}
			\centering
			\includegraphics[width=1\textwidth]{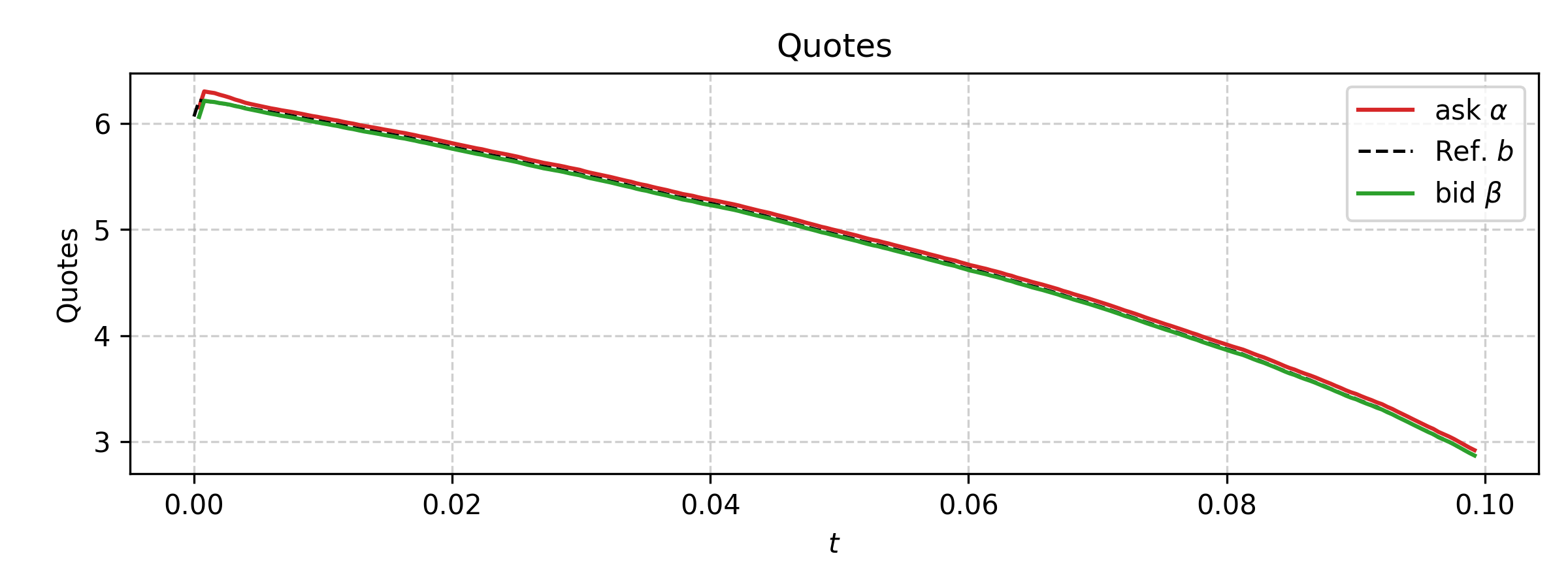} 
			\caption{Average quoting strategy ($I_0 = -100$ and low liquidity).}
			\label{FIG:market_making_linear_quotes_liqu_constraints}
		\end{subfigure}
		\hfill
		\begin{subfigure}[t]{0.49\textwidth}
			\centering
			\includegraphics[width=1\textwidth]{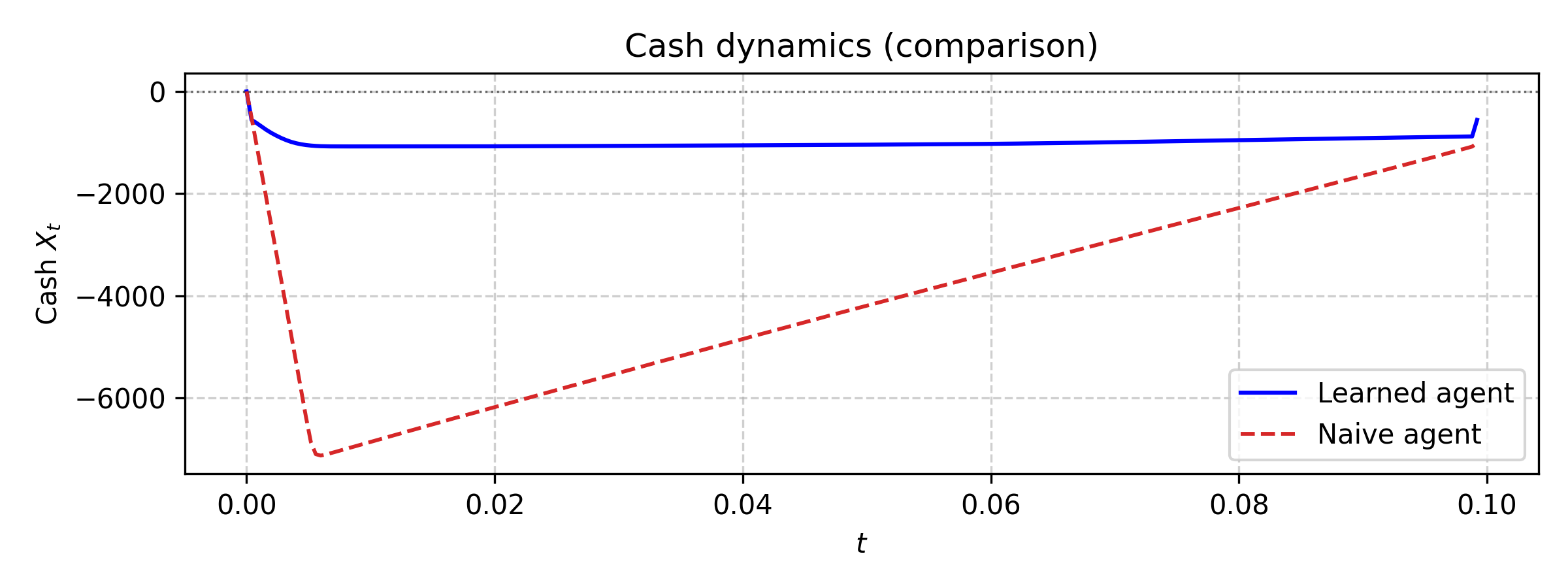}
			\caption{Average cash trajectory of the market maker ($I_0 = -100$ and low liquidity).}
			\label{FIG:market_making_linear_pnl_liqu_constraints}
		\end{subfigure}
		\caption{Quoting strategy and cash dynamics under liquidity constraints.}
	\end{figure}
	
\begin{figure}[H]
		\centering
		\includegraphics[width=0.6\textwidth]{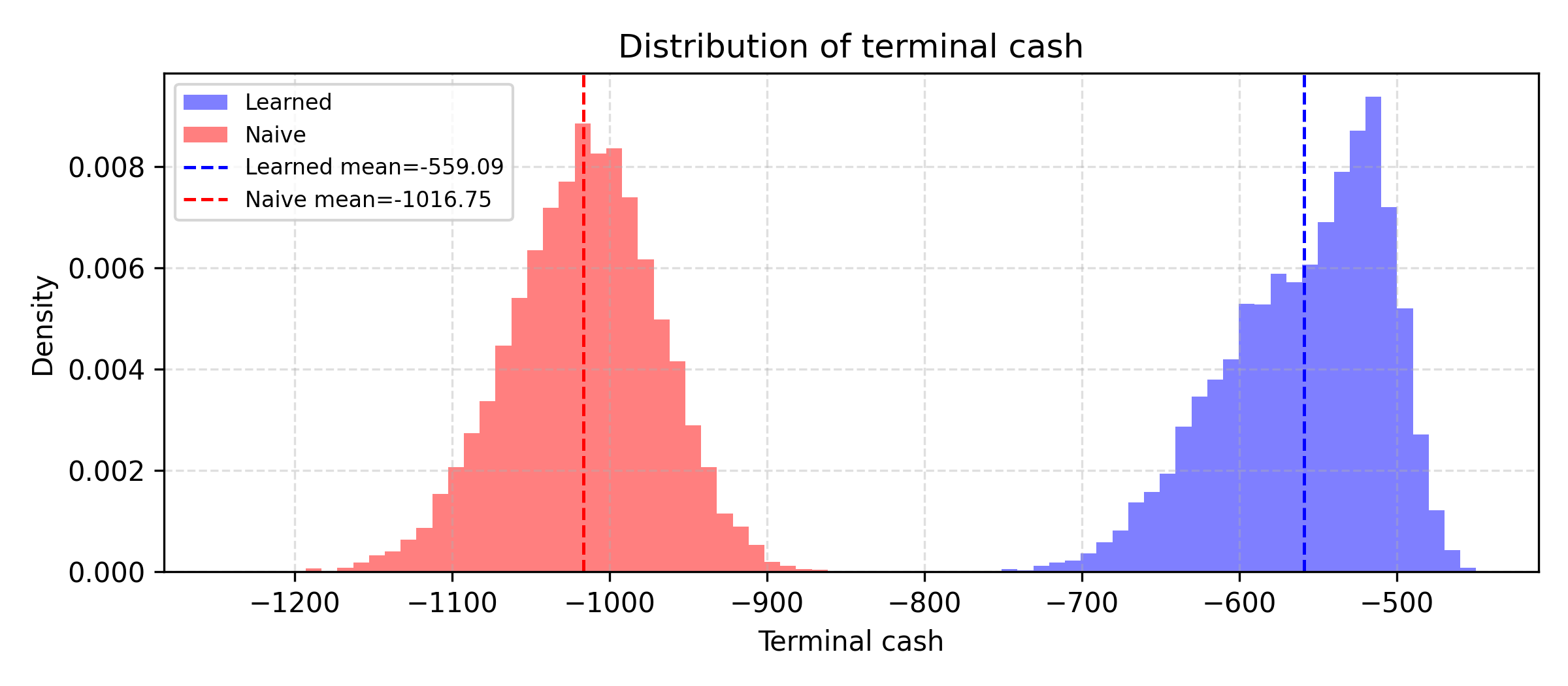} 
		\caption{Terminal cash distribution under liquidity constraints.}
		\label{FIG:market_making_linear_impact_liqu_constraints_terminal_pnl_dist}
	\end{figure}

	As the horizon progresses, the agent gradually reduces its short option position while maintaining a controlled level of exposure. 
	Rather than enforcing full inventory neutrality, it converges toward an optimal compromise between residual risk and execution costs, reflecting the limited liquidity of the market.
	The cash dynamics, displayed in Figure~\ref{FIG:market_making_linear_pnl_liqu_constraints}, highlight the cost of operating under liquidity constraints. 
	The learned agent initially incurs losses due to partial hedging and adverse execution conditions. 
	However, by prioritizing inventory unwinding through adapted quotes and limiting costly underlying trades, it stabilizes its cash and significantly improves its terminal outcome.
	\begin{figure}[H]
		\centering
		\begin{subfigure}[t]{0.49\textwidth}
			\centering
			\includegraphics[width=1\textwidth]{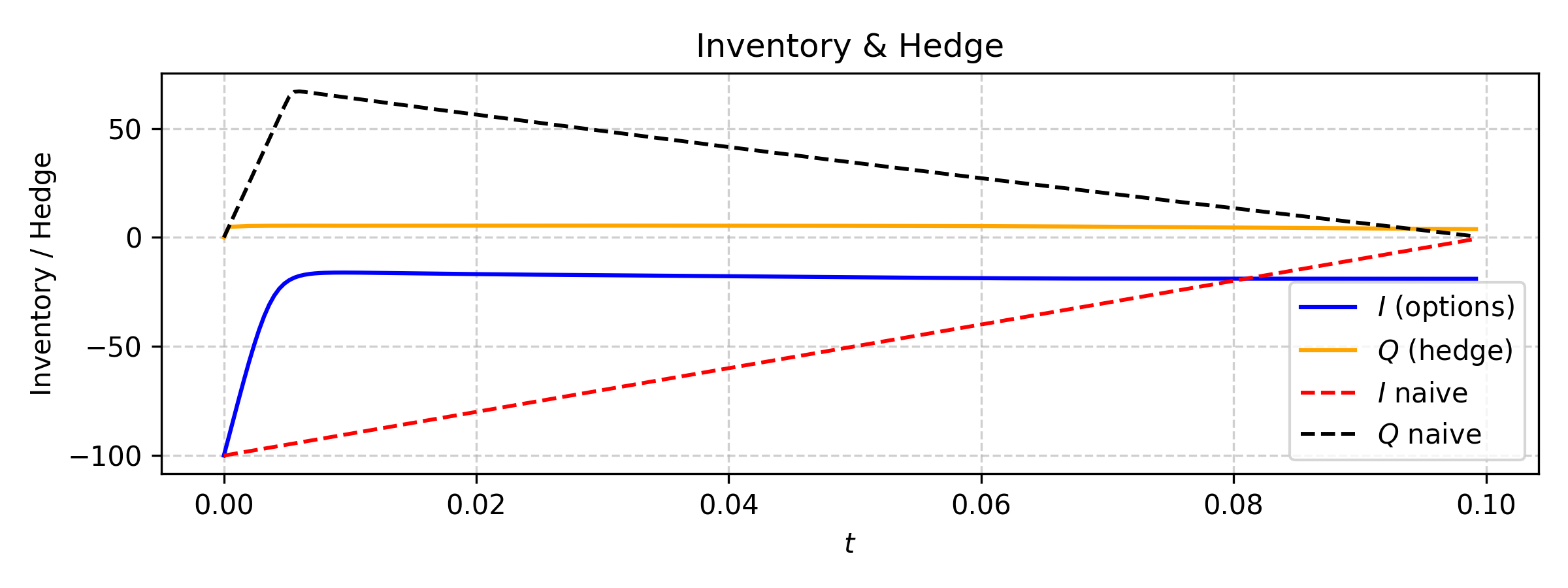} 
			\caption{Average option inventory and hedging position over time ($I_0 = -100$ and low liquidity).}
			\label{FIG:market_making_inventories_liqu_constraints}
		\end{subfigure}
		\hfill
		\begin{subfigure}[t]{0.49\textwidth}
			\centering
			\includegraphics[width=1\textwidth]{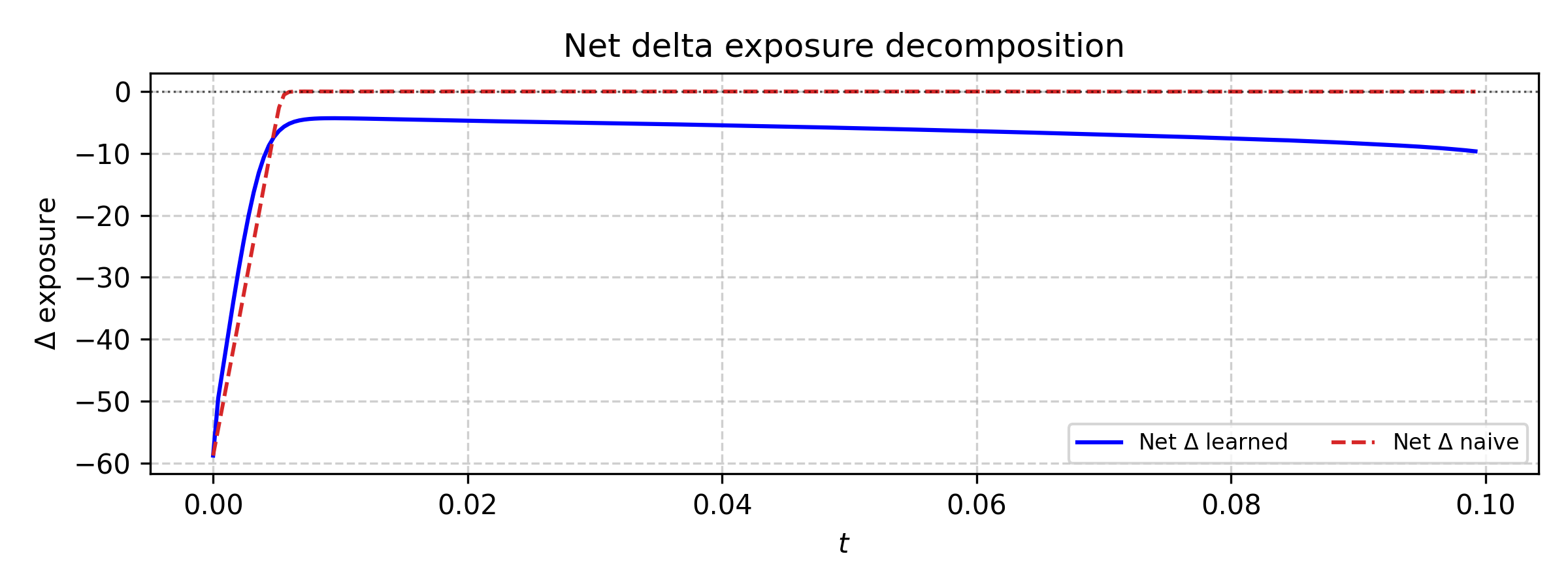} 
			\caption{Average net delta decomposition over time ($I_0 = -100$ and low liquidity).}
			\label{FIG:market_making_linear_impact_liqu_constraints_net_delta}
		\end{subfigure}
		\caption{Inventory and net delta dynamics under liquidity constraints.}
	\end{figure}
	In contrast, the naive strategy attempts to hedge aggressively despite the low liquidity, enforcing rapid delta neutralization at the expense of execution costs. 
	Figure~\ref{FIG:market_making_linear_impact_liqu_constraints_net_delta} shows that this leads to excessive trading in the underlying and inefficient risk management.
	By comparison, the learned policy maintains a controlled residual exposure, reflecting an optimal compromise between hedging precision and execution costs under limited liquidity. \\
	Turning to performance, Table~\ref{tab:perf_liquidity_constraints} and Figure~\ref{FIG:market_making_linear_impact_liqu_constraints_terminal_pnl_dist} show a substantial improvement of the learned policy over the naive benchmark across all metrics.
	\begin{table}[H]
		\centering
		\begin{tabular}{l|ccccc}
			\toprule
			Agent & Mean & Std & Median & Q5 & Q95 \\
			\midrule
			Learned & $-559.09$ & $50.84$ & $-550.19$ & $-652.56$ & $-491.38$ \\
			Naive   & $-1016.75$ & $47.55$ & $-1014.87$ & $-1097.62$ & $-941.55$ \\
			\bottomrule
		\end{tabular}
		\caption{Terminal cash distribution for the liquidity constraints configuration.}
		\label{tab:perf_liquidity_constraints}
	\end{table}
	The learned strategy achieves significantly higher terminal P\&L and lower dispersion. 
	This performance gain is driven by its ability to jointly control inventory and execution under liquidity constraints, rather than enforcing immediate risk neutrality.
	In particular, the learned agent accepts a temporary and controlled exposure, while using asymmetric quoting to reduce its option inventory at lower cost. 
	The naive policy, by contrast, follows a rigid hedging approach, leading to excessive trading costs and a significantly worse terminal distribution.
	This experiment highlights a key structural insight: under liquidity constraints, optimal market making departs from classical delta-hedging principles and instead relies on a joint control of inventory, execution, and residual risk.
	
	\subsubsection{Training stability and clipping diagnostics}
	
	In addition to performance, we assess the stability of the learning procedure and the role of constraint enforcement during training. Table~\ref{tab:training_diagnostics} summarizes the corresponding diagnostics across all configurations.
	The hedge penalty at convergence increases with the magnitude of the initial inventory imbalance, reflecting the residual cost of managing large exposures under market impact. In every case, the incentive penalty vanishes at convergence, confirming that the learned policies consistently meet the activity requirement.
	\begin{table}[H]
		\centering
		\small
		\begin{tabular}{lccc}
			\toprule
			Configuration & Hedge pen. & \makecell{Clip rate\\(peak)} & \makecell{Clip rate\\(avg. last 10 ep.)}  \\
			\midrule
			$I_0 = -100$           & $17.15$  & $0.16\%$  & $0.02\%$ \\
			$I_0 = -75$            & $10.46$  & $0.12\%$  & $0.02\%$ \\
			$I_0 = -50$            & $5.59$   & $0\%$     & $0\%$ \\
			$I_0 = -25$            & $2.11$   & $0\%$     & $0\%$ \\
			$I_0 = 0$ (baseline)   & $<0.01$  & $0\%$     & $0\%$ \\
			$I_0 = +25$            & $1.79$   & $0\%$     & $0\%$ \\
			$I_0 = +50$            & $6.05$   & $0\%$     & $0\%$ \\
			$I_0 = +75$            & $5.07$   & $0.12\%$  & $0.03\%$ \\
			$I_0 = +100$           & $10.38$  & $0.19\%$  & $0.12\%$ \\
			\midrule
			Shifted intensities    & $<0.01$  & $25.26\%$ & $0\%$ \\
			Low liquidity          & $39.06$  & $6.59\%$  & $<0.01\%$ \\
			\bottomrule
		\end{tabular}
		\caption{Training diagnostics: converged hedge penalty and clipping rates across configurations.}
		\label{tab:training_diagnostics}
	\end{table}
	The clipping rate remains at zero or below $0.2\%$ for all standard-liquidity configurations, confirming that the soft-clipping 
	mechanism does not affect the converged policies. In the shifted-intensities setting, a transient spike reaching $25\%$ occurs 
	during epochs $60$--$115$, where the agent explores aggressive hedging strategies before learning to control its inventory through quoting; the rate drops to zero well before convergence. In the low-liquidity regime, the peak rate of $6.6\%$ reflects the difficulty of hedging a large short position under severe depth constraints, but rapidly declines to less than $0.01\%$ at convergence, confirming that the learned policy adapts to the liquidity environment.
	
	\printbibliography[heading=bibintoc, title={Bibliography}]
	
	\appendix
	\section{Technical proofs}
	\label{APPENDIX:proofs}
	
	\subsection{Stability and moment estimates}
	\label{SUBSEC:stability_and_moments_estimaties}
	We establish stability and moment bounds for the Hawkes intensities and the state 
	variables, which are then propagated to the price, spread, and resilience dynamics. 
	These estimates provide the probabilistic control required to prove boundedness of the value function.
	
	\begin{Lemma}[Moment bounds for the option inventory]
		\label{LEM:market_making_inventory_moments}
		Consider the inventory process defined in Equation~\eqref{EQN:market_maker_option_inventory_dynamics} satisfying Assumption~\ref{ASS:market_making_option_intensities}.
		Then, there exist constants $C_{1,T}^{(I)}, C_{2,T}^{(I)}>0$, depending only on $T$ and on  $(\overline\lambda^b, \overline\lambda^a)$, such that
		\begin{equation}
			\label{eq:inventory_moments_bound}
			\sup_{t\in[0,T]}\E\big[|I_t|\big] \le C_{1,T}^{(I)}\big(1 + |i|\big), \qquad \sup_{t\in[0,T]}\E\big[|I_t|^2\big] \le C_{2,T}^{(I)}\big(1+i^2\big).
		\end{equation}
	\end{Lemma}
	
	\begin{proof}
		By \eqref{EQN:market_maker_option_inventory_dynamics}, the inventory satisfies $I_t=i+N_t^{b}-N_t^{a}$ for $t\in[0,T]$. Let $\Lambda_t^{b}$ and $\Lambda_t^{a}$ be the compensators of the counting processes and define the compensated martingales $M_t^{b}:=N_t^{b}-\Lambda_t^{b}$ and $M_t^{a}:=N_t^{a}-\Lambda_t^{a}$. 
		Assumption~\ref{ASS:market_making_option_intensities} gives $0\le\lambda^{b}\le\overline\lambda^{b}$ and $0\le\lambda^{a}\le\overline\lambda^{a}$, hence $0\le\Lambda_t^{b}\le \overline\lambda^{b}t$ and $0\le\Lambda_t^{a}\le \overline\lambda^{a}t$. For the first moment,
		\begin{align*}
			\E[|I_t|] &\le |i|+\E[N_t^{a}]+\E[N_t^{b}] \\
			&=|i|+\E[\Lambda_t^{a}]+\E[\Lambda_t^{b}] \\
			&\le |i|+t(\overline\lambda^{a}+\overline\lambda^{b}),
		\end{align*}
		and taking $\sup_{t\in[0,T]}$ yields
		\begin{equation*}
			\sup_{t\in[0,T]}\E[|I_t|]\le |i|+T(\overline\lambda^{a}+\overline\lambda^{b}) \le C_{1,T}^{(I)}(1+|i|).
		\end{equation*}
		
		For the second moment, $(a+b+c)^2\le 3(a^2+b^2+c^2)$ gives $|I_t|^2\le 3 i^2 + 3(N_t^{a})^2 + 3(N_t^{b})^2$.
		Writing $N_t^{k}=M_t^{k}+\Lambda_t^{k}$ ($k\in\{a,b\}$) and using $(a+b)^2\le 2a^2+2b^2$,
		\begin{equation*}
			\E[(N_t^{k})^2]\le 2 \E[(M_t^{k})^2]+2 \E[(\Lambda_t^{k})^2].
		\end{equation*}
		Since $\langle M^{k}\rangle_t=\Lambda_t^{k}$, the martingale isometry yields $\E[(M_t^{k})^2]=\E[\Lambda_t^{k}]\le \overline\lambda^{k}t$, while $\E[(\Lambda_t^{k})^2]\le (\overline\lambda^{k}t)^2$. Therefore
		\begin{equation*}
			\E[(N_t^{k})^2]\le 2 \overline\lambda^{k}t+2 (\overline\lambda^{k}t)^2,
		\end{equation*}
		and consequently
		\begin{equation*}
			\sup_{t\in[0,T]}\E[|I_t|^2]
			\le 3 i^2 + 6(\overline\lambda^{a}+\overline\lambda^{b}) T
			+ 6\big((\overline\lambda^{a})^2+(\overline\lambda^{b})^2\big) T^2
			\le C_{2,T}^{(I)}(1+i^2),
		\end{equation*}
		This proves \eqref{eq:inventory_moments_bound} and the square–integrability of $(I_t)_{t\in[0,T]}$.
	\end{proof}
	
	\begin{Lemma}[Moment bounds for Hawkes processes with interventions]
		\label{LEM:market_making_hawkes_bounds_with_intervention}
		Consider the intensity process $(\lambda_t^\pm)_{t\ge 0}$ defined in \eqref{EQN:market_making_full_intensity_bid}--\eqref{EQN:market_making_full_intensity_ask}, with parameters satisfying Assumption~\ref{ASS:market_making_hawkes_stability}, and let $(N_t^\pm)_{t\ge 0}$ be the associated counting process. 
		Assume the intervention processes $(H_t^\pm)_{t\ge 0}$ are adapted, nondecreasing, c\`adl\`ag,
		$H_0^\pm=0$, and admit bounded first moment on $[0,T]$, i.e.
		\begin{equation*}
			\overline H_T^\pm = \sup_{t\in[0,T]} \E[H_t^\pm] < \infty.
		\end{equation*}
		Then there exists two constants $C_{1,T}^{(H^\pm)}, \widetilde C_{1,T}^{(H^\pm)} > 0$, depending only on $T$, the model parameters $(\theta, \mu, \kappa)$, and the intervention moments $\overline H_T^\pm$, such that
		\begin{equation}
			\label{EQN:hawkes_bound_with_intervention}
			\sup_{t \in [0, T]} \E[\lambda_t^\pm] \le C_{1,T}^{(H^\pm)}\big(1+\lambda_0\big), \quad \sup_{t\in[0,T]} \E[N_t^\pm] \le \widetilde C_{1,T}^{(H^\pm)} \big(1 + \lambda_0^\pm\big).
		\end{equation}
		
		If in addition, we suppose that
		\begin{equation*}
			\overline H_{2, T}^{\pm} = \sup_{t\in[0,T]}\E[H_t^2]<\infty.
		\end{equation*}
		Then there exist constants $C_{2,T}^{(H^\pm)},\ \widetilde C_{2,T}^{(H^\pm)}>0$,
		depending only on $T$, the model parameters $(\theta,\mu,\kappa)$, and the intervention
		moments $\overline H_T^\pm, \overline H_{2, T}^{\pm}$, such that
		\begin{equation*}
			\sup_{t\in[0,T]}\E\!\big[(\lambda_t^\pm)^2\big]\ \le\ C_{2,T}^{(H^\pm)} \left(1+\lambda_0^2\right),
			\qquad
			\sup_{t\in[0,T]}\E\!\big[(N_t^\pm)^2\big]\ \le\ \widetilde C_{2,T}^{(H^\pm)} \left(1 + (\lambda_0^\pm)^2\right).
		\end{equation*}
	\end{Lemma}
	
	\begin{proof}
		For simplicity, we drop the $\pm$ superscripts, but the proof is valid for both cases.
		Let $M_t := N_t - \int_0^t \lambda_udu$ be the compensated martingale associated with $N$.
		From \eqref{EQN:market_making_full_intensity_bid}--\eqref{EQN:market_making_full_intensity_ask}, the intensity satisfies the linear SDE
		\begin{equation*}
			d\lambda_t = \theta\mu dt - (\theta-\kappa)\lambda_t dt + \kappa dM_t + \kappa dH_t,
		\end{equation*}
		with $\lambda_0\ge 0$. By variation of constants, taking expectations and noting that $\E\!\left[\int_0^t dM_u\right]=0$,
		\begin{equation*}
			\E[\lambda_t]
			= e^{-(\theta-\kappa)t}\lambda_0
			+ \frac{\theta\mu}{\theta-\kappa}\big(1-e^{-(\theta-\kappa)t}\big)
			+ \kappa\E\!\left[\int_0^t e^{-(\theta-\kappa)(t-u)}dH_u\right].
		\end{equation*}
		Thanks to Assumption~\ref{ASS:market_making_hawkes_stability} we have $\theta - \kappa > 0$ and this allows us to prove the first claim:
		\begin{align*}
			\Expect{\lambda_t} \leq \lambda_0 + \frac{\theta \mu}{\theta - \kappa} + \kappa \overline H_T \leq C_{1,T}^{(H)} (1 + \lambda_0).
		\end{align*}
		Furthermore, for $t\in[0,T]$,
		\begin{equation*}
			\E[N_t] =\int_0^t \E[\lambda_u]du \leq \frac{\lambda_0}{\theta-\kappa} + \frac{\theta\mu}{\theta-\kappa}t + \kappa \int_0^t \E\!\left[\int_0^u e^{-(\theta-\kappa)(u-s)}dH_s\right]du.
		\end{equation*}
		Applying Tonelli–Fubini for nonnegative integrands pathwise (since $H$ is nondecreasing),
		\begin{equation*}
			\int_0^t \left(\int_0^u e^{-(\theta-\kappa)(u-s)}dH_s\right)du
			= \int_0^t \left(\int_s^t e^{-(\theta-\kappa)(u-s)}du\right) dH_s \le \frac{H_t}{\theta-\kappa}.
		\end{equation*}
		Taking expectations, recalling that $\sup_{u\in[0,T]}\E[H_u] = \overline H_T$ and maximizing over $t\in[0,T]$,
		\begin{equation*}
			\sup_{t\in[0,T]}\E[N_t]
			\le \frac{\theta\mu}{\theta-\kappa}T + \frac{\lambda_0}{\theta-\kappa} + \frac{\kappa}{\theta-\kappa}\overline H_T
			\le \widetilde C_{1,T}^{(H)} \big(1+\lambda_0\big),
		\end{equation*}
		for some constant $\widetilde C_{1,T}^{(H)}$ depending only on $T$, $(\theta,\mu,\kappa)$, and the intervention
		moments $\overline H_T$.
		This proves \eqref{EQN:hawkes_bound_with_intervention}.
		For the square-integrability we start by calculating
		\begin{align*}
			\mathbb{E} \left[ \lambda_t^2 \right] &= \mathbb{E} \Bigg[ \lambda_0^2 e^{-2(\theta - \kappa)t} + \left( \theta \mu \Int{0}{t} e^{-(\theta - \kappa)(t - u)} du \right)^2 + 2 \lambda_0 \theta \mu \Int{0}{t} e^{-(\theta - \kappa)(2t - u)} du \\
			&\qquad + \left( \kappa \Int{0}{t} e^{-(\theta - \kappa)(t - u)} d M_u \right)^2 + \left( \kappa \Int{0}{t} e^{-(\theta - \kappa)(t - u)} d H_u \right)^2 \\
			&\qquad + 2 \lambda_0 \kappa \Int{0}{t} e^{-(\theta - \kappa)(2t-u)} dH_u + 2 \left( \theta \mu \Int{0}{t} e^{-(\theta - \kappa)(t - u)} du \right) \left( \kappa \Int{0}{t} e^{-(\theta - \kappa)(t - u)} dH_u \right) \\
			&\qquad +2\left( \kappa \Int{0}{t} e^{-(\theta - \kappa)(t - u)} d M_u \right)\left( \kappa \Int{0}{t} e^{-(\theta - \kappa)(t - u)} d H_u \right)  \Bigg].
		\end{align*}
		Applying Cauchy--Schwarz and using $\theta-\kappa>0$ (Assumption~\ref{ASS:market_making_hawkes_stability}), which implies $e^{-(\theta-\kappa)(\cdot)}\le 1$, we obtain
		\begin{align*}
			\mathbb{E} \left[ \lambda_t^2 \right] &\leq \lambda_0^2 + (\theta \mu)^2 t^2 + 2\lambda_0 \theta \mu t + \kappa^2 \Expect{M_t^2} + \kappa^2 \Expect{H_t^2} + 2\lambda_0 \kappa \Expect{H_t} \\
			&\qquad + 2\theta \mu \kappa t \Expect{H_t} + 2 \kappa^2 \sqrt{\Expect{M_t^2}} \sqrt{\Expect{H_t^2}}.
		\end{align*}
		We now absorb the cross terms by Young’s inequality and Jensen:
		\begin{equation*}
			2\lambda_0\theta\mu t\le \lambda_0^2+(\theta\mu t)^2,\qquad
			2\lambda_0\kappa \E[H_t]\le \lambda_0^2+\kappa^2(\E[H_t])^2\le \lambda_0^2+\kappa^2 \E[H_t^2],
		\end{equation*}
		and 
		\begin{equation*}
			2\theta\mu\kappa t \E[H_t]\le (\theta\mu t)^2+\kappa^2 \E[H_t^2],\qquad
			2\kappa^2\sqrt{\E[M_t^2]}\sqrt{\E[H_t^2]}\le \kappa^2 \E[M_t^2]+\kappa^2 \E[H_t^2].
		\end{equation*}
		Plugging these bounds above yields
		\begin{equation*}
			\mathbb{E}\!\big[\lambda_t^2\big]
			\le 3\lambda_0^2 + 3(\theta\mu)^2 t^2
			+ 2\kappa^2 \mathbb{E}[M_t^2]
			+ 4\kappa^2 \mathbb{E}[H_t^2].
		\end{equation*}
		Since $\mathbb{E}[M_t^2]=\mathbb{E}\!\big[\int_0^t \lambda_u du\big]=\mathbb{E}[N_t]$, the first-moment bound from~\eqref{EQN:hawkes_bound_with_intervention} gives $\sup_{t\le T}\mathbb{E}[M_t^2]\le \widetilde C_{1,T}^{(H)}(1+\lambda_0)$. 
		Together with $\sup_{t\le T}\mathbb{E}[H_t^2]\le \overline H_{2,T}<\infty$ and $t\le T$ we conclude that
		\begin{equation*}
			\sup_{t\in[0,T]}\mathbb{E}[\lambda_t^2] \le C_{2,T}^{(H)} (1+\lambda_0^2),
		\end{equation*}
		for some constant $C_{2,T}^{(H)}$ depending only on $T$, $(\theta,\mu,\kappa)$, and the intervention
		moments $\overline H_T,\overline H_{2,T}$.
		Finally, for the counting process we write $N_t = M_t + \Lambda_t$ with $\Lambda_t:=\int_0^t\lambda_u du$.
		Then $N_t^2\le 2M_t^2+2\Lambda_t^2$, hence
		\begin{equation*}
			\mathbb{E}[N_t^2]\le 2 \mathbb{E}[M_t^2]+2 \mathbb{E}[\Lambda_t^2].
		\end{equation*}
		By the martingale isometry, $\mathbb{E}[M_t^2]=\mathbb{E}[\Lambda_t]\le \widetilde C_{1,T}^{(H)}(1+\lambda_0)$,
		and by Cauchy--Schwarz,
		\begin{equation*}
			\mathbb{E}[\Lambda_t^2]=\mathbb{E}\!\Big[\Big(\int_0^t \lambda_u du\Big)^2\Big]
			\le t\int_0^t \mathbb{E}[\lambda_u^2] du
			\le t^2 \sup_{0\le u\le t}\mathbb{E}[\lambda_u^2]
			\le T^2 C_{2,T}^{(H)}(1+\lambda_0^2).
		\end{equation*}
		Therefore,
		\begin{equation*}
			\sup_{t\in[0,T]}\mathbb{E}[N_t^2]\ \le\ \widetilde C_{2,T}^{(H)} (1+\lambda_0^2),
		\end{equation*}
		for some finite constant $\widetilde C_{2,T}^{(H)}$ depending only on $T$, $(\theta,\mu,\kappa)$, and the intervention moments.
	\end{proof}
	
	Lemma~\ref{LEM:market_making_hawkes_bounds_with_intervention} provides uniform first and second-moment control of the Hawkes intensities and their counting processes, which are the sole source of randomness in the model. 
	Combined with the integrated representations of $P$, $D$, and $S$, these bounds yield uniform moment estimates for the full state on $[0,T]$, ensuring integrability of all terms and supporting the finiteness of the value function.
	
	\begin{Proposition}[Moment bounds for the state process]
		\label{PROP:market_making_state_moment_bounds}
		Let $(E_u)_{u \ge 0}:=(P_u,D_u,S_u,\lambda_u^-,\lambda_u^+)$ denote the market state process with
		initial condition $E_0=e:=(p,d,s,\lambda_0^-,\lambda_0^+)$.
		We assume Assumptions~\ref{ASS:market_making_right_truncated_depth} and
		\ref{ASS:market_making_hawkes_stability} hold.
		Assume moreover that the intervention processes $(H_t^\pm)_{t\ge0}$ are adapted, nondecreasing, càdlàg, satisfy $H_0^\pm=0$, and admit bounded first moments on $[0,T]$, i.e.
		\begin{equation*}
			\overline H_T^\pm = \sup_{t\in[0,T]}\E[H_t^\pm]<\infty.
		\end{equation*}
		Then there exists a finite constant $C_{1,T}^{(E)}$, depending only on $T$, the
		model parameters and on $\overline H_T^\pm$, such that,
		\begin{equation*}
			\sup_{u\in[0,T]}\E\!\left[|P_u|+|D_u|+S_u+\lambda_u^-+\lambda_u^+\right] \le C_{1,T}^{(E)}\bigl(1+\|e\|\bigr).
		\end{equation*}
		If, in addition,
		\begin{equation*}
			\overline H_{2,T}^\pm = \sup_{t\in[0,T]}\E\!\big[(H_t^\pm)^2\big] < \infty,
		\end{equation*}
		then there exists a finite constant $C_{2,T}^{(E)}$, depending only on $T$, the	model parameters and on $\overline H_T^\pm,\overline H_{2,T}^\pm$, such that,
		\begin{equation*}
			\sup_{u\in[0,T]}\E\!\left[|P_u|^2+|D_u|^2+S_u^2+(\lambda_u^-)^2+(\lambda_u^+)^2\right]
			\le C_{2,T}^{(E)}\bigl(1+\|e\|^2\bigr).
		\end{equation*}
	\end{Proposition}
	
	\begin{proof}
		Set $U_\star:=\max\{U_A,U_B\}$. By  Assumption~\ref{ASS:market_making_right_truncated_depth} and  Proposition~\ref{PROP:market_making_convex_concave_cost_functions},
		the price impacts per single trade/impulse are uniformly bounded by $U_\star$. \\
		\textbf{Resilient component.}
		Using the integrated representation of Equation~\eqref{EQN:resilient_impact_dynamics}, 
		obtained via variation of constants, and noting that each jump has magnitude at most $U_\star$,
		\begin{equation}
			\label{EQN:market_making_bound_resilient_part}
			|D_u| \le e^{-ru}|d| + \frac{1-\eta}{2} \int_0^u e^{-r(u-s)} d\Big(N_s^+ + N_s^- + H_s^+ + H_s^-\Big) U_\star .
		\end{equation}
		Taking expectations and dropping the exponential yields
		\begin{equation*}
			\E[|D_u|] \le |d| + \frac{1-\eta}{2}U_\star
			\Big(\E[N_u^+]+\E[N_u^-]+\E[H_u^+]+\E[H_u^-]\Big).
		\end{equation*}
		\textbf{Spread.}
		Similarly, from \eqref{EQN:spread_dynamics},
		\begin{equation}
			\label{EQN:market_making_bound_spread}
			S_u \le \delta + e^{-\rho u}(s-\delta) +
			\int_0^u e^{-\rho(u-s)} d\Big(N_s^+ + N_s^- + H_s^+ + H_s^-\Big) U_\star,
		\end{equation}
		hence
		\begin{equation*}
			\E[S_u] \le \delta + s + U_\star\Big(\E[N_u^+]+\E[N_u^-]+\E[H_u^+]+\E[H_u^-]\Big).
		\end{equation*}
		\textbf{Mid-price.}
		From Equation~\ref{EQN:mid_price_dynamics}, $P$ is the sum of its permanent jump part and $D$.
		Using again the bound $U_\star$ for each jump,
		\begin{equation}
			\label{EQN:market_making_bound_mid_price}
			|P_u| \le |p| + |D_u| + \frac{\eta}{2}U_\star\Big(N_u^+ + N_u^- + H_u^+ + H_u^-\Big),
		\end{equation}
		and thus
		\begin{equation*}
			\E[|P_u|] \le |p| + \E[|D_u|] + \frac{\eta}{2}U_\star\Big(\E[N_u^+]+\E[N_u^-]+\E[H_u^+]+\E[H_u^-]\Big).
		\end{equation*}
		Combining the three displays for $\E[|D_u|]$, $\E[S_u]$, and $\E[|P_u|]$, inserting the bounds on $\E[H_u^\pm]$ and using bounds of $\E[N_u^\pm]$ and $\E[\lambda_u^\pm]$ obtained in Lemma~\ref{LEM:market_making_hawkes_bounds_with_intervention}, we obtain
		\begin{equation*}
			\sup_{u\in[0,T]} \E\!\left[|P_u|+|D_u|+S_u+\lambda_u^-+\lambda_u^+\right]
			\le C_{1,T}^{(E)} \Big(1+|p|+|d|+s+\lambda_0^-+\lambda_0^+\Big)
			= C_{1,T}^{(E)} \big(1+\|e\|\big),
		\end{equation*}
		for a constant $C_{1,T}^{(E)} < \infty$ depending only on $T$, the model parameters and on $\sup_{u\in[0,T]}\E[H_u^\pm]$. This proves the first claim. \\
		Now, for the square integrability we use the integrated representations together with $e^{-r(\cdot)},e^{-\rho(\cdot)}\le 1$ and Young's inequality, we bound each component on $[0,T]$. \\ 
		\textbf{Resilient component.}
		From \eqref{EQN:market_making_bound_resilient_part} we obtain:
		\begin{equation*}
			| D_u |^2 \leq 2d^2 + 2 \left(\frac{1 - \eta}{2}\right)^2 U_\star^2 \left( \int_0^u d\Big(N_s^+ + N_s^- + H_s^+ + H_s^-\Big) \right)^2
		\end{equation*}
		Using $(x_1+x_2+x_3+x_4)^2 \le 4\sum_{i=1}^4 x_i^2$ and taking expectations,
		\begin{equation*}
			\E[|D_u|^2] \le 2d^2 + 8\!\left(\frac{1-\eta}{2}\right)^{\!2} U_\star^2
			\Big(\E[(N_u^+)^2]+\E[(N_u^-)^2]+\E[(H_u^+)^2]+\E[(H_u^-)^2]\Big).
		\end{equation*}
		\textbf{Spread.}
		From \eqref{EQN:market_making_bound_spread} and $s\ge \delta$,
		\begin{equation*}
			S_u^2 \leq 2s^2 + 2 U_\star^2 \left( \int_0^u d\Big(N_s^+ + N_s^- + H_s^+ + H_s^-\Big) \right)^2
		\end{equation*}
		and therefore
		\begin{equation*}
			\E[S_u^2] \le 2s^2 + 8U_\star^2 
			\Big(\E[(N_u^+)^2]+\E[(N_u^-)^2]+\E[(H_u^+)^2]+\E[(H_u^-)^2]\Big).
		\end{equation*}
		\textbf{Mid-price.}
		From \eqref{EQN:market_making_bound_mid_price},
		\begin{equation*}
			| P_u |^2 \leq 4p^2 + 4D_u^2 + 2\left(\frac{\eta}{2}\right)^2 U_\star^2 \left( \int_0^u d\Big(N_s^+ + N_s^- + H_s^+ + H_s^-\Big) \right)^2
		\end{equation*}
		and taking expectations,
		\begin{equation*}
			\E[P_u^2] \le 4p^2 + 4 \E[|D_u|^2] + 8\!\left(\frac{\eta}{2}\right)^{\!2} U_\star^2 
			\Big(\E[(N_u^+)^2]+\E[(N_u^-)^2]+\E[(H_u^+)^2]+\E[(H_u^-)^2]\Big).
		\end{equation*}
		Finally, by Lemma~\ref{LEM:market_making_hawkes_bounds_with_intervention} we have the uniform bounds
		\begin{equation*}
			\sup_{u\le T}\E\!\big[(N_u^\pm)^2\big]\le \widetilde C_{2,T}^{(H^\pm)}\big(1+(\lambda_0^\pm)^2\big)
			\quad\text{and}\quad
			\sup_{u\le T}\E\!\big[(\lambda_u^\pm)^2\big]\le C_{2,T}^{(H^\pm)}\big(1+(\lambda_0^\pm)^2\big),
		\end{equation*}
		while by assumption $\sup_{u\le T}\E[(H_u^\pm)^2]\le \overline H_{2,T}^\pm<\infty$. 
		Plugging these estimates into the bounds obtained above for $D$, $S$, and $P$, and enlarging constants if necessary, we conclude that there exists a finite constant $C_{2,T}^{(E)}$ such that
		\begin{equation*}
			\sup_{u\in[0,T]}\E\!\Big[|P_u|^2+|D_u|^2+S_u^2+(\lambda_u^-)^2+(\lambda_u^+)^2\Big]	\le C_{2,T}^{(E)}\Big(1+p^2+d^2+s^2+(\lambda_0^-)^2+(\lambda_0^+)^2\Big).
		\end{equation*}
		This completes the proof of the second–order bound.
	\end{proof}
	
	\subsection{Proofs of the value-function bounds}
	
	Building on the previous results, we now establish the main statements that ensure the well-posedness of the value function. We begin by proving a global lower bound.
	
	\begin{Proposition}[Lower bound for the value function]
		\label{PROP:market_making_lower_bound_value_function}
		For all $(t,q,i,e) \in[0,T] \times\R\times\Z\times(\R^+)^5$, there exists a constant $C_T^{(-)}>0$, depending only on $T$ and on the model parameters, such that
		\begin{equation*}
			v(t,q,i,e)\ \ge\ -C_T^{(-)}\bigl(1+q^2+i^2+\|e\|^2\bigr).
		\end{equation*}
	\end{Proposition}
	
	\begin{proof}[Proof of Proposition~\ref{PROP:market_making_lower_bound_value_function}]
		Fix an admissible strategy with no impulses in the underlying and constant option quotes $\beta \equiv 0$ and $\alpha \ge 0$. Then $Q_u\equiv q$ for all $u\in[t,T]$, and the order–flow term $\int_t^T[\alpha\lambda^a-\beta\lambda^b]du= \alpha\int_t^T\lambda^adu$ is nonnegative. From \eqref{EQN:market_making_value_function} and \eqref{EQN:liquidation_function},
		\begin{equation*}
			v(t,q,i,e)\ \ge\ \E\Big[-\int_t^T(g+h)(u,Q_u,I_u,E_u)du\ +\ I_T\varphi(P_T)\ -\ P_A\Big(P_T+\tfrac12 S_T,\ |q|\Big)\Big].
		\end{equation*}
		By Assumption~\ref{ASS:market_making_penalties}, $g$ and $h$ satisfy a quadratic growth bound; since $Q_u\equiv q$,
		\begin{equation*}
			-\int_t^T(g+h)du \ge -C\int_t^T\big(1+q^2+I_u^2+\|E_u\|^2\big)du.
		\end{equation*}
		Lemma~\ref{LEM:market_making_inventory_moments} and Proposition~\ref{PROP:market_making_state_moment_bounds} with $H_u^\pm \equiv 0$ for all $u \in [t, T]$, yield
		\begin{equation}\label{eq:penalty-lb}
			\E\Big[-\int_t^T(g+h)du\Big] \ge -\left( C_T^{(g)} + C_T^{(h)} \right)\big(1+q^2+i^2+\|e\|^2\big).
		\end{equation}
		The payoff has linear growth, so $I_T\varphi(P_T)\ge -C^{(\varphi)}\big(|I_T|+|I_T||P_T|\big)$. Taking expectations and applying Cauchy–-Schwarz together with the same moment bounds gives
		\begin{equation}\label{eq:option-lb}
			\E\big[I_T\varphi(P_T)\big]\ \ge\ -C_T\big(1+i^2+\|e\|^2\big).
		\end{equation}
		For the execution cost, the integral representation \eqref{EQN:market_making_PA_rep} and the bound $\Phi_A^{-1}\le U_A$ imply
		\begin{equation*}
			P_A\left( P_T+\tfrac12 S_T,|q|\right) \le\ \left( P_T+\tfrac12 S_T \right)|q|+U_A|q|.
		\end{equation*}
		Taking expectations, and since $|q|$ is deterministic under the chosen strategy yields
		\begin{equation*}
			\E\Big[P_A\big(P_T+\tfrac12 S_T,|q|\big)\Big]\ \le\ |q|\E\big[P_T+\tfrac12 S_T\big]\ +\ U_A|q|.
		\end{equation*}
		Proposition~\ref{PROP:market_making_state_moment_bounds} gives $\E[P_T+S_T/2]\le C_T(1+\|e\|)$, hence
		\begin{equation}\label{eq:PA-ub}
			\E\Big[P_A\big(P_T+\tfrac12 S_T,|q|\big)\Big]\ \le\ C_T|q|(1+\|e\|)\ +\ U_A|q|
			\ \le\ C_T\big(1+q^2+\|e\|^2\big),
		\end{equation}
		where the last inequality uses $|q|\le 1+q^2$.
		
		Combining \eqref{eq:penalty-lb}, \eqref{eq:option-lb}, and \eqref{eq:PA-ub} yields
		\begin{equation*}
			v(t,q,i,e)\ \ge\ -C_T^{(-)}\big(1+q^2+i^2+\|e\|^2\big),
		\end{equation*}
		for a constant $C_T^{(-)}$ depending only on $T$ and the model parameters.
	\end{proof}
	
	We now establish an upper bound for the value function, which implies its finiteness.
	
	\begin{Proposition}[Quadratic-growth of the value function]
		\label{PROP:market_making_upper_bound_value_function}
		The value function defined in Equation~\ref{EQN:market_making_value_function} is well-defined for all  
		\begin{equation*}
			(t, q, i, e) \in [0,T]\times \R\times \Z \times (\R^+)^5.
		\end{equation*}
		Moreover, there exists a generic constant $C_T^{(+)} > 0$, depending only on the time horizon $T$, model parameters, such that
		\begin{equation}
			v(t, q, i, e) \leq C_T^{(+)} \left(1 + |i|^2 + \|e\|^2\right)
		\end{equation}
	\end{Proposition}
	
	\begin{proof}[Proof of Proposition~\ref{PROP:market_making_upper_bound_value_function}]
		Let us recall that any admissible strategy ensures that:
		\begin{equation*}
			\sup_{t\in[0,T]}\E[H_t^\pm]<\infty, \qquad \sup_{t\in[0,T]}\E[H_t^2]<\infty,
		\end{equation*}
		hence we can apply all results of  Section~\ref{SUBSEC:stability_and_moments_estimaties}.
		
		By definition of the value function and positivity of $g$, $h$ $P_A$ and $c$:
		\begin{align*}
			v\left(t, q, i, e\right) &\leq \underset{\gamma \in \Ac}{\sup} \mathbb{E} \Bigg[ \Int{t}{T} \left[ \alpha_u \lambda^a - \beta_u \lambda^b \right](u, \alpha_u, \beta_u, E_u) du \\ 
			&\quad + \Sum{i=1}{+\infty} \Ind{\nu_i \in [t, T]} \left(P_B(P_{\nu_i} - S_{\nu_i}/2, \xi_i^-) \right) + L(T, Q_T, I_T, P_T, S_T) \Bigg].
		\end{align*}
		We first derive a strategy–independent bound for the option order–flow term. Under Assumptions~\ref{ASS:market_making_ref_price_linear_growth} and \ref{ASS:market_making_option_intensities}, for any admissible $(\alpha,\beta)$ and $t\le T$,
		\begin{align*}
			\E \Big[ \Int{t}{T}(\alpha_u\lambda^a-\beta_u\lambda^b)&(u,\alpha_u,\beta_u,E_u)du \Big] = \E \Big[\Int{t}{T}(\lambda^a(u,\alpha_u,\beta_u,E_u)(\alpha_u - b(u, E_u))  \\
			&\quad - \lambda^b(u,\alpha_u,\beta_u,E_u)(\beta_u - b(u, E_u)) + b(u, E_u) (\lambda^a - \lambda^b)(u,\alpha_u,\beta_u,E_u) )du \Big] \\
			&\leq \E \Big[ \Int{t}{T} ( C^{(\lambda)} (1 + \| E_u \|)  + b(u, E_u) (\lambda^a - \lambda^b)(u,\alpha_u,\beta_u,E_u) ) du \Big] \\
			&\leq \E \Big[  \Int{t}{T} (C^{(\lambda)} (1 + \| E_u \|)  +  (\overline \lambda^a + \overline \lambda^b ) C^{(b)} (1 + \| E_u \|) ) du \Big] 
		\end{align*}
		By the state first-moment estimate of Proposition~\ref{PROP:market_making_state_moment_bounds}, we conclude that there exists $C_T^{(\lambda, b)} > 0$ such that
		\begin{equation*}
			\sup_{\alpha, \beta} \Expect{\Int{t}{T}(\alpha_u\lambda^a-\beta_u\lambda^b)(u,\alpha_u,\beta_u,E_u)du} \leq C_T^{(\lambda, b)} ( 1 + \|e\|).
		\end{equation*}
		Now, for the liquidation function. By Lemma~\ref{LEM:market_making_positivity_quotes}, $P_T-\tfrac12 S_T\ge0$ a.s.
		Using finite depth (Assumption~\ref{ASS:market_making_right_truncated_depth}) and
		Lemma~\ref{PROP:market_making_convex_concave_cost_functions}, we have
		\begin{equation*}
			0 \le P_B\!\left(P_T-\tfrac12 S_T,\min\{Q_T,\Phi_B(B_T)\}\right)
			\le \Phi_B(U_B) (P_T-\tfrac12 S_T).
		\end{equation*}
		Moreover, $-P_A(\cdot)\le 0$, so for an upper bound we may drop the ask term. With the linear growth of $\varphi$,
		\begin{equation*}
			\big|I_T \varphi(P_T)\big|\le C^{(\varphi)}\big(|I_T|+|I_T| |P_T|\big).
		\end{equation*}
		Therefore
		\begin{align*}
			\mathbb{E}\big[L(T,Q_T,I_T,P_T,S_T)\big]
			&\le \Phi_B(U_B) \mathbb{E}[P_T-\tfrac12 S_T]
			+ C^{(\varphi)} \mathbb{E}\!\big[ |I_T|+|I_T| |P_T| \big].
		\end{align*}
		Apply the elementary inequalities $|x|\le \tfrac12(1+x^2)$ and Young’s inequality $|xy|\le \tfrac12(x^2+y^2)$ to obtain
		\begin{equation*}
			\mathbb{E}[P_T-\tfrac12 S_T] \le \tfrac12 + \tfrac12 \mathbb{E}[P_T^2] + \tfrac14 \mathbb{E}[S_T^2],
			\qquad
			\mathbb{E}\!\big[ |I_T|+|I_T| |P_T| \big]
			\le \tfrac12 + \mathbb{E}[I_T^2] + \tfrac12 \mathbb{E}[P_T^2].
		\end{equation*}
		Hence
		\begin{equation*}
			\mathbb{E}\big[L(T,Q_T,I_T,P_T,S_T)\big]
			\le K \Big(1 + \mathbb{E}[I_T^2] + \mathbb{E}[P_T^2] + \mathbb{E}[S_T^2]\Big).
		\end{equation*}
		By Lemma~\ref{LEM:market_making_inventory_moments} and by Proposition~\ref{PROP:market_making_state_moment_bounds}, enlarging constants we conclude that
		\begin{equation*}
			\mathbb{E}\big[L(T,Q_T,I_T,P_T,S_T)\big] \le C_T^{(L)} \big(1 + i^2 + \|e\|^2\big).
		\end{equation*}
		Finally as $P_B(b, q) \leq b q$:
		\begin{align*}
			v\left(t, q, i, e\right) &\leq \underset{\gamma \in \Ac}{\sup} \mathbb{E} \Bigg[ \Int{t}{T} \left[ \alpha_u \lambda^a - \beta_u \lambda^b \right](u, \alpha_u, \beta_u, E_u) du \\ 
			&\quad + \Sum{i=1}{+\infty} \Ind{\nu_i \in [t, T]} \left(P_B(P_{\nu_i} - S_{\nu_i}/2, \xi_i^-) \right) + L(T, Q_T, I_T, P_T, S_T) \Bigg] \\
			&\leq \underset{\gamma \in \Ac}{\sup} \mathbb{E} \Bigg[ C_T^{(\lambda, b)} ( 1 + \|e\|) + \Phi_B(U_B) \Sum{i=1}{+\infty} \Ind{\nu_i \in [t, T]} P_{\nu_i} + C_T^{(L)} \big(1 + i^2 + \|e\|^2\big) \Bigg] \\
			&\leq C_T^{(\lambda, b)} ( 1 + \|e\|) + \Phi_B(U_B) \Expect{H_t^-} \left( 1 + \|e\| \right) + C_T^{(L)} \big(1 + i^2 + \|e\|^2\big)
		\end{align*}
		Hence:
		\begin{equation*}
			v(t, q, i, e) \leq C_T^{(+)} \left( 1 + i^2 + \|e\|^2 \right)
		\end{equation*}
		Which shows that $v(t, q, i, e) < +\infty$. Recalling that, by Proposition~\ref{PROP:market_making_lower_bound_value_function} we also have $v(t, q, i, e) > -\infty$ proves that the value function is well defined. 
	\end{proof}
	
\end{document}